\documentclass[lettersize,journal]{IEEEtran}
\usepackage{bm}
\usepackage{amsmath}
\usepackage{amsthm}
\usepackage{extarrows}
\usepackage{amssymb}
\usepackage{graphicx}
\usepackage{mathrsfs}
\usepackage{color}
\usepackage[dvipsnames]{xcolor}
\usepackage{enumerate}
\usepackage{bookmark}
\graphicspath{{figure}}
\usepackage{setspace}
\usepackage{multirow}
\usepackage{subfigure}
\usepackage{float}
\usepackage{booktabs}
\usepackage{algorithm}
\usepackage{subcaption}
\usepackage{caption}
\usepackage{makecell}
\usepackage{algpseudocode}
\usepackage{cite}
\usepackage{framed} 
\usepackage{scalerel}
\usepackage{tikz}
\usetikzlibrary{matrix}
\hypersetup{hidelinks}

\newcommand{\diag}{\mr{diag}}

\newcommand{\mr}{\mathrm} 

\newcommand{\BE}{\begin{equation}}
\newcommand{\EE}{\end{equation}}
\newcommand{\BS}{\begin{subequations}}
\newcommand{\ES}{\end{subequations}}
\renewcommand{\bf}{\bm}

\newcommand{\tabincell}[2]{\begin{tabular}{@{}#1@{}}#2\end{tabular}}
\newtheorem{theorem}{Theorem}
\newtheorem{proposition}{Proposition}
\newtheorem{assumption}{Assumption}
\newtheorem{definition}{Definition}

\newtheorem{lemma}{Lemma}
\newtheorem{corollary}{Corollary} 
\newtheorem{conjecture}{Conjecture}

\renewcommand{\arraystretch}{1.4}
\allowdisplaybreaks \allowdisplaybreaks[1]

\graphicspath{{Figures/}}
\begin{document}

\title{Random Multiplexing} 

\author{Lei~Liu, Yuhao~Chi, Shunqi~Huang, and Zhaoyang~Zhang
\thanks{Lei Liu and Zhaoyang Zhang are with the Zhejiang Provincial Key Laboratory of Multi-Modal Communication Networks and Intelligent Information Processing, College of Information Science and Electronic Engineering, Zhejiang University, Hangzhou 310007, China, and Lei Liu is also with the State Key Laboratory of Integrated Services Networks, Xidian University, Xi’an 710071, China (e-mail: \{lei\_liu, ning\_ming\}@zju.edu.cn).}
\thanks{Yuhao Chi is with the State Key Laboratory of Integrated Services Networks, School of Telecommunications Engineering, Xidian University, Xi'an, 710071, China (e-mail: yhchi@xidian.edu.cn).}
\thanks{Shunqi Huang is with the School of Information Science, Japan Advanced Institute of Science and Technology (JAIST), Nomi 923-1292, Japan (e-mail: shunqi.huang@jaist.ac.jp).}
\thanks{The source code of this work is publicly available at
\href{https://github.com/LeiLiu-s-Lab/Random-Multiplexing}{\textcolor{blue}{GitHub}}.}
} 
 
\maketitle

\begin{abstract} 
As wireless communication applications evolve from traditional multipath environments to high-mobility scenarios like unmanned aerial vehicles, multiplexing techniques have advanced accordingly. Traditional single-carrier frequency-domain equalization (SC-FDE) and orthogonal frequency-division multiplexing (OFDM) have given way to emerging orthogonal time-frequency space (OTFS) and affine frequency-division multiplexing (AFDM). These approaches exploit specific channel structures—e.g., Toeplitz-structured multipath channel matrix for OFDM and SC-FDE or doubly selective channels for OTFS and AFDM—to diagonalize or sparsify the effective channel, thereby enabling low-complexity detection. However, their reliance on these structures significantly limits their robustness in dynamic, real-world environments. To address these challenges, this paper studies a random multiplexing technique that is decoupled from the physical channels, thereby enabling its application to arbitrary norm-bounded and spectrally convergent channel matrices. Random multiplexing achieves statistical fading-channel ergodicity for transmitted signals by constructing an equivalent input-isotropic channel matrix in the random transform domain. It guarantees the asymptotic replica MAP bit-error rate (BER) optimality of AMP-type detectors for linear systems with arbitrary norm-bounded, spectrally convergent channel matrices and signaling configurations, under the unique fixed point assumption. A low-complexity cross-domain memory AMP (CD-MAMP) detector is considered for random multiplexing systems, leveraging the sparsity of the time-domain channel and the input isotropy of the equivalent channel. Optimal power allocations are derived to minimize the replica MAP BER and maximize the replica constrained capacity of random multiplexing systems, respectively. The optimal coding principle and replica constrained-capacity optimality of CD-MAMP detector are investigated for random multiplexing systems. Additionally, the versatility of random multiplexing in diverse wireless applications is explored. Numerical results are presented to validate the theoretical findings.
\end{abstract}


\section{Introduction} 
With the rapid development of wireless applications, wireless channels have become increasingly complex, driving the continuous evolution of wireless technology to ensure high-rate high-reliability communications. Orthogonal frequency division multiplexing (OFDM) is employed in 4G and 5G communication systems\cite{tse2005fundamentals}, which introduces a cyclic prefix to convert the multipath channels into equivalent cyclic Toeplitz matrices, ensuring parallel transmission of signals on orthogonal subcarriers in the frequency domain while avoiding inter-symbol interference (ISI). Nevertheless, in emerging wireless applications, such as high-mobility communications (e.g., high-speed railways~\cite{HighSpeed}, low Earth orbit satellites~\cite{LEO}, and unmanned aerial vehicles~\cite{UAVs}, underwater communications~\cite{Underwater}, and integrated sensing and communication~\cite{ISCA_Survey}), device mobility causes the wireless channels to be affected by the additional Doppler effect, resulting in doubly selective channels. This compromises the subcarrier orthogonality in OFDM, leading to a substantial decline in performance.

To address this issue, orthogonal time frequency space (OTFS)~\cite{OTFS1} and affine frequency division multiplexing (AFDM) \cite{AFDM} techniques have been proposed in recent years to constructing sparse or nearly sparse equivalent channel matrices to suppress the ISI. Although considerable progress has been made, the development of low-complexity and high-reliability detection algorithms capable of achieving maximum \emph{a posteriori} (MAP) BER performance for OTFS and AFDM systems remains an open challenge. State-of-the-art low-complexity and replica-optimal signal recovery algorithms, such as approximate message passing (AMP) \cite{AMP2009}, orthogonal AMP (OAMP) \cite{MaAcess2017}, vector AMP (VAMP) \cite{Rangan2019TIT}, and memory AMP (MAMP) \cite{LeiMAMP}, offer potential solutions for these scenarios. However, their theoretical analyses and coding designs are generally predicated on assumptions of independent and identically distributed (IID) or right-unitarily invariant channel matrices \cite{AMP2009,MaAcess2017,Rangan2019TIT,LeiMAMP,liu2021capacity,LeiOptOAMP,Code_MAMP}. In practical applications, channel distributions often deviate from these assumptions, resulting in performance degradation. 

In a nutshell, existing multiplexing, signal detection, and coding design techniques are heavily dependent on specific  assumptions regarding channel matrix structures, such as cyclic Toeplitz matrices of static multipath channels in OFDM\cite{tse2005fundamentals} and single-carrier frequency-domain equalization (SC-FDE) \cite{SC-FDE}, doubly selective channels for OTFS\cite{OTFS1} and AFDM\cite{AFDM}, which substantially restricts their applicability to  complex and dynamic real-world wireless channels. 

\subsection{Multiplexing Techniques}
In static multipath channels, the design principle of OFDM \cite{tse2005fundamentals} and SC-FDE \cite{SC-FDE} is to avoid ISI by converting the equivalent frequency-domain channels into multiple parallel single-input single-output (SISO) channels through orthogonalization, thereby enabling signal recovery using minimum mean-square error (MMSE) detection. However, the orthogonality no longer holds in time-varying doubly-selective channels. To address this issue, OTFS multiplexes information symbols over the delay-Doppler domain, enabling the receiver to mitigate ISI by separating signals based on time delay and Doppler shift\cite{OTFS1}. Independently, in AFDM, inverse discrete affine Fourier transform (IDAFT) modulates information symbols in the time-frequency domain, effectively separating signals at the path in terms of time delay and Doppler shift\cite{AFDM}. Both AFDM and OTFS result in sparse or nearly sparse equivalent channel matrices, which facilitates the design of low-complexity signal detection and channel estimation algorithms. Nevertheless, OTFS and AFDM can achieve near-full and full diversity, respectively, when utilizing maximum likelihood (ML) detectors, albeit with prohibitively high complexity. However, guaranteeing the full diversity advantage with practical low-complexity detection algorithms remains an open challenge. To tackle these challenges, the interleave frequency division multiplexing (IFDM) has been proposed \cite{IFDM} (see also earlier closely related works in \cite{EST, EST-EQ}). It utilizes an inverse fast Fourier transform (IFFT) cascaded with a random interleaver to construct an equivalent dense and input-isotropic channel matrix. On this basis, IFDM can effectively cope with the doubly selective fading in time-varying multipath channels, ensuring reliable signal transmission. However, in IFDM, the interleaved-IDFT matrix’s deterministic structure precludes channel-adaptive randomization—a necessity for 6G’s dynamic environments. 

\subsection{Power Allocation Schemes}
In wireless communication systems, when channel state information (CSI) is available to the transceiver, power allocation optimization is typically used to minimize symbol (or bit) error probability or maximize the transmission rate. The traditional approach involves decomposing the wireless channel into multiple parallel single-input single-output (SISO) channels using singular value decomposition (SVD), followed by power allocation to each sub-channel to maximize capacity. For Gaussian signaling, Gaussian waterfilling is employed\cite[Sec. 10.4]{Cover1990}, while mercury waterfilling is applied for discrete signaling~\cite{Lozano2006}. Although the SVD-based power allocation scheme is simple in design, it requires independent channel encoders and decoders with different rates to be assigned to each sub-channel, resulting in a highly complex practical system implementation. To address this difficulty, a power allocation solution based on a specific detector design is explored~\cite{Caire2002,Caire2004,MIMO_rate,EST-EQ}. In \cite{Caire2002, Caire2004}, for Gaussian multiple access channels, power allocation is optimized for iterative multiuser detectors via linear programming to improve BER performance. For MIMO channels, in \cite{MIMO_rate}, a power allocation optimization scheme is proposed based on iterative linear minimum mean-square error (LMMSE) detection and discrete signaling, aiming to achieve the maximum achievable rate of the MIMO system. On this basis, a single code is designed that achieves a significantly higher rate than the multi-code design for the SVD-based waterfilling scheme, while maintaining comparable encoding and decoding complexity for both schemes. Additionally, an optimal power allocation scheme is proposed for a \emph{belief propagation} (BP)-based iterative receiver, an early version of the OAMP/VAMP\footnote{In this paper, OAMP and VAMP are collectively referred to as OAMP/VAMP due to their underlying equivalence.} receiver \cite{MaAcess2017, Rangan2019TIT}, in energy-spreading-transform (EST)-based MIMO systems \cite{EST-EQ}. The aim is to achieve the target BER while minimizing transmit power. Meanwhile, it is noted that EST can be considered a special case of the random multiplexing investigated in this paper. However, SVD-based optimal power allocation in \cite{Cover1990,Lozano2006} does not guarantee optimal MAP BER (see Fig.~\ref{fig:RM_PA}), only capacity optimal, but relies on multiple AWGN capacity-achieving channel codes with different rates. Moreover, detector-based optimal power allocation in \cite{MIMO_rate, EST-EQ} requires high-complexity LMMSE detection, which is challenging to implement in large-scale systems due to its prohibitive complexity.
\begin{figure*}[t]\vspace{-0.4cm}
    \centering
    \includegraphics[width =1\textwidth]{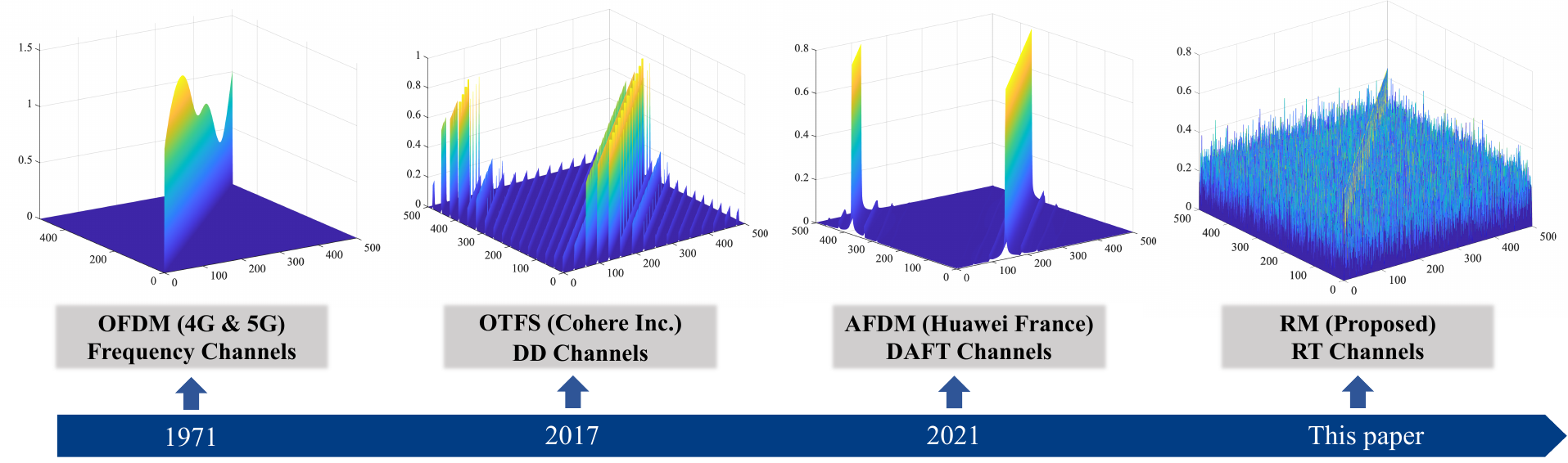}
    \caption{A comparison of the equivalent  channel matrices $\bf{H}_{{\rm{output}}\times{\rm{input}}}^{\rm{eqv}}=\bf{\Xi
    }^{\rm{H}}\bf{H}\bf{\Xi}$ of OFDM\cite{tse2005fundamentals} ($\bf{\Xi}=\bf{F}^{\rm{H}}$) in static multipath channels, and OTFS ($\bf{\Xi}={\bf{F}}^{\rm{H}}\otimes \bf{I}$)\cite{OTFS1}, AFDM ($\bf{\Xi}=\bf{\Lambda}_{c_1}^{\mr{H}}\bf{F}^{\mr{H}}_{{N}}\ \bf{\Lambda}_{c_2}^{\mr{H}}$) \cite{AFDM} and the random multiplexing (RM) in Definition~\ref{Def:RUP} in time-varying doubly-selective channels, where $\bf{H}$ denotes the time-domain channels,  $\bf{\Xi}$ the multiplexing matrix, and $\bf{\Lambda}_{c_i}\triangleq\mr{diag}(e^{-j2\pi c_in^2}, n=0, \cdots\!, N-1)$, $i=1,2$.}
    \label{fig:RMcomp}
\end{figure*}

\subsection{Detection Algorithms}
AMP-type algorithms have been extensively studied for signal recovery in linear systems, owing to their low complexity and high efficiency in high-dimensional settings. In particular, the low-complexity AMP algorithm has been shown to be Bayes-optimal via scalar state evolution (SE) analysis\cite{reeves2019replica,barbier2020mutual}, but it easily diverges when applied to non-independent and identically distributed (IID) measurement matrices. To overcome this limitation, a convolutional AMP was proposed for general rotationally invariant matrices\cite{CAMP}, but it suffers from slow convergence and instability under high condition numbers. Alternatively, a long-memory AMP has been shown to be effective for right-unitarily invariant matrices~\cite{fan2022approximate}. Another line of work is OAMP/VAMP, whose Bayes optimality under right-unitarily invariant matrices has been established via replica methods \cite{Kabashima2006,Tulino2013}, and later proven under specific spectral constraints \cite{zhoufanTIT,Barbier2018ISIT}. However, their reliance on high-complexity LMMSE detectors limits scalability. To address this, MAMP replaces LMMSE with a low-complexity memory matched filter (MF), retaining Bayes optimality. Independently, a warm-start conjugate gradient VAMP (WS-CG-VAMP) was proposed in \cite{WS-CG-VAMP}, leveraging conjugate gradient methods for similar complexity reduction. The convergence of OAMP/VAMP and MAMP with optimized damping has been established in \cite{LeiMAMP, LMOAMP, liu2022sufficient}.  Moreover, AMP-type algorithms have been shown to exhibit universal behavior across a wide range of random matrices, including IID, right-unitarily invariant, sign/permutation-invariant, certain Wigner-resolvent, and delocalized orthogonal ensembles~\cite{Rishabh2024,wang2024universality}.

More recently, AMP-type algorithms have been applied to signal recovery in multiplexing systems. For instance, a cross-domain (CD) OAMP/VAMP, inspired by the turbo compressed sensing algorithm~\cite{TurboCS}, was proposed for OTFS~\cite{OTFS-OAMP}. Additionally, a delay-Doppler OAMP/VAMP (DD-OAMP/VAMP) was introduced in \cite{OAMP/VAMP}. However, these methods struggle to exploit the sparsity of channel matrices in the time or delay-Doppler domain because of the matrix multiplication and inversion required in LMMSE detection. To this end, a low-complexity DD-MAMP detector has been proposed in\cite{MAMPOTFSconf}, incorporating a memory MF to leverage the channel sparsity in delay-Doppler domain. Nevertheless, it neglects the greater sparsity inherent in time-domain channels, leaving a promising avenue unexplored. Besides, the Bayesian optimality of widely used AMP-type algorithms~\cite{AMP2009, reeves2019replica, barbier2020mutual, CAMP, fan2022approximate, MaAcess2017, Rangan2019TIT, Kabashima2006,Tulino2013, zhoufanTIT,Barbier2018ISIT, LeiMAMP, LMOAMP, liu2022sufficient, WS-CG-VAMP, semi-random} relies on specific assumptions on the channel matrices, such as IID and right-unitarily invariant  matrices. However, in real-world wireless applications, the channel matrices, including the equivalent channel matrices of common multiplexing schemes such as SC-FDE, OFDM, OTFS, and AFDM, generally do not meet these assumptions required for AMP-type algorithms to achieve optimal performance. Consequently, AMP-type detectors suffer from significant performance degradation, as shown in Fig.~\ref{fig:RM_withoutmd} and Fig.~\ref{fig:RM_diffm}. 

\subsection{Achievable Rate Analysis and Channel Coding Design}
Channel coding design is crucial for ensuring reliable communication and approaching the channel capacity in wireless systems. Most modern coding designs are aimed at classical binary erasure channels and binary-input additive white Gaussian noise (AWGN) channels\cite{ryan2009channel}, such as  Turbo codes \cite{Turbo_Codes}, low-density parity-check (LDPC) codes \cite{LDPC_TIT}, spatially coupled LDPC codes \cite{SC-LDPC}, Polar codes \cite{Polar_Code}, sparse regression codes \cite{SR_codes0,SR_code1,Ramji_SRcode}. It should be noted that all of these well-designed channel codes focus on single-input-single-output (SISO) channels but are suboptimal in linear systems~\cite{liu2021capacity,LeiOptOAMP,Code_MAMP}.

To address this challenge, existing literature explored the channel coding optimization for joint AMP-type detection and decoding in linear systems\cite{liu2021capacity,LeiOptOAMP,Code_MAMP}. In~\cite{liu2021capacity}, the potential constrained capacity optimality of the AMP receiver for coded linear systems with arbitrary input and IID channel matrices is proven \cite{reeves2019replica, barbier2020mutual}, under the assumption that the AMP receiver’s SE is correct and has a unique fixed point. The potential capacity optimality of OAMP/VAMP for right-unitarily invariant matrices and the corresponding optimal coding principle are shown in \cite{LeiOptOAMP}. To address OAMP/VAMP’s high complexity, the replica capacity-optimal MAMP receiver and its coding principle are established in \cite{Code_MAMP}. Note that when practical wireless channels do not satisfy the IID or right-unitarily invariant conditions, the conclusions of the above studies no longer hold, presenting an ongoing challenge.

\subsection{Contributions}
To address the aforementioned challenges, we study a random multiplexing framework that unifies and extends the principles of IFDM \cite{IFDM} and EST \cite{EST,EST-EQ}. This random multiplexing framework employs a random matrix for multiplexing, allowing it to \emph{be applied to arbitrary norm-bounded and spectrally convergent channel matrices while ensuring that the equivalent matrices belong to the universality class $\mathscr{U}$ \cite{Rishabh2024}.} Unlike conventional constellation modulation that focuses on mapping symbol-level information bits to complex constellation points, the random multiplexing studied in this paper focuses on the mapping of high-dimensional complex symbol vectors. Furthermore, random multiplexing can be regarded as a general terminology for existing precoding techniques that reshape the equivalent channel distribution to satisfy the universality class condition, even in practical wireless scenarios. Therefore, random multiplexing preserves the performance limits of linear systems, including the replica MAP BER and constrained channel capacity. As shown in Fig.~\ref{fig:RMcomp}, the  equivalent channel matrices of random multiplexing is more densely stable compared to existing OFDM\cite{tse2005fundamentals}, OTFS\cite{OTFS1}, and AFDM\cite{AFDM}, making it easier to ensure reliable signal transmission. To avoid the high complexity of signal detection on equivalent dense matrices, a cross-domain MAMP detection has been considered, which leverages both the sparsity of the time-domain channels and the input isotropy of the equivalent channel matrices. Meanwhile, when CSI is available at the transceiver, power allocation optimization strategies are proposed for random multiplexing systems with arbitrary norm-bounded and spectrally convergent channel matrices, as well as arbitrary discrete signaling, aiming to achieve the replica MAP BER and maximize the constrained capacity, respectively. On this basis, the optimal coding principle and replica capacity optimality of random multiplexing systems are investigated. The main contributions of this paper are summarized as follows:
\begin{enumerate}
    \item This paper studies a random multiplexing technique that performs a unitary random transformation on the discrete signal vector prior to transmission. The transformation operates independently of both the signal vector and the channel matrix, in contrast to conventional multiplexing schemes that inherently couple the multiplexing matrix with the channel characteristics. This decoupling in random multiplexing simplifies system design and greatly enhances flexibility. Unlike conventional multiplexing schemes that rely on channel sparsification to enable low-complexity detection, random multiplexing establishes an equivalent fully dense and input-isotropic channel matrix. This ensures that every signal symbol homogeneously undergoes sufficient statistical channel fading, thereby enhancing overall system performance. Moreover, random multiplexing preserves both the replica MAP BER and the replica constrained capacity of linear systems. The random multiplexing framework includes two key types of random matrices: (i) Haar distributed matrices, and (ii) permutation-invariant matrices, ensuring both the theoretical validity and asymptotic optimality (achieving replica MAP-BER performance) for AMP detection.
    \item Based on the state evolution analysis for linear systems, we demonstrate that the power allocation to minimize the replica MAP BER is a bilevel problem (with a concave maximization inner problem), and the power allocation to maximize the replica constrained capacity is a concave maximization problem. Efficient algorithms for solving the optimal power allocations are provided. Compared with conventional Gaussian/mercury waterfilling \cite{Lozano2006,Cover1990}, the proposed scheme achieves a lower MAP BER. In contrast to the BER-minimizing power allocation based on high-complexity iterative LMMSE detector \cite{MIMO_rate,EST-EQ}, this paper develops power allocation for low-complexity CD-MAMP receivers. Furthermore, we extend our analysis beyond BER minimization to investigate the replica constrained capacity-optimal power allocation.
    \item The versatility of random multiplexing in various wireless applications is explored, including its low-complexity random transform, higher spectral efficiency, adaptability to other multiplexing schemes, and support for compressed and spread random multiplexing.
    \item Numerical results show that in correlated time-varying multipath MIMO channels, random multiplexing can achieve BER and block-error rate (BLER) performance gains of up to $2\sim 10$ dB compared to existing schemes (e.g., OFDM/OTFS/AFDM with well-designed SISO codes), under both uniform and optimized power allocation, as well as optimal coding. 
\end{enumerate}
Part of the results in this paper has been published in\cite{RMISIT}. In this paper, we present additional key properties of random multiplexing, detailed proofs, a discussion of potential applications, the design of optimal codes, achievable rate analysis, and further numerical results.

\subsection{Related Works}
\subsubsection{Comparison with Existing Multiplexing Schemes}
Existing OFDM\cite{tse2005fundamentals}, SC-FDE\cite{SC-FDE}, OTFS\cite{OTFS1}, and AFDM\cite{AFDM} rely on constructing diagonalized or sparse equivalent channel matrices for static or time-varying multipath channels to mitigate ISI at the receiver, thereby facilitating signal recovery. However, due to the non-orthogonal structure of the equivalent channel matrices, state-of-the-art (SOTA) low-complexity signal detection methods suffer from significant performance degradation (see Fig. \ref{fig:RM_diffm}). In contrast, EST \cite{EST, EST-EQ} and IFDM \cite{IFDM} enhance signal detection performance by constructing an equivalent input-isotropic channel matrix  through interleaved FFT/IFFT, allowing the signal to undergo sufficient statistical channel fading. This paper explores random multiplexing based on a general random transform, which generalizes both IFDM and EST. Additionally, the theoretical optimality of random multiplexing—namely, replica MAP-BER and replica constrained capacity—are provided.  Moreover, random multiplexing offers a generalized framework that decouples the multiplexing matrix from the channel matrix, going beyond the conventional channel-coupled  OFDM\cite{tse2005fundamentals}, OTFS\cite{OTFS1},  AFDM\cite{AFDM}, and IFDM \cite{IFDM}.

\subsubsection{Connection to the Universality Class $\mathscr{U}$ in \cite{Rishabh2024}} 
The universal behavior of AMP-type algorithms is established in \cite{Rishabh2024} for a broad class of universality class $\mathscr{U}$, including linear transformations of IID matrices, right-unitarily invariant matrices, permutation-invariant matrices, among others. A critical limitation, however, arises for permutation-invariant matrices: The analysis in \cite{Rishabh2024} requires channel matrices $\bf{A}$ to satisfy a restrictive structural condition—specifically, their singular value decompositions must obey $\bf{A} = \bf{U}_{A}\bf{\Sigma}_{A}\bf{V}^{\rm H}_{A}$ with $\bf{V}_{A}=\bf{I}$ (i.e., the right singular vectors form the identity matrix). In this paper, we overcome this limitation by extending the applicability of permutation-invariant matrices to more general off-diagonal-sum constrained channel matrices, a broader class that subsumes the earlier restrictive case. Furthermore, we conjecture that these results may generalize to arbitrary norm-bounded and spectrally convergent channel matrices, suggesting a potentially universal framework beyond currently verifiable cases.

\subsubsection{Comparison with Existing Detection Algorithms} 
State-of-the-art AMP-type signal recovery algorithms (e.g., OAMP\cite{MaAcess2017}, VAMP\cite{Rangan2019TIT}, and MAMP\cite{LeiMAMP}) have been studied and applied to existing multiplexing systems, including CD/DD-OAMP/VAMP~\cite{OTFS-OAMP,OAMP/VAMP} and DD-MAMP\cite{MAMPOTFSconf}. However, these algorithms require the channel matrix to exhibit a specific right-unitary invariance property. This requirement is typically unmet by practical communication channels, resulting in significant performance degradation (see Fig. \ref{fig:RM_withoutmd}). Random multiplexing ensures that the equivalent channel matrix consistently satisfies the input isotropy, regardless of the distribution of the original channel. This guarantees the validity of the AMP-type detection algorithms. Specifically, a low-complexity CD-MAMP detector is considered for random multiplexing systems, which is extended from the specific interleaved-IDFT matrix applicable in IFDM \cite{IFDM} and interleaved block-sparse transform (IBST) matrix\cite{IBST} to the more general unitary random transformation matrices, including but not limited to Haar distributed matrices and permutation-
invariant matrices  $\bf{\Xi}_{\rm PI} \equiv \bf{\Pi U D}$, where $\bm{D} = \diag\big\{[\mr{e}^{\mr{i}\theta_1}, \cdots\!, \mr{e}^{\mr{i}\theta_N}]\big\}$ is a uniformly random phase matrix with $\theta_{1:N} \overset{\mr{i.i.d.}}{\sim} \mathsf{Unif}\big\{[0, 2\pi)\big\}$, $\bf{\Pi}$ is a uniformly random permutation matrix independent of $\bm{D}$, and $\bf{U}$ is a delocalized deterministic unitary matrix (e.g., DFT, Hadamard–Walsh, DCT) satisfying $\|\bf{U}\|_{\max}\lesssim N^{-1/2+\epsilon}$ for any $\epsilon>0$.  By exploiting the isotropy of the transform-domain channel, CD-MAMP preserves the desirable properties, including rigorous state evolution analysis, replica Bayes optimality, and low-complexity implementation. Furthermore, CD-MAMP introduces a cross-domain processing mechanism that leverages the time-domain channel sparsity and fast transformation, lowering the complexity to $O(\mathcal{K}N + N\log N)$, where $\mathcal{K}$ is the the number of non-zero elements per row in time-domain channel matrix $\bf{H}$, without sacrificing performance. These make CD-MAMP a competitive solution for signal detection in random multiplexing systems, delivering both  replica Bayes-optimal performance and computational efficiency.

\subsubsection{Distinct from Existing Power Allocation Methods} 
The conventional power allocation methods use SVD channel decomposition followed by water filling of the power\cite{Cover1990,Lozano2006}, but they overlook the need for multiple capacity-achieving codes to achieve the channel capacity, which is practically prohibitive. Meanwhile, the state-of-the-art detector-based power allocation methods in \cite{MIMO_rate, EST-EQ} are constrained by the high-complexity iterative receivers involving LMMSE estimation and limited to the replica MAP BER minimization. To this end, leveraging the advantages of random multiplexing, this paper proposes optimal power allocation strategies for low-complexity CD-MAMP detectors, aiming to minimize the replica MAP BER and maximize the replica constrained capacity, respectively.

\subsubsection{Distinct from Well-Designed Channel Codes}
Conventional channel coding designs focus on overcoming the effects of channel noise to achieve the SISO AWGN channel capacity\cite{ryan2009channel}, but they remain strictly suboptimal in linear systems\cite{liu2021capacity,LeiOptOAMP,Code_MAMP}. While AMP-type detectors achieve capacity optimality under right-unitary invariant matrices \cite{liu2021capacity,LeiOptOAMP,Code_MAMP}, this assumption rarely holds in practical wireless systems. In contrast, random multiplexing ensures the input isotropy of the equivalent channel matrices, integrating the low-complexity CD-MAMP detection to effectively address this gap by enabling accurate performance characterization via state evolution, while maintaining replica Bayesian optimality. This establishes the optimal coding principle and replica-capacity optimality to be applied to linear systems with arbitrary norm-bounded and spectrally convergent channel matrices, as well as arbitrary signaling schemes.

\subsection{Notations}
Lowercase letters denote scalars and boldface lowercase letters denote vectors. $[\cdot]^{\rm{T}}$ and $[\cdot]^{\rm{H}}$ denote transpose and conjugate transpose operations, respectively. $\bm{I}$ denotes the identity matrix of appropriate size. $I(\bm{x},\bm{y})$ denotes mutual information between $\bm{x}$ and $\bm{y}$. $|\mathcal{S}|$ is the cardinality of a set $\mathcal{S}$. ${\rm tr}(\bf{A})$ and ${\mr{det}}(\bf{A})$ are the trace and the determinant of $\bf{A}$, respectively. $\|\bf{a}\|$ denotes the $\ell_2$-norm of a vector $\bf{a}$. $\mr{E}\{\cdot\}$ denotes the expectation over all random variables included in the brackets. $\mr{E}\{a|b\}$ denotes the conditional expectation of $a$ for given $b$. $\rm{mmse}\{a|b\}$ represents $\mr{E}\{|a-E\{a|b\}|^2|b\}$. $\mathcal{CN}(\bf{\mu},\bf{\Sigma})$ represents the circularly-symmetric Gaussian distribution with mean $\bf{\mu}$ and covariance matrix $\bf{\Sigma}$. $[N]$ denotes the set $\{1, \cdots\!, N\}$. $\|\bm{A}\|_{\max} \equiv \max_{i,j} |A_{i,j}|$ denotes the max norm, $\|\bm{A}\|_2$ denotes the spectral norm. $f^{-1}(\cdot)$ denotes the inverse function of $f(\cdot)$, and $1/f(\cdot)$ or $[f(\cdot)]^{-1}$ denotes the multiplicative inverse. We write $a_N \lesssim b_N$ to indicate that
\begin{align}
    a_N \leq C b_N, \quad \forall N \geq N_0,
\end{align}
for some constants $C > 0$ and $N_0 > 0$. 
\begin{definition}[Spectrally Convergent Matrix]\label{def:spectrally_convergent}
   A matrix $\bf{A}$ is said to be spectrally convergent if the empirical spectral distribution of $\bf{A}^{\rm H}\bf{A}$ converges to a compactly supported probability distribution. Specifically, for any fixed $k \in \mathbb{N}^*$, 
    \begin{align}
        {\rm tr}[(\bf{A}^{\rm H} \bf{A})^k] /N \rightarrow \int \lambda^k \mu( {\rm d}\lambda) \quad {\rm as}\; N \rightarrow \infty,
    \end{align}
    where $\mu$ denotes a compactly supported probability distribution on $[0, \infty)$.
\end{definition}

\subsection{Paper Outline}
This paper is organized as follows. Section \ref{Sec:preliminaries} presents the preliminaries. Section \ref{Sec:RP} introduces the random multiplexing and its advantages. In Section \ref{Sec:MAMP}, we consider CD-MAMP detection for random multiplexing. Section \ref{Sec:PA} focuses on optimal power allocation techniques, while Section \ref{Sec:code} introduces the optimal coding principle for random multiplexing. The extended applications of random multiplexing are explored in Section \ref{Sec:dis}, followed by numerical results in Section \ref{Sec:sim}.

\section{Preliminaries}\label{Sec:preliminaries}
This section introduces the norm-bounded spectrally convergent linear systems and the mercury/waterfilling scheme. 

\subsection{Norm-Bounded and Spectrally Convergent Linear Systems}  
Consider a linear system with power allocation: 
\BE\label{Eqn:linear_sys}
    \bm{y}  = \bm{HP} \bm{x} + \bm{n}={\bm A}\bm{x} + \bm{n}, 
\EE
where $\bm A=\bm{HP}$, $\bf{H}\in\mathbb{C}^{M \times N}$ is a channel matrix, $\bm{P} \in \mathbb{C}^{N \times N}$  a power allocation matrix subject to ${\rm tr}\{\bm{P}\bm{P}^{\rm H}\}=P_{\rm sum}$, $\bm{y} \in \mathbb{C}^{M \times 1}$  a vector of observations, $\bf{x} \in \mathbb{C}^{N \times 1}$  a vector to be estimated, and $\bm{n} \in \mathbb{C}^{M \times 1}$  a noise vector. When $\bf{P} = \bf{I}$, it reduces to the uniform power allocation. Without loss of generality, the average powers of $\bf{P}$ and $\bf{x}$ are normalized, i.e., $P_{\rm sum}=N$ and $\frac{1}{N}\|\bf{x}\|^2\overset{\rm a.s.}{=}1$, where $\overset{\rm a.s.}{=}$ denotes almost sure convergence. Let ${\rm snr} = \sigma^{-2}$ represent the transmit-signal-noise ratio (SNR). In addition, we impose the following assumptions on $\bf{A}$, $\bf{x}$, and $\bf{n}$.  

\begin{assumption} \label{ASS:Model}
We consider a large-scale linear system that $M,N\to\infty$ with a fixed $\delta=M/N$. The measurement matrix $\bf{A}$ is spectrally convergent (see Definition \ref{def:spectrally_convergent}) and has a bounded spectral norm satisfying $\|\bf{A}\|_2 \lesssim 1$. The entries of $\bf{x}$ are IID distributed, i.e., $\bf{x}\overset{\rm i.i.d.}{\sim} P_{x}$. The noise vector is IID Gaussian, i.e., $\bm{n}\sim\mathcal{CN}(\bf{0},\sigma^2\bf{I})$ for some $\sigma>0$.  In addition, $\bf{x}$,  $\bf{n}$ and $\bf{A}$ are mutually independent.
\end{assumption}

Note that, unlike AMP-type algorithms, which typically rely on specific isotropy assumptions for $\bf{A}$  (e.g., IID or unitary invariance), Assumption \ref{ASS:Model} significantly relaxes the constraints on the matrix $\bf{A}$.  Specifically, it allows $\bf{A}$ to be any norm-bounded and spectrally convergent matrices, thereby encompassing a broad spectrum of practical applications. These include, but are not limited to, inter-symbol interference (ISI) equalization, compressed sensing, multiple-input multiple-output (MIMO), multiple-access channels (MAC), random access, channel estimation, channel coding, etc\cite{CE-MIMO,NF-URA}. This generality makes the framework applicable to a wide range of real-world problems beyond the limitations of traditional AMP-based approaches.

\subsection{{MAP BER and Constrained Channel Capacity}}
Let $\mathcal{R}_{\bf{R}}(\cdot)$ denote the {R-transform} with a Hermitian matrix $\bm{R}$, as defined in \cite{Tulino2004}:
\BS\label{Eqn:R_transform}
\begin{align}
    \mathcal{R}_{\bf{R}}(w) =  \mathcal{S}^{-1}_{\bf{R}}(-w)-1/w,  
\end{align}
where $\mathcal{S}^{-1}_{\bf{R}}(\cdot)$ is the inverse function of the Stieltjes transform:
\begin{align}
    \mathcal{S}_{\bm{R}}(w) = \frac{1}{N}{\rm tr}\left\{(\bm{R}-w\bm{I})^{-1}\right\}. 
\end{align}\ES
The MMSE and BER metrics are widely recognized as important criteria in signal processing and communication. The theoretical MMSE and symbol-wise MAP BER of a linear system in \eqref{Eqn:linear_sys} under individually optimal detection can be predicted by the replica method as follows.
\begin{lemma}[Replica MMSE and MAP BER]\label{Lem:replic-MMSE}
Under Assumptions~\ref{ASS:Model}, the MMSE of the linear system in \eqref{Eqn:linear_sys} can be predicted by solving the fixed-point equation for $v^*$:
\begin{align}\label{Eqn:replicaMMSE}
    {\rm mmse}^{-1}(v^*) = {\sigma^{-2}} \cdot \mathcal{R}_{\bf{R}}\left( -\sigma^{-2}v^* \right),
\end{align}
which is derived from the replica method \cite{Kabashima2006,Tulino2013}. Correspondingly, the replica MAP BER is
\begin{align}
    {\rm BER}^* =Q_{\mathcal{S}}(\rho^*), \label{Eqn:replicaBER}
\end{align}
where ${\rho^*}={\rm mmse}^{-1}(v^*)$, ${\rm mmse}^{-1}(\cdot)$ denotes the inverse function of 
\BS\label{Eqn:mmsetol}
\begin{align}
    v = {\rm mmse}(\rho) \equiv {\mr{E}}\big\{|\hat{x}_{\rm post}-x|^2\big\} \label{Eqn:mmsedef}
\end{align}
with the {a posteriori} mean 
\begin{align}
    \hat{x}_{\mr{post}}={\mr{E}}\big\{x|\sqrt{\rho}x+z, x\sim P_{x}, z\sim \mathcal{CN}({0},1)\big\}, \label{Eqn:postmean}
\end{align}
\ES
and $Q_{\mathcal{S}}(\rho)$ denotes the MAP demodulation BER for $\sqrt{\rho}x+z$ under the signal constellation constraint $x\in \mathcal{S}$ (See Appendix~\ref{App:MAP_BER} for details).
\end{lemma} 
\textbf{Remark}: The replica method is heuristic, based on an unjustified exchange of limits and an unsupported replica symmetry assumption. If the fixed point is unique, it is conjectured that the replica MMSE equals the true MMSE for right unitarily-invariant matrices $\bm{A}$. Currently, this conjecture has only proven for IIDG matrices \cite{reeves2019replica,barbier2020mutual} and a specific sub-class of rotationally-invariant matrices \cite{Barbier2018ISIT, zhoufanTIT}.
\begin{assumption}[Unique Fixed Point]\label{Asp:SEfixed} 
Assume that the fixed-point equation in \eqref{Eqn:replicaMMSE}, derived from the replica method \cite{Kabashima2006,Tulino2013} for the linear systems in \eqref{Eqn:linear_sys}, has a unique solution $v^*$. 
\end{assumption}

Channel capacity is a common metric for performance limit in communication systems. The following lemma, proven in \cite{Guo2005}, establishes the connection between MMSE and the constrained capacity of a SISO AWGN channel given $P_x$.
\begin{lemma}[Scalar I-MMSE\cite{Guo2005}]\label{Lem:S-I-MMSE}
The constrained capacity of a SISO-AWGN channel is 
\BE\label{Eqn:C_mmse}
    C_{\rm SISO}({\rm snr}) = I\big({x}; \sqrt{{\rm snr}}\,x+z\big) = \int_{0}^{{\rm snr}} {\rm mmse}(\rho) d\rho.
\EE
\end{lemma}

Based on Lemmas~\ref{Lem:replic-MMSE} and \ref{Lem:S-I-MMSE}, the replica constrained capacity of a   linear system in \eqref{Eqn:linear_sys} is presented in the following lemma.
\begin{lemma}[Replica Constrained Capacity\cite{Barbier2018ISIT,Kabashima2006,Tulino2013}]\label{Lem:Re_cap}
Under Assumptions \ref{ASS:Model} and \ref{Asp:SEfixed}, the constrained capacity (per transmit symbol) of the linear system in \eqref{Eqn:linear_sys}, as predicted by the replica method, is given by:
\BE\label{Eqn:Replica_C}
C_{\rm MIMO}  =  \int_0^{v^*{\rm snr}} \!\!\!\! \mathcal{R}_{\bf{R}}(-z)dz + C_{\rm SISO}(\rho^*) - \rho^*v^*,
\EE
where $\mathcal{R}_{\bf{R}}(\cdot)$ is the {R-transform} given in \eqref{Eqn:R_transform} with $\bf{R}=\bf{A}^{\rm H}\bf{A}$, and $C_{\rm SISO}(\cdot)$ is given in \eqref{Eqn:C_mmse}.
\end{lemma}

\subsection{Key Challenges}
The existing researches on the optimal transceiver faces the following two challenges, depending on whether CSI is available at the transmitter:

\subsubsection{CSI Known at Transceivers} 
Traditional approaches decompose wireless channels into parallel SISO channels via SVD, enabling capacity to be achieved using multiple SISO capacity-achieving codes rather than a single code. This would bring an extremely high-complexity challenge, rendering it impractical in numerous practical communication systems. Furthermore, although \emph{mercury/waterfilling} can maximize the constrained capacity of the underloaded linear systems \cite{Lozano2006}, it does not guarantee achieving the MAP BER when considering only the BER performance of detection. As a result, \emph{optimizing power allocations to minimize the MAP BER and achieve the maximal constrained capacity using a single code remains an open challenge.}

\subsubsection{CSI Known Only at Receiver}
For practical large-scale communication systems, acquiring the accurate CSI at the transmitter is difficult and expensive, while the complexity of implementing SVD and water filling incurs significant computational overhead, particularly for low-cost terminal devices. Additionally, when the channel experiences rapid and frequent changes, repeatedly acquiring CSI and performing SVD for each channel change becomes impractical. In practice, it is commonly assumed that CSI is unknown at the transmitter, in which conventional techniques such as SVD and waterfilling cannot be employed. In such cases, the optimal strategy is to adopt an identical uniform power distribution, i.e., $x_i\sim P_x$ and ${\rm E}\{|x_i|^2\}=1$ for $ i \in [N]$. Nevertheless, \emph{the design of multiplexing, detection, encoding, and decoding algorithms with practical complexity that approach the replica MAP BER or replica constrained capacity of the linear system in \eqref{Eqn:linear_sys} remains an open problem.}

\section{Random Multiplexing}\label{Sec:RP}
In this section, we introduce random multiplexing and its intrinsic advantages as a practical solution for linear systems.  

\subsection{Universal Multiplexing} 
\begin{figure}[t!]
    \centering
    \includegraphics[width = 0.4\textwidth]{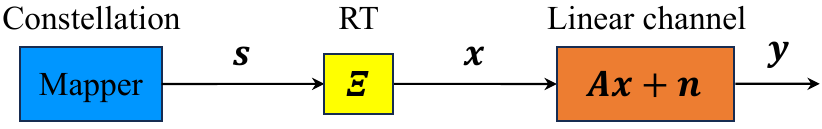}
    \caption{The linear system with random multiplexing, where the mapper corresponds to the constellation constraint $\bf{s}\overset{\rm i.i.d.}{\sim} P_{s}$, and $\bf{\Xi}$ denotes the RT matrix that is independent of the signal vector $\bf{s}$, the measurement matrix $\bf{A}$, and the channel noise $\bf{n}$.}
    \label{fig:RUP_model}
\end{figure}

The signal vector $\bm{s}\in \mathbb{C}^{N\times 1}$ is modulated by the transform matrix $\bm{\Xi}\in \mathbb{C}^{N\times N}$, i.e., 
\BE\label{Eqn:RUP}
\bf{x} = \bf{\Xi} \bf{s},
\EE   
where entries of $\bf{s}$ are IID distributed, i.e., $\bf{s}\overset{\rm i.i.d.}{\sim} P_{s}$. As illustrated in Fig.~\ref{fig:RUP_model}, a linear multiplexing system with power allocation can be described by:
\BS\label{Eqn:RUP_sys}
\begin{align}
    &\bm{y}   
        = \bm{A}\underbrace{\bm{\Xi s}}_{\bf{x}} + \bm{n},\\
    &{\rm s.t.}\quad \bf{s}\overset{\rm i.i.d.}{\sim} P_{s}.  
\end{align}
\ES 
As stated in Assumption \ref{ASS:Model}, we assume throughout this paper that $\bm{A}$ is spectrally convergent and has a bounded spectral norm $\|\bm{A}\|_2 \lesssim 1$. This means that the empirical spectral distribution of $\bm{A}^{\rm H}\bm{A}$ converges to a compactly supported distribution, and the largest singular value of $\bm{A}$ is finite.

\begin{definition}[Universal Multiplexing]\label{Def:UniModu}
We refer to $\bf{\Xi} \bf{s}$ as universal multiplexing of the signal $\bf{s}$, if the equivalent channel matrix $\bm{A\Xi}$ lies in the universality class $\mathscr{U}$ \cite{Rishabh2024}, defined as: $\bm{A\Xi} = \bm{JD}$, where\footnote{Strictly speaking, if $\bm{J}$ is random but independent of $\bm{D}$, and almost surely satisfies the conditions in items 2 and 3, then it follows that $\bm{J}\bm{D}$ almost surely lies in $\mathscr{U}$. For brevity, we do not distinguish this case from the deterministic one in this paper.}
\begin{enumerate}
    \item $\bm{D} = \diag\big\{[\mr{e}^{\mr{i}\theta_1}, \cdots\!, \mr{e}^{\mr{i}\theta_N}]\big\}$ is a uniformly random phase matrix with $\theta_{1:N} \overset{\mr{i.i.d.}}{\sim} \mathsf{Unif}\big\{[0, 2\pi)\big\}$.
    \item $\bm{J}$ is a deterministic spectrally convergent matrix with a bounded spectral norm $\|\bm{J}\|_2 \lesssim 1$.
    \item  For any fixed $k \in \mathbb{N}^*$, $\epsilon > 0$, 
    \begin{align} \label{eq:univ-class-moment-convergence}
        \Big\| (\bm{J}^{\rm H} \bm{J})^k - \frac{{\rm tr}[(\bm{J}^{\rm H} \bm{J})^k]}{N} \bm{I}_{N} \Big\|_{\max} \lesssim N^{-1/2+\epsilon}.
    \end{align}  
\end{enumerate}
\end{definition} 
\textbf{Remark:} The universality class $\mathscr{U}$, introduced in \cite{Rishabh2024}, plays a pivotal role by ensuring that the dynamics of AMP-type algorithms can be accurately tracked via state evolution. While the original definition of $\mathscr{U}$ in \cite{Rishabh2024} is formulated over the real field, we extend it to the complex field in Definition \ref{Def:UniModu}. This extension involves replacing real-field concepts with their natural complex counterparts: (i) transposes are replaced with Hermitian transposes, and orthogonal matrices with unitary ones; (ii) the uniformly random sign matrix $\bm{S}$ (i.e., a diagonal matrix with IID entries from $\mathsf{Unif}\{\pm 1\}$) is generalized to the uniformly random phase diagonal matrix $\bm{D}$. It has been shown in \cite{Rishabh2024} that, over the real field, $\mathscr{U}$ ensures that the error vectors in OAMP/VAMP are asymptotically IID Gaussian and thus ensures the accuracy of SE. In this work, we assume that this property also holds in the complex setting.
\begin{lemma}[IID Matrices\cite{Rishabh2024}]\label{Lem:IID}
Suppose that $\bm{A}$ is spectrally convergent and has a bounded spectral norm $\|\bm{A}\|_2 \lesssim 1$. Then, $\bm{A}\bm{\Xi}_{\mr{IID}} \in \mathscr{U}$, where the entries of $\bm{\Xi}_{\mr{IID}}$ are independently drawn from a circularly symmetric distribution with mean zero and variance $1/N$.
\end{lemma}

\subsection{Random Multiplexing}
The multiplexing matrices $\bm{\Xi}$ in Definition \ref{Def:UniModu} are broadly defined, covering a wide variety of matrices. In current multiplexing systems, unitary matrices are commonly used as they preserve the channel capacity. For greater flexibility, we further require that $\bm{\Xi}$ be independent of the channel and signal. These considerations lead us to random multiplexing as defined below.
\begin{definition}[Random Multiplexing]\label{Def:RUP} 
We refer to $\bf{\Xi} \bf{s}$ as random multiplexing of the signal $\bf{s}$ if:
\begin{enumerate}
    \item $\bm{\Xi}$ is a unitary random matrix, satisfying $\bm{\Xi}^{\rm H}\bm{\Xi} = \bm{I}$, and is independent of $\{\bm{A}, \bm{s},\bm{n}\}$. 
    \item The equivalent channel matrix $\bm{A\Xi}$ belongs to the universality class $\mathscr{U}$, i.e., $\bm{A\Xi} \in \mathscr{U}$. 
\end{enumerate}
\end{definition} 

Based on Definition \ref{Def:RUP}, we refer to $\bm{\Xi}$ as the random transform (RT) matrix and $\bm{\Xi}\bm{s}$ as the RT of $\bm{s}$. Accordingly, $\bm{\Xi}^{\rm H}$ and $\bm{\Xi}^{\rm H}\bm{s}$ are the inverse random transform (IRT) matrix and the IRT, respectively. Subsequently, we introduce several well-defined classes of the RT matrices $\bm{\Xi}$.
\begin{theorem}[Permutation-Invariant Matrices] \label{The:PIM}
Suppose that $\bf{A}$ is spectrally convergent, has a bounded spectral norm $\|\bm{A}\|_2~\lesssim~1$, and satisfies the off-diagonal-sum condition: for any fixed $k \in \mathbb{N}^*, \epsilon > 0$, 
\begin{align}\label{Eqn:off_diag}
    \Big|\textstyle\sum_{i, j \in [N], i \neq j}\big[(\bm{A}^{\mr{H}}\bm{A})^k\big]_{i, j}\Big| \lesssim N^{1/2+\epsilon}. 
\end{align}
Then, $\bm{A}\bm{\Xi}_{\rm PI}\in \mathscr{U}$, where  $\bm{\Xi}_{\rm PI} \equiv \bf{\Pi U D}$ is permutation-invariant, satisfying:
\begin{enumerate}  
    \item $\bm{D} = \diag\big\{[\mr{e}^{\mr{i}\theta_1}, \cdots\!, \mr{e}^{\mr{i}\theta_N}]\big\}$ is a uniformly random phase matrix with $\theta_{1:N} \overset{\mr{i.i.d.}}{\sim} \mathsf{Unif}\big\{[0, 2\pi)\big\}$.
    \item $\bm{\Pi}$ is a uniformly random permutation matrix independent of $\bm{D}$.
    \item $\bm{U}$ is a delocalized deterministic unitary matrix satisfying $\|\bf{U}\|_{\max}\lesssim N^{-1/2+\epsilon}$ for any $\epsilon>0$.
\end{enumerate} 
\end{theorem}
\begin{IEEEproof}
    See Appendix \ref{App:PIM}.
\end{IEEEproof}
In the real case, Theorem \ref{The:PIM} directly reduces to the formulation corresponding to the original definition of $\mathscr{U}$ over the real field in \cite{Rishabh2024}. The result of \cite[Lemma 3]{Rishabh2024} is a special case of Theorem~\ref{The:PIM}, where the right singular vectors of $\bm{A}$ form the identity matrix. For ease of reference, we restated it as the following corollary.
\begin{corollary}\label{Cor:PIM} 
    Let $\bm{A} = \bm{U}_{A}\bm{\Sigma}_{A}\bm{V}^{\mr{H}}_{A}$ be the singular value decomposition of $\bm{A}$, where $\bm{V}_{A}=\bm{I}$ (i.e., the right singular vectors form the identity matrix). Suppose that $\bm{A}$ is spectrally convergent and satisfies $\|\bm{\Sigma}_A\|_2 \lesssim 1$. Then, $\bm{A\Xi}_{\rm PI}\in \mathscr{U}$.
\end{corollary}
For Haar-distributed matrices, a subclass of the permutation-invariant matrices, the off-diagonal-sum condition on $\bm{A}$ in \eqref{Eqn:off_diag} is not necessary, yielding the following lemma.
\begin{lemma}[Haar-Distributed Matrices\cite{Rishabh2024}]\label{Lem:Haar}
Suppose that $\bm{A}$ is spectrally convergent and has a bounded spectral norm $\|\bm{A}\|_2 \lesssim 1$. Then, $\bm{A}\bm{\Xi}_{\mr{Haar}} \in \mathscr{U}$, where the random transform matrix $\bm{\Xi}_{\mr{Haar}}$ is Haar distributed, i.e., $\bm{\Xi}_{\rm Haar}\sim {\rm Unif}(\mathbb{U}(N))$, where $\mathbb{U}(N)$ denotes the unitary group.     
\end{lemma}
\begin{IEEEproof}
    Let $\bm{A} = \bm{U}_{A}\bm{\Sigma}_{A}\bm{V}^{\mr{H}}_{A}$. 
    The Haar measure is distributionally invariant to left or right multiplication by any independent unitary matrices. Hence, $\bm{A}\bm{\Xi}_{\mr{Haar}} = \bm{U}_A\bm{\Sigma}_A\widetilde{\bm{\Xi}}$, where $\widetilde{\bm{\Xi}} = \bm{V}_A^{\mr H}\bm{\Xi}_{\mr{Haar}}$ is Haar distributed. Furthermore, $\widetilde{\bm{\Xi}}$ has the same distribution as $\bm{\Pi}\widetilde{\bm{\Xi}}\bm{D}$ for any independent permutation $\bm{\Pi}$ and diagonal phase matrix $\bm{D}$. In addition, for any $\epsilon > 0$, $\|\widetilde{\bm{\Xi}}\|_{\max} \lesssim N^{-1/2+\epsilon}$ with probability $1$ \cite{jiang2005maxima}. Therefore, we apply Corollary \ref{Cor:PIM} to complete the proof.
\end{IEEEproof}
In practice, we typically set $\bm{D}=\bm{I}$ and choose $\bm{U}$ to be fast transform matrices for the permutation-invariant matrices in Theorem \ref{The:PIM}. That is, $\bm{\Xi}_{\rm PT}\equiv\bm{\Pi T}$, where $\bm{T}$ denotes fast transform matrices, such as discrete Fourier transform (DFT), Hadamard-Walsh transform (WHT), or discrete cosine transform (DCT), interleaved block-sparse transform\cite{IBST}, $\bm{F}^{\mr{H}}_L\otimes\bm{I}_K$ in OTFS with the normalized IDFT matrix $\bm{F}^{\mr{H}}_L$ and $N=LK$, $\bf{\Lambda}_{c_1}^{\mr{H}}\bf{F}^{\mr{H}}\bf{\Lambda}_{c_2}^{\mr{H}}$ in AFDM with $\bf{\Lambda}_{c_i}\triangleq\mr{diag}(e^{-\mr{j}2\pi c_in^2}, n=0, \cdots\!, N-1)$, $i=1,2$, and in orthogonal chirp division multiplexing (OCDM) with $c_1=c_2=\tfrac{1}{2N}$. A suitable unitary matrix is selected in accordance with practical application requirements and hardware complexity constraints.
\begin{itemize}
    \item \emph{CSI Known Only at Receiver}: $\bf{\Xi}_{\rm PT}$ represents a special case of that in Theorem \ref{The:PIM} with $\bf{P}=\bf{I}$. 
    \item \emph{CSI Known at Transceivers}: $\bf{\Xi}_{\rm PT}$ represents a special case of that in Corollary \ref{Cor:PIM} with the optimal form of $\bm{P}$ (See Theorem \ref{The:Opt_P} in Section \ref{Sec:PA}).
\end{itemize}
In addition, $\bf{\Xi}$ can be also constructed using multi-layer permutation-invariant matrices, expressed as 
\BE
\bf{\Xi}_{\mr{ML}}= {\bm{\Pi}_1\bm{T}_1\bm{\Pi}_1\bm{T}_2\cdots\bm{\Pi}_L\bm{T}_L},
\EE
where each pair $\{\bm{\Pi}_l, \bm{T}_l\}$ is randomly selected\cite{EST}. However, it remains unproven that the equivalent matrix $\bm{A\Xi}_{\mr{ML}}$ lies in the universality class $\mathscr{U}$. 

\subsection{Intrinsic Advantages of Random Multiplexing}\label{Sec:RP_B}

\subsubsection{Decoupling from the Channel Matrix}
Conventional multiplexing schemes, such as OFDM, SC-FDE, OTFS, and AFDM, inherently couple the multiplexing matrix with the channel characteristics. In contrast, random multiplexing decouples the multiplexing matrix $\bf{\Xi}$ from the channel matrix $\bf{H}$, i.e., $\bf{\Xi}$ is independent of $\bf{H}$. This decoupling enables random multiplexing to be applied to arbitrary norm-bounded and spectrally convergent channel matrices. 

\subsubsection{Asymptotic Performance Limits} The introduction of RT matrix $\bm \Xi$ in \eqref{Eqn:RUP_sys} does not change the eigenvalues of $\bf{A}^{\rm H}\bf{A}$. As a result, the random multiplexing in \eqref{Eqn:RUP_sys} does not lead to any performance or rate loss compared to the original linear systems in \eqref{Eqn:linear_sys}.  

\begin{proposition}[Preservation of Replica Limits]\label{Prop:Limit}
Random  multiplexing  preserves the replica MAP BER and constrained capacity of the original linear systems in \eqref{Eqn:linear_sys}, including the Gaussian capacity with Gaussian signaling.
\end{proposition}
\begin{IEEEproof}
    Following Lemmas \ref{Lem:replic-MMSE} and \ref{Lem:Re_cap}, both the replica MAP BER and the replica constrained capacity of a linear system are determined by the Stieltjes transform. Define $\bf{R}\equiv(\bf{A\Xi})^{\rm H}\bf{A\Xi}$. We have
    \BS
    \begin{align}\label{Eqn:ST_equal}
        \mathcal{S}_{\bf{R}}(w) &= \frac{1}{N}{\rm tr}\left\{(\bf{R}-w\bf{I})^{-1}\right\}\\
        & = \frac{1}{N}{\rm tr}\left\{(\bf{A\Xi})^{\rm H}\bf{A\Xi}-w\bf{I})^{-1}\right\}\label{Eqn:ST_RP}\\
        & = \frac{1}{N}{\rm tr}\left\{\bf{A}^{\rm H}\bf{A}-w\bf{I})^{-1}\right\},\label{Eqn:ST}
    \end{align} 
    \ES
    where \eqref{Eqn:ST} follows $\bf{\Xi}^{\rm H}\bf{\Xi}=\bf{I}$. Therefore, the Stieltjes transform of the random multiplexing system is identical to that of  the original linear system.  Consequently, both the replica MAP BER and the replica constrained capacity remain unchanged in the random multiplexing system. 
\end{IEEEproof}

\subsubsection{Low-Complexity and Asymptomatically Optimal Receiver} Existing multiplexing  matrices $\bf{\Xi}$ and channel matrices $\bf{H}$ are typically highly structured, making it challenging for $\bf{A}\mathbf{\Xi}$ to exhibit the required input isotropy. This lack of isotropy hampers the development of efficient signal recovery algorithms \cite{OTFS-OAMP, OAMP/VAMP, MAMPOTFSconf}, with the exception of OFDM in time-invariant channels, where $\mathbf{\Xi}^{\rm H}\bm{H}\mathbf{\Xi}$ forms an orthogonal matrix. As shown in Fig.~\ref{fig:RMcomp}, the equivalent channel matrices of OFDM, OTFS, and AFDM exhibit diagonalized and specific sparsification structures in static and time-varying multipath channels, respectively. Specifically, the OTFS multiplexing matrix is $\bm{\Xi}_{\rm OTFS}=\bm{F}_{L}^{\rm{H}} \otimes \bm{I}_{K}$ with $N=LK$~\cite{OTFS1}, the AFDM matrix is $\bm{\Xi}_{\rm AFDM}=\bf{\Lambda}_{c_2}\bf{F}_{{N}}\bf{\Lambda}_{c_1}$ with $\bf{\Lambda}_{c_i}\triangleq\text{diag}(e^{-j2\pi c_in^2}, n=0, \cdots\!, N-1)$, $i=1,2$~\cite{AFDM}, a multiplexing matrix $\bm{\Xi}$ is a sparse Walsh-Hadamard matrix\cite{SparsePrecode}, and the ISI channel matrix $\bf{H}$ is a Toeplitz matrix. As a result, existing AMP-type algorithms in \cite{MaAcess2017,Rangan2019TIT,CAMP,fan2022approximate,mondelli2021pca,WS-CG-VAMP,LeiMAMP} suffer significant performance losses and are unable to approach the  MAP BER or the  constrained capacity. Thanks to the random multiplexing in \eqref{Eqn:RUP}, the effective channel matrix $\bf{A\Xi}$ belongs the universality class $\mathscr{U}$, facilitating the utilization of AMP-type receivers, which asymptotically achieve the replica MAP-BER or the replica constrained capacity of the linear systems in \eqref{Eqn:linear_sys}.

\begin{figure*}[t]\vspace{-0.3cm}
\centering  
\subfigure[Noise pattern of asymmetric noise $\bar{\bm n}$ in \eqref{Eqn:2D_bpsk}. $\bf{\Xi}=\bm{V}_A$.]
{\includegraphics[width=0.3\textwidth]{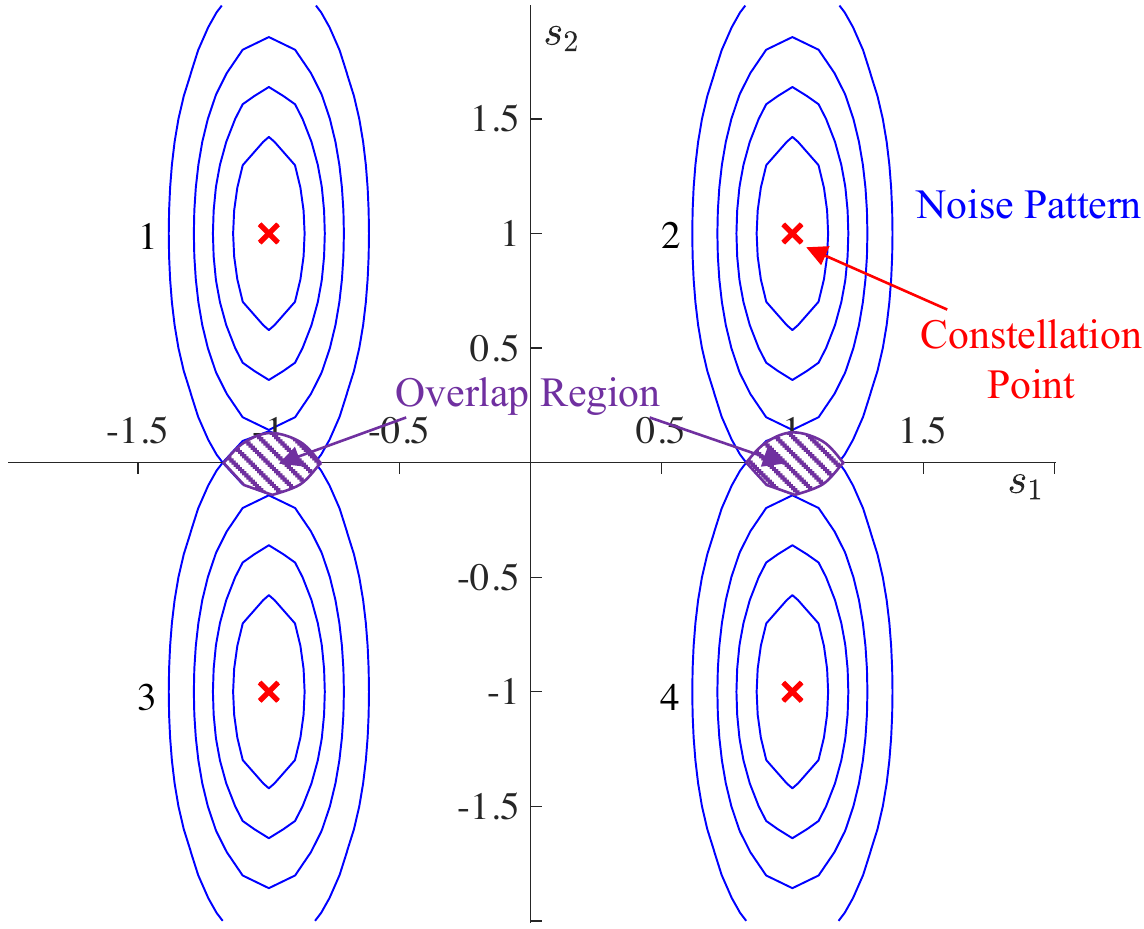}\label{fig:asynoise}}
\hfill
\subfigure[Noise pattern of symmetric (i.e., input-isotropic) noise $\bar{\bm n}$ in \eqref{Eqn:2D_bpsk}. $\bf{\Xi}=\bm{V}_A\bf{\Theta}$, $\bf{\Theta}=\begin{bmatrix}\cos\frac\pi4, -\sin\frac\pi4; \sin\frac\pi4, \cos\frac\pi4\end{bmatrix}$. ]
{ 
\includegraphics[width=0.3\textwidth]{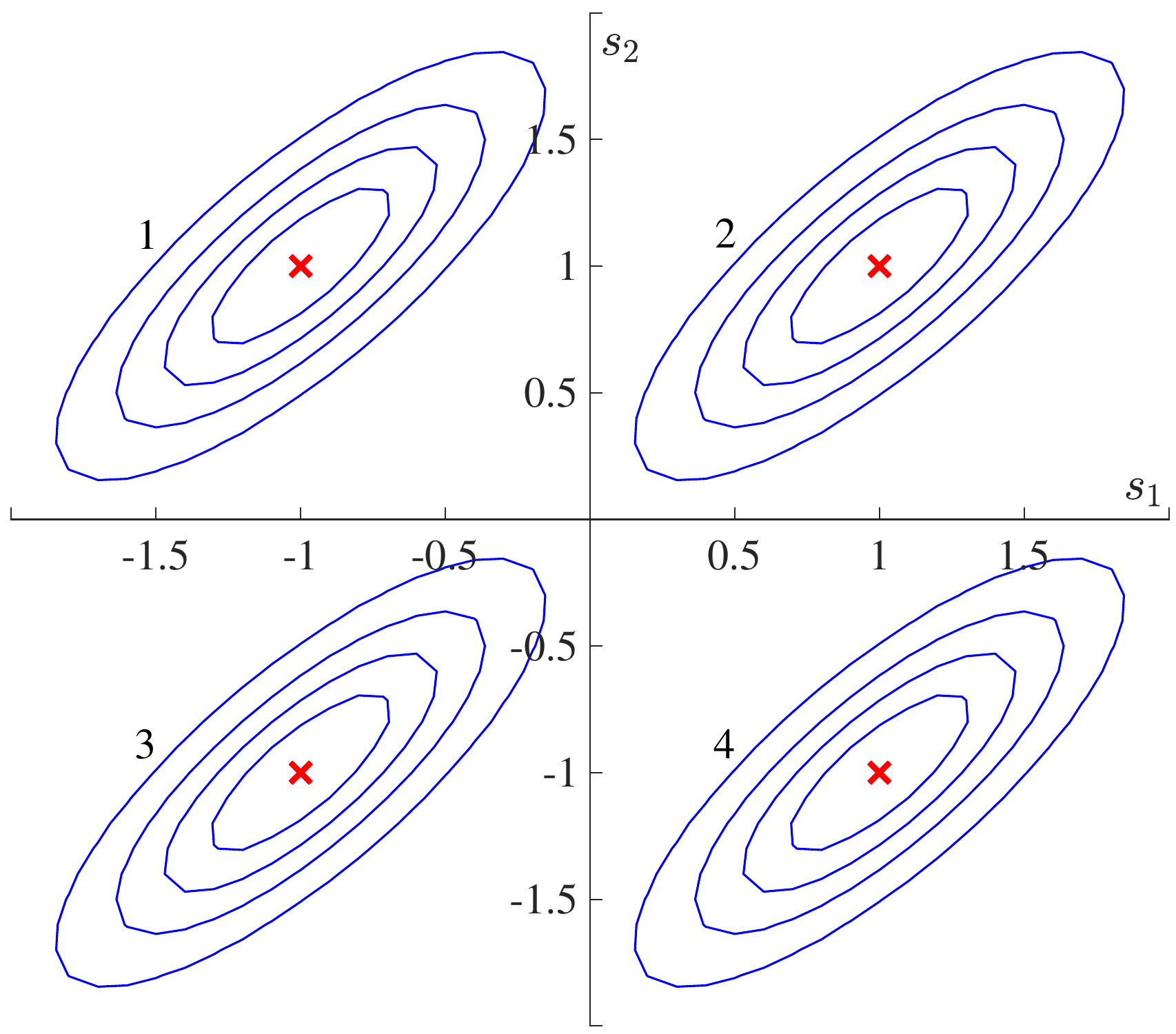}\label{fig:synoise}
}
\hfill
\subfigure[BER comparisons of the 2D orthogonal multiplexing and 45-degree-rotation multiplexing with BPSK signaling and ML detection.]
{\includegraphics[width=0.3\textwidth]{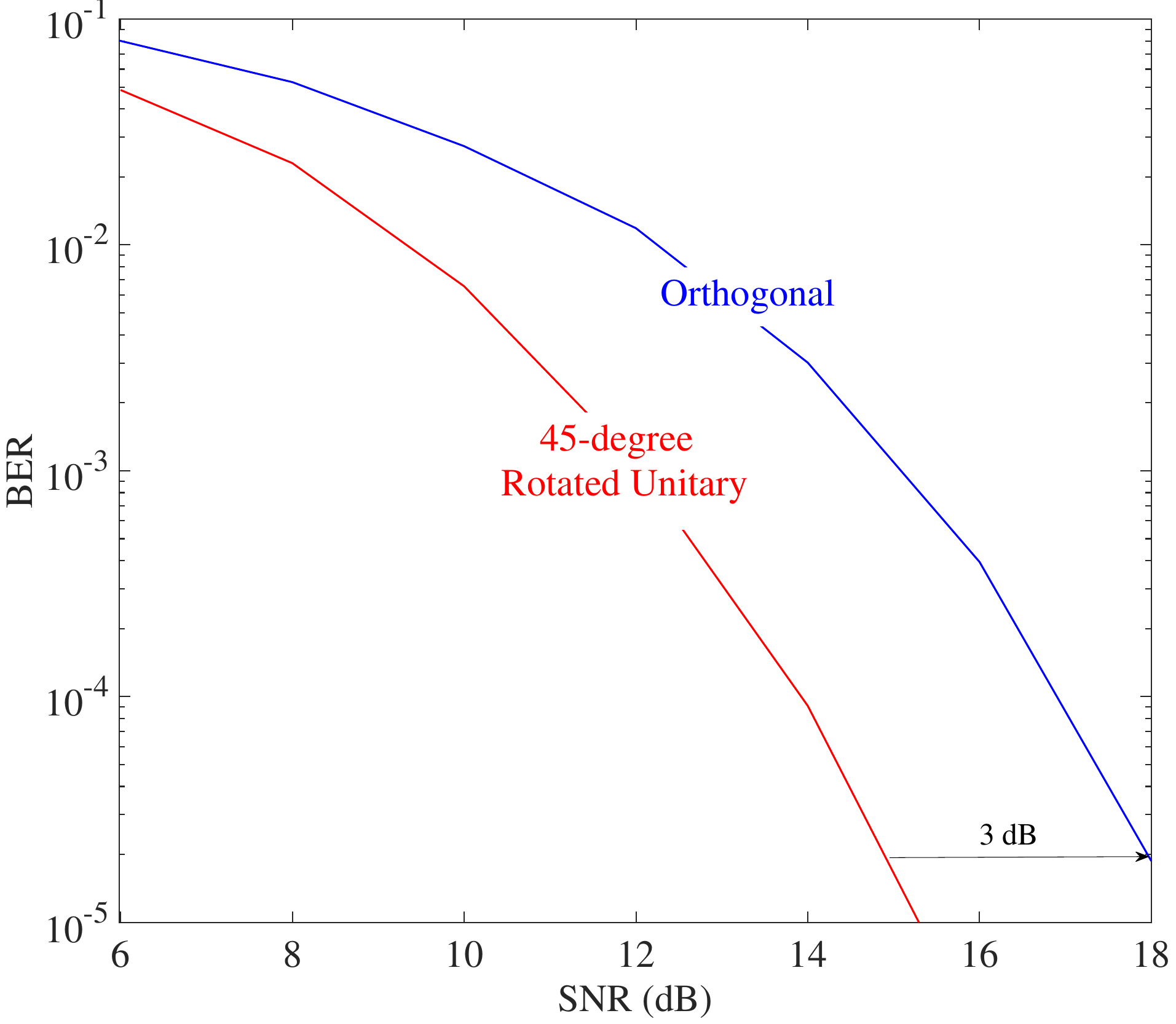}\label{fig:ML_2D}}
\caption{Visualization and BER comparison under ML detection for 2D BPSK linear systems with orthogonal and unitary multiplexing.} \label{Fig:2Dexmp}
\end{figure*}
\begin{lemma}[Replica MAP BER/Capacity Optimality]\label{Prop:opti}
  In random  multiplexing systems, we have $\bm{A\Xi}\in \mathscr{U}$. Suppose that the Assumptions \ref{ASS:Model} and \ref{Asp:SEfixed} hold. Then,
 \begin{itemize}
     \item  OAMP/VAMP \cite{MaAcess2017,Rangan2019TIT}, long-memory AMP\cite{fan2022approximate}, PCA-initialized AMP \cite{mondelli2021pca}, and MAMP \cite{LeiMAMP} can achieve the replica MAP BER in \eqref{Eqn:replicaBER}, and
     \item OAMP/VAMP and MAMP, utilizing optimal Lipschitz-continuous decoding (as defined in \eqref{Eqn:opt_code}), can achieve the replica constrained capacity in \eqref{Eqn:Replica_C} \cite{LeiOptOAMP,Code_MAMP}.
 \end{itemize} 
\end{lemma} 

\subsubsection{Diversity}  
In linear systems with random multiplexing, the effective channel matrix $\bf{A}\bf{\Xi}$ exhibits enhanced stability, enabling each element of $\bf{s}$ to experience all channel fading effects in $\bf{A}$ due to the input isotropy of $\bf{\Xi}$. Consequently, random multiplexing can potentially attain maximum diversity, thereby improving signal recovery performance. We have not conducted a thorough analysis of the diversity of random multiplexing. However, as demonstrated in Lemma \ref{Prop:opti}, \emph{its replica MAP BER optimality suggests that random multiplexing is capable of achieving maximum diversity}. Notably, while MAP BER optimality is sufficient to guarantee maximum diversity, the converse does not hold: maximum diversity does not necessarily imply MAP BER optimality.

\subsection{Illustrative Examples}\label{Sec:RP_C}
To visually demonstrate the advantages of multiplexing matrix $\bm \Xi$, we provide experimental analyses of two-dimensional and high-dimensional random multiplexing linear systems. \emph{For simplicity, let $M=N$ and $\bm A$ be invertible}. Accordingly, \eqref{Eqn:RUP_sys} is rewritten as   
\BE\label{Eqn:2Dlnear}
\bm{y} =\bm{A}\bm{x}+\bm{n}= \bm{U}_A \bm{\Sigma}_{A}\bm{V}_A^{\rm H}\bm{\Xi s} + \bm{n},
\EE
where $\bm A = \bm{U}_A \bm{\Sigma}_{A} \bm{V}_A^{\rm H}$ and $\bf{x}=\bm{\Xi s}$. Equivalently, we have
\begin{align}\label{Eqn:2D_bpsk}
\bar{\bm{y}} ={\bm{\Xi}^{\rm H}}\bm{V}_A\bm{\Sigma}_{A}^{-1}\bm{U}_A^{\rm H}{{\bm{y}}} =\bf{s} + \underbrace{{\bm{\Xi}^{\rm H}}\bm{V}_A\bm{\Sigma}_{A}^{-1} \bm{U}_A^{\rm H}{\bf{n}}}_{=\bar{\bf{n}}},  
\end{align} 
where $\bar{\bf{n}} \!\sim \! \mathcal{CN}(\mathbf{0},\sigma^2{\bm{\Xi}^{\rm H}}\bm{V}_A\bm{\Sigma}_{A}^{-2}\bm{V}_A^{\rm H} {\bm{\Xi}})$. Based on \eqref{Eqn:2D_bpsk}, we have the following observations.

\emph{1) Example 1:} Consider a 2D linear communication system with  BPSK signal vector $\bf{s}=[s_1, s_2]^{\rm T}$ in \eqref{Eqn:2D_bpsk},  where  $\sigma^2=0.1$ and $\bm{\Sigma}_{A}={\rm diag}\{3/2, 1/2\}$. We compare the orthogonal and unitary modulations as follows.
\begin{itemize}
\item \emph{2D Orthogonal Multiplexing (e.g., OFDM)}: Assuming $\bf{\Xi}=\bm{V}_A$, we obtain an asymmetric noise vector $\bar{\bf{n}}=\bm{\Sigma}_{A}^{-1}\tilde{\bf{n}}=[2/3\tilde{n}_1, 2\tilde{n}_2]^{\rm T}$ in \eqref{Eqn:2D_bpsk}. That is, $s_1$ is subject to reduced noise, whereas $s_2$ experiences increased noise. Consequently, the overall signal recovery performance is affected by the performance of $s_1$, leading to deterioration. In Fig. \ref{fig:asynoise}, we visualize the contour lines of the received signals $\bar{\bf{y}}$ affected by asymmetric effective noise $\bar{\bf{n}}$, referred to as noise patterns. In the $s_1$ direction, the noise patterns are far apart, making them easy to distinguish. Nevertheless, in the $s_2$ direction, the noise patterns overlap, resulting in poor signal recovery performance.
\item \emph{2D Unitary Multiplexing}: Consider $\bf{\Xi}=\bm{V}_A\bf{\Theta}$, where $\bf{\Theta} = [\cos\frac\pi4, -\sin\frac\pi4; \sin\frac\pi4, \cos\frac\pi4]$ is a 45-degree rotation matrix.  As shown in Fig.~\ref{fig:synoise}, we then get an effective noise vector $\bar{\bf{n}}= \bf{\Theta} [2/3\tilde{n}_1, 2\tilde{n}_2]^{\rm T}$ with symmetric with covariance matrix $\sigma^2 \bf{\Theta} \bm{\Sigma}_{A}^{-2} \bf{\Theta}^{\rm T} =[0.222, 0.177; 0.177,0.222]$. Consequently, the noise pattern of each constellation point is sufficiently distinct from those of the other constellation points, enhancing the overall performance.
\end{itemize}

\begin{figure*}[t]
\centering
\subfigure[$N=8$]{
\begin{minipage}[t]{0.335\textwidth}
\centering
\includegraphics[width=\textwidth]{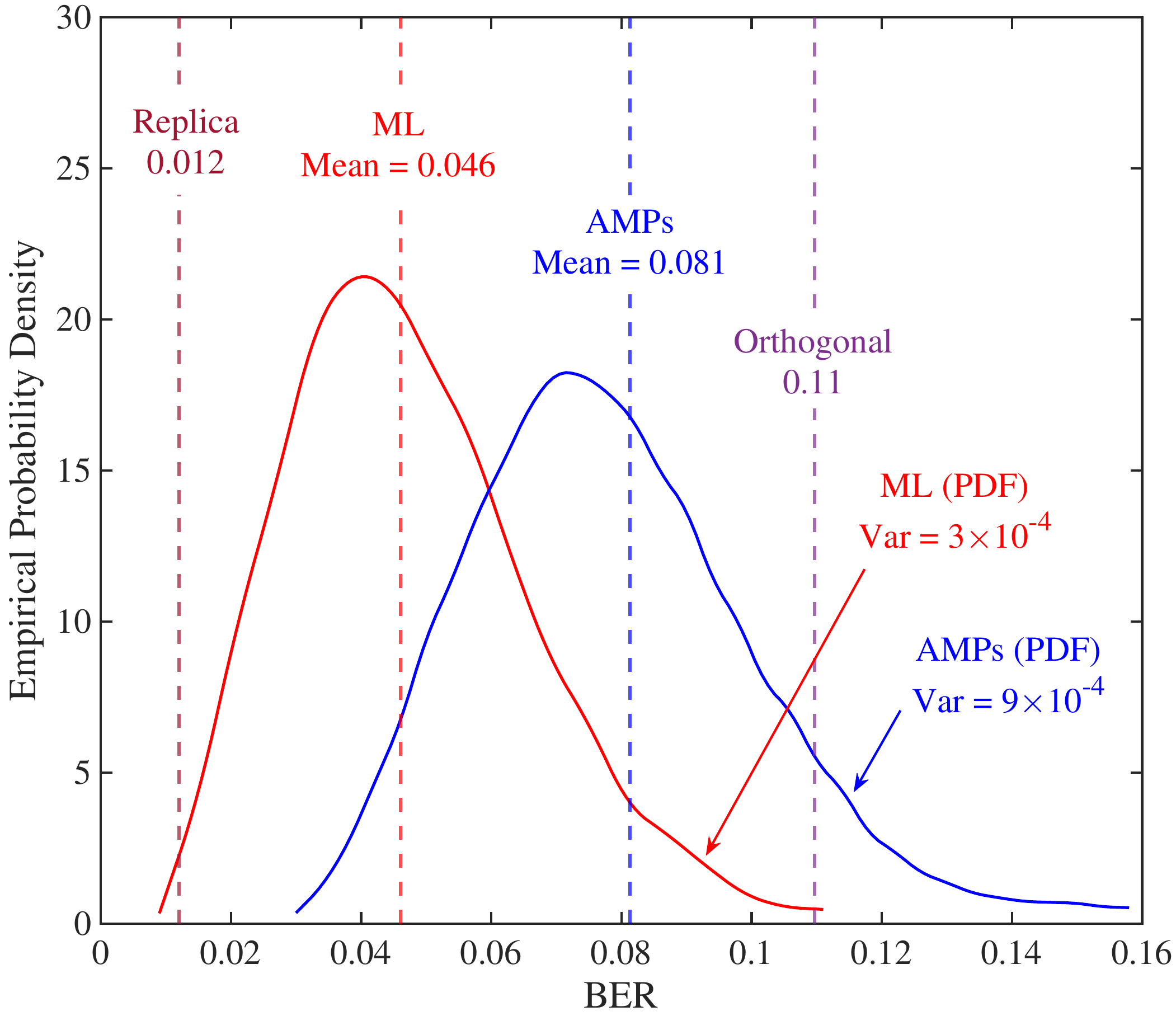}
\end{minipage}%
}%
\subfigure[$N=32$]{
\begin{minipage}[t]{0.325\textwidth}
\centering
\includegraphics[width=\textwidth]{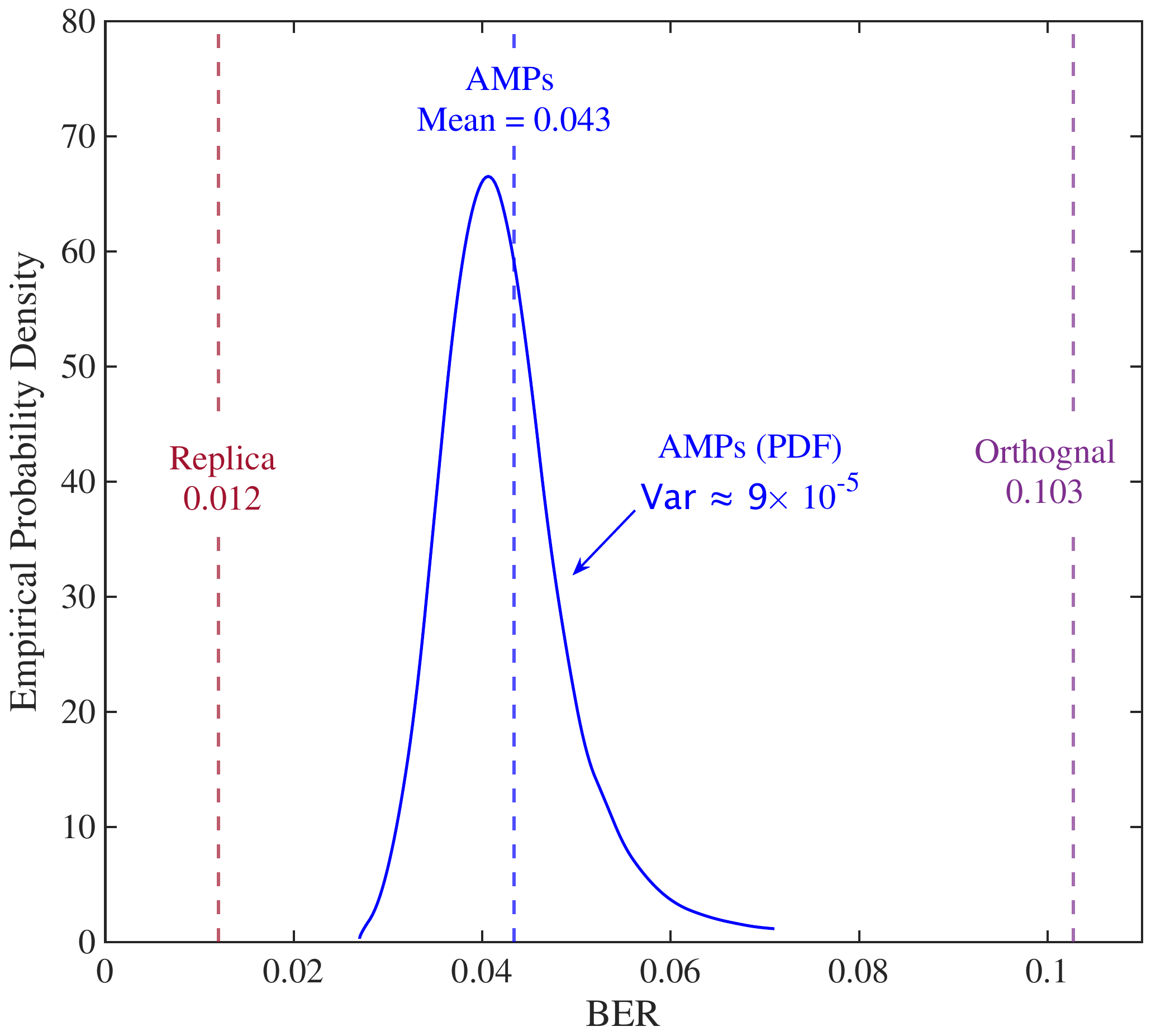}
\end{minipage}%
}%
\subfigure[$N=512$]{
\begin{minipage}[t]{0.335\textwidth}
\centering
\includegraphics[width=\textwidth]{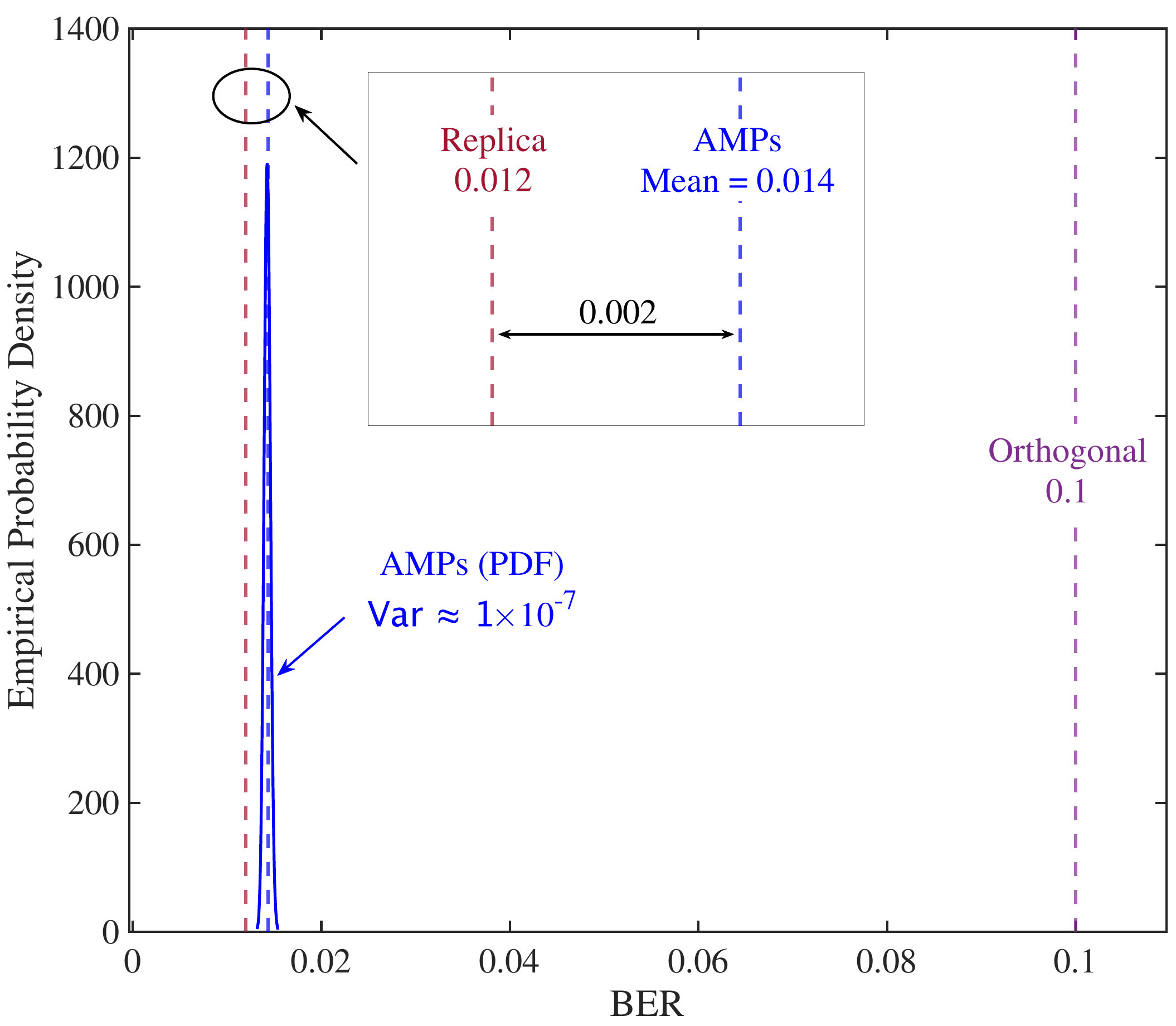}
\end{minipage}
} 
\centering
\caption{BER comparisons for 1) orthogonal multiplexing $\bm{y} = \bm{\Sigma}_A \bm{x} + \bm{n}$ using an element-wise MMSE detector, 2) unitary multiplexing $\bm{y} = \bm{\Sigma}_A\bm{\Xi} \bm{x} + \bm{n}$ using ML and AMP-type detectors. The $\bm{\Sigma}_A$ and $\bm{\Xi} = \bm{V}_A^{\rm H}$ are obtained from the SVD on $10^5$ normalized IID Gaussian matrices $\bm{A}$. For each $\bm{A}$, the BER is averaged over $10^4$ Monte Carlo simulations for each detector. The system dimension is set to $N \in \{ 8, 32, 512\}$.}\label{Fig:RUP_high}
\end{figure*}
Fig.~\ref{fig:ML_2D} shows that the BER of the 45-degree rotation multiplexing with ML detection outperforms that of the 2D orthogonal multiplexing.  

\emph{Challenge:} For 2D linear systems, a suitable $\bm \Xi$ can be obtained through meticulous design. However, for high-dimensional linear systems, finding a appropriate $\bm \Xi$ is challenging. We investigate the statistical characteristics of a randomly selected $\bm \Xi$ using the experimental setup outlined below for high-dimensional linear systems.

\emph{2) Example 2:} We compare orthogonal and unitary modulations to study the impacts of a randomly chosen $\bm \Xi$ in high-dimensional situations, in which $\bm \Xi = \bm{V}_A$ is set for orthogonal multiplexing  and $10^5$ RT matrices $\bm{\Xi}$ are randomly generated\footnote{A RT matrix $\bm{\Xi}$ can be constructed as follows: First, randomly generate a normalized IID Gaussian matrix $\bf{A}$. Then, conduct the SVD decomposition of $\bf{A}$, i.e., $\bf{A}=\bm{U}_A\bm{\Sigma}_{A}\bm{V}_A^{\rm H}$. Finally, set $\bm{\Xi}$ as $\bm{V}_A^{\rm H}$.} for unitary multiplexing. As illustrated in Fig. \ref{Fig:RUP_high}, the following trends emerge as the system size increases.  
\begin{itemize} 
    \item \emph{Replica MAP BER:} Based on Lemma~\ref{Lem:replic-MMSE},  ``Replica" denotes the average BER performance limit predicted by the replica method for the linear systems. 
    \item  \emph{Converged BER Distribution:} For the vast majority of unitary matrices, the BERs of the low-complexity AMP detector are concentrated near its mean. 
    \item \emph{Near-Optimality of AMP:} The BERs of the AMP detector of the random multiplexing approach the replica limit, with the average performance of the AMP detector being $0.014$, close to the replica limit of $0.012$, with only a negligible distance of $0.002$. 
    \item \emph{ML Performance:} Due to the prohibitive computational complexity of the ML detector, the BER curves for the ML detector are not provided for $N=\{32,512\}$. However, its performance is  bounded between the replica limit and the performance of the AMP detector. As $N$ increases, AMP approaches ML and then approaches the replica limit. 
    \item \emph{Poor BER of Orthogonal Multiplexing:} The BERs of orthogonal multiplexing with MMSE detection are poor, with marginal improvement as $N$ grows. 
\end{itemize}
In conclusion, the above observations demonstrate that the BER performance of linear systems with randomly generated unitary matrices, when paired with ML or AMP-type detectors, approaches the replica limit with high probability as $N$ increases. That is, {\emph{as long as the system is sufficiently large, the probability of a randomly generated unitary matrix being good (approaching the replica limit) is extremely high.}} This effectively solves the difficulty of finding good unitary multiplexing matrices in high-dimensional systems. 

Beyond the intuitive explanation above, theoretical analyses have demonstrated that random multiplexing achieves both the replica minimum MSE/BER and the constrained capacity in linear systems \cite{MaAcess2017,Rangan2019TIT,CAMP,fan2022approximate,mondelli2021pca,WS-CG-VAMP,LeiMAMP,LeiOptOAMP,Code_MAMP}. Refer to Lemma \ref{Prop:opti} in Section \ref{Sec:RP_B} for more details. It is important to note that the replica optimality established in \cite{MaAcess2017,Rangan2019TIT,CAMP,fan2022approximate,mondelli2021pca,WS-CG-VAMP,LeiMAMP,LeiOptOAMP,Code_MAMP} relies on the assumption that $\bf{A}$ is right-unitarily invariant, which does not hold for practical channel matrices. This assumption is relaxed through the random multiplexing technique in this paper.

\emph{Discussions:} The essence of random multiplexing aligns with the random codebook selection in Shannon’s capacity theorem~\cite{Cover1990}. In high-dimensional systems, a randomly selected codebook is asymptotically capacity-achieving, while well-designed codebooks often underperform in practice. Unlike random coding, which suffers from the absence of efficient optimal decoding algorithms, the transmitted signals in the linear system with random multiplexing can be effectively recovered using low-complexity AMP-type algorithms.  

\section{Cross-Domain MAMP Detection}\label{Sec:MAMP}
Signal detection in the linear systems with random multiplexing presents a significant challenge. In this section, we introduce a general cross-domain message passing detection framework, featuring two Bayes-optimal detectors: CD-OAMP/VAMP for theoretical analysis and CD-MAMP for practical low-complexity implementations. Critically, random multiplexing ensures the transform-domain channel matrix exhibits the requisite input isotropy, placing it in the universality class $\mathscr{U}$\cite{Rishabh2024}---a necessary condition for the efficacy of both detectors.

\subsection{Problem Formulation}
We rewrite the random multiplexing in \eqref{Eqn:RUP_sys} as
\BS\label{Eqn:RUP_dect}
\begin{align}
    \text{Linear constraint}\; \Gamma: \; & \bf{y}=\bf{A}\bf{x}+\bf{n}, \\ 
    \text{Random transform}\; T:\; &\bf{x}=\bf{\Xi}\bf{s}, \\
    \text{Nonlinear constraint}\; \Phi: \; & \bf{s} \sim P_S(\bf{s}).
\end{align}
\ES
The goal is to find the MMSE estimate of $\bf{s}$. That is, its MSE converges to \cite{kay1993fundamentals}
\BE\label{Eqn:mmse_s}
\mr{mmse}\{\bf{s}|\bf{y},\bf{A}, \bf{\Xi}, \Gamma, T, \Phi\} \equiv \frac{1}{N} \mr{E}\{||\hat{\bf{s}}_{\rm{post}}-\bf{s}||^2\},
\EE
where $\hat{\bf{s}}_{\rm{post}}=\mr{E}\{\bf{s}|\bf{y},\bf{A}, \bf{\Xi},T,  \Gamma, \Phi\}$ is the \emph{a-posteriori} mean of $\bf{s}$. In the special case where $\bf{s}$ is Gaussian, the standard LMMSE detector achieves optimality. However, for arbitrary non-Gaussian $\bf{s}$ and general $\bf{A}$, finding the optimal solution is generally NP-hard \cite{verdu1984optimum}. 

Existing linear detectors, such as ZF and LMMSE, are suboptimal as they ignore the a priori information of the signal $\bm{s}$. For this reason, nonlinear iterative detectors, such as the factor graph-based message passing algorithms, are widely employed \cite{OTFS_GMP}. Nevertheless, when the measurement matrix $\bf{A}$ is dense, it often induces correlation problems, leading to performance degradation. To overcome this limitation, a sequence of advanced AMP-type algorithms, including OAMP/VAMP\cite{MaAcess2017,Rangan2019TIT}, CAMP~\cite{CAMP}, long-memory AMP\cite{fan2022approximate},  PCA initialized AMP~\cite{mondelli2021pca}, WS-CG-VAMP\cite{WS-CG-VAMP}, and MAMP\cite{LeiMAMP}, have been developed. In particular, a necessary condition for these algorithms to achieve the replica MAP BER is that the equivalent channel matrix $\bf{A \Xi}$ exhibits sufficient input isotropy (i.e., belongs to the universality class $\mathscr{U}$ \cite{Rishabh2024}). However, the wireless channel matrices or the equivalent channel matrices in existing multicarrier systems (e.g., OFDM\cite{tse2005fundamentals}, OTFS\cite{OTFS1}, and AFDM\cite{AFDM}) exhibit strong structural properties. This discrepancy often leads to significant performance degradation of AMP-type algorithms in practical scenarios. Random multiplexing is a simple yet effective solution to ensure that $\bf{A\Xi}$ exhibits sufficient input isotropy, thereby allowing AMP-type algorithms to achieve the replica optimality under Assumptions~\ref{ASS:Model} and \ref{Asp:SEfixed}.

\begin{figure}[t!]\vspace{-0.2cm}
    \centering
    \includegraphics[width = 0.4\textwidth]{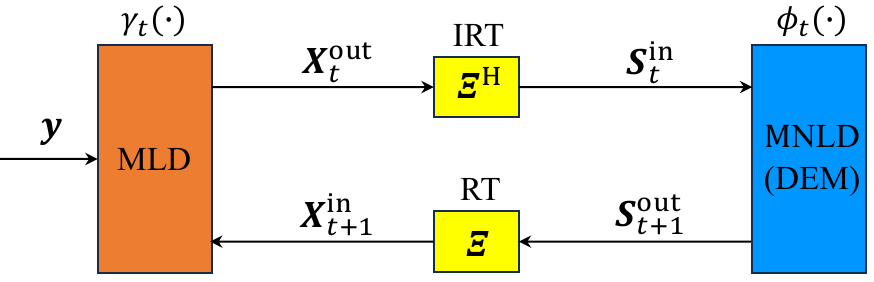}
    \caption{The CD-MAMP framework for the linear system with random multiplexing, where $t$ denotes the iteration index, $\bf{\Xi}$ the RT,  $\bf{\Xi}^{\rm H}$ the IRT, and DEM the demodulation.}
    \label{fig:MIP_CD}
\end{figure}
\subsection{Cross-Domain MAMP Framework} 
The CD-MAMP  framework and its state evolution (SE) are presented for random multiplexing systems as follows, generalizing the approach in \cite{IFDM,LeiMAMP}. 

\subsubsection{CD-MAMP Framework}
As shown in Fig.~\ref{fig:MIP_CD}, the CD-MAMP detector consists of a memory linear detector (MLD) $\gamma_t(\cdot)$, the random transform (RT) and its inverse (IRT), and a memory nonlinear detector (MNLD) $\phi_t(\cdot)$, which correspond to the linear constraint $\Gamma$, the RT $T$, and nonlinear constraint $\Phi$ in \eqref{Eqn:RUP_dect}, respectively. Specifically, MLD is designed to fully exploit the sparsity of the time-domain channels for low-complexity recovery of the time-domain signal $\bf{x}$, while MNLD is tailored to incorporate the \emph{a priori} information of the symbol-domain signal vector $\bf{s}$. Then, an iterative refinement process is performed between MLD and MNLD via the RT and IRT, culminating in the recovery of the signal $\bf{s}$. A key feature of both MLD and MNLD is the use of orthogonalization to ensure that the input and output estimation errors remain orthogonal. This prevents error correlation during iterative processing. The detailed formulation of the CD-MAMP framework is presented below.

Starting with iteration index $t=1$ and $\bf{X}_1^{\rm{in}}=\bm{0}$,
\BS\label{Eqn:MIP}
\begin{align} 
\!\!\!\!\!\!{\text {MLD}:}& \;\; \bf{x}_t^{\rm{out}} \!=\! \gamma_t(\bf{X}_t^{\rm{in}}) \!=\! \bf{\mathcal{Q}}_t\bf{y}\!+ \!\textstyle\sum_{i=1}^{t}{\bf{\mathcal{P}}}_{t,i}\bf{x}_i^{\rm{in}},\label{Eqn:MIP_LE}\\
\!\!\!\!\!\! {\text {IRT}:}& \;\;    \bf{s}^{\rm in}_t =\bf{\Xi}^{\rm H}\bf{x}^{\rm out}_t, \\
\!\!\!\!\!\!{\text {MNLD}:}&  \;\;  \bf{s}_{t+1}^{\rm out} = \phi_t (\bf{S}^{\rm in}_t),   \\
\!\!\!\!\!\!{\text {RT}:}& \;\;    \bf{x}_{t+1}^{\rm{in}}=\bf{\Xi}\bf{s}^{\rm out}_{t+1},    
\end{align}\ES 
where $\bf{X}_t^{\rm{in}}=\{\bf{x}_{1}^{\rm{in}}, \ldots, \bf{x}_{t}^{\rm{in}}\}$, $\bf{X}_t^{\rm{out}}=\{\bf{x}_{1}^{\rm{out}},$ $ \ldots, \bf{x}_{t}^{\rm{out}}\}$, $\bf{X}_t^{\rm{in}}$ and $\bf{x}_t^{\rm{out}}$ denote the input and output estimates of MLD, $\bf{S}_t^{\rm{in}}=\{\bf{s}_{1}^{\rm{in}}, \ldots, \bf{s}_{t}^{\rm{in}}\}$, $\bf{S}_t^{\rm{out}}=\{\bf{s}_{1}^{\rm{out}}, \ldots, \bf{s}_{t}^{\rm{out}}\}$,  $\bf{S}_t^{\rm{in}}$ and $\bf{s}_t^{\rm{out}}$ denote the input and output estimates of MNLD,
$\bf{\mathcal{Q}}_t\bf{A}$ and $\{\bf{\mathcal{P}}_{t, i}\}$ are polynomials in $\bf{A}^{\rm H}\bf{A}$. 
To ensure orthogonality in MLD, ${\bf{Q}}_t$ and $\{{\bf{P}}_{t,i}\}$ are required to satisfy:
\BE\label{Eqn:QP}
\tfrac{1}{N} {\rm tr} \big\{\bf{\mathcal{Q}}_t\bf{A}  \big\} =1,\;\; {\rm and} \;\; {\rm tr} \big\{ \bf{\mathcal{P}}_{t,t'}\big\} =0, \; t'=1,\dots,t.
\EE
We assume that the norms of $\bf{\mathcal{Q}}_t$ and $\bf{\mathcal{P}}_{t}$ are finite. Thus, $\gamma_t(\cdot)$ is Lipschitz-continuous. Furthermore, we assume that  $\phi_t(\cdot)$ is Lipschitz-continuous and divergence-free, i.e.,
\BE\label{Eqn:diverg}
   {\rm E} \{ {\partial \phi_t(\bf{S}^{\rm in}_t)}/{\partial  \bf{s}^{\rm in}_{t'}}\}=0, \quad\forall t'=1,\dots, t,
\EE
which ensures the orthogonality in MNLD.

\subsubsection{State Evolution (SE)}
Suppose that Assumption \ref{ASS:Model} holds. For the system with random multiplexing in \eqref{Eqn:RUP_sys}, the asymptotic IID Gaussianity result presented in \cite[Theorem 3]{Rishabh2024} (along with closely related earlier works in \cite{CAMP, KeigoISIT19, LeiMAMP}) allows us to evaluate the output covariances of $\gamma_t(\cdot)$ and $\phi_t(\cdot)$ in \eqref{Eqn:MIP} by the SE functions given by: As $N\to \infty$,
\BS\label{Eqn:IIDG}
\begin{align}
&v_{t,t'}^{\gamma}\! \!\overset{\rm a.s.}{=}\! \tfrac{1}{N}{\rm E}\Big\{\big[\gamma_t\big(\bf{X}\! +\!\bf{W}_{t})\!-\!\bf{x}\big]^{\rm H}\big[\gamma_{t'}\big(\bf{X}\!+\!\bf{W}_{t'}\big)\!-\!\bf{x}\big]\Big\},\\
&v_{t\!+\!1,t'\!+\!1}^{\phi}\! \!\overset{\rm a.s.}{=}\! 
\tfrac{1}{N}{\rm E}\Big\{\big[\phi_t\big(\bf{S}\!+\!\bf{Z}_{t} )\!-\!\bf{s}\big]^{\rm H} \big[\phi_{t'}\big(\bf{S}\!+\!\bf{Z}_{t'}\big)\!-\!\bf{s}\big]\Big\},
\end{align}
\ES
where $\bf{X}=\bf{x}\cdot \bf{1}^{\rm{T}}$ and $\bf{S}=\bf{s}\cdot\bf{1}^{\rm{T}}$ with $\bf{1}$ denoting an all-ones vector of proper size, $\bf{W}_t=[\bf{w}_1,\cdots\!,\bf{w}_t]$ and $\bf{Z}_t=[\bf{z}_1,\cdots\!,\bf{z}_t]$ are independent of $\bf{x}$ and $\bf{s}$, with IID Gaussian columns and jointly Gaussian rows that satisfy $\bf{w}_t\sim \mathcal{CN}(\bf{0}, v_{t,t}^{\phi}\bf{I})$ with ${\rm E}\{\bf{w}_t(\bf{w}_{t'})^{\rm H}\}\!=\!v_{t,t'}^{\phi}\bf{I}$, and $\bf{z}_t\!\!\sim\!\mathcal{CN}(\bf{0},v_{t,t}^{\gamma}\bf{I})$ with ${\rm E}\{\bf{z}_t(\bf{z}_{t'})^{\rm H}\}\!=\!v_{t,t'}^{\gamma}\bf{I}$. Since the RT matrix $\bf{\Xi}$ is unitary, the covariance matrices $\bf{V}_t^{\gamma}\equiv[v_{i,j}^{\gamma}]_{t\times t}$ and $\bf{V}_t^{\phi}\equiv[v_{i,j}^{\phi}]_{t\times t}$ remain invariant under the RT and IRT in \eqref{Eqn:MIP}. Hence, we rewrite the SE of CD-MAMP in \eqref{Eqn:IIDG} as: Let $\bf{v}_t^{\gamma}\equiv[v_{t,1}^{\gamma},\cdots\!, v_{t,t}^{\gamma}]^{\rm T}$ and $\bf{v}_t^{\phi}\equiv[v_{t,1}^{\phi},\cdots\!, v_{t,t}^{\phi}]^{\rm T}$. Starting with $v_{1,1}^{\phi}=1$,
\BS\label{Eqn:MAMP_SE0}
\begin{align}
    \bf{v}_t^{\gamma}&=\gamma_{\text{SE}}(\bf{V}_t^{\phi}), \\
     \bf{v}_{t+1}^{\phi}&=\phi_{\text{SE}}(\bf{V}_t^{\gamma}).
\end{align}
\ES  
The SE provides a methodology for analyzing and optimizing CD-MAMP receivers, characterizing fundamental performance limits such as achievable rates, MAP BER, and constrained capacity.  Beyond theoretical analysis, SE plays a pivotal role in system design by facilitating optimal power allocation, guiding channel coding strategies, and demonstrating the constrained-capacity optimality of CD-MAMP receivers.

\subsection{Cross-Domain OAMP/VAMP Detector}
The CD-OAMP/VAMP detector is proposed in~\cite{OTFS-OAMP,TurboCS} that can be regarded as a special case of CD-MAMP framework, i.e., the current output estimates of $\gamma_t(\bf{x}_t^{\rm{in}})$ and $\phi_t(\bf{s}_t^{\rm{in}})$ in \eqref{Eqn:MIP} depend only on the current input estimate, where the LMMSE detector $\hat{\gamma}_t(\bf{x}_t^{\rm{in}})$ is employed in the linear detector (LD) and the MMSE demodulator $\hat{\phi}_t(\bf{s}_t^{\rm{in}})=\mr{E}\{\bf{s}|\bf{s}_t^{\rm{in}}=\bf{s} + \sqrt{v_t^\gamma}\bf{z},\bf{s}\sim P_{\bf{s}}\}$ is employed in the nonlinear detector (NLD) \cite{MaAcess2017}. The detailed process is as follows: Starting with $t=1$ and $\bf{x}_1^{\rm in}=\bf{0}$,
\BS\label{Eqn:OAMP/VAMP}\begin{align}
\!\!\!\!\!\!\!{\text{LD}:}& \; \bf{x}^{\rm out}_t  =  \gamma_t \left(\bf{x}_t^{\rm{in}}\right)=  \tfrac{1}{ {\epsilon}^\gamma_t}     \hat{\gamma}_t \left(\bf{x}_t^{\rm{in}}\right) + \bf{x}_t^{\rm{in}} , \label{Eqn:OAMP/VAMP_LE}\\
\!\!\!\!\!\!\!{\text{IRT}:}& \; \bf{s}^{\rm in}_t =\bf{\Xi}^{\rm H}\bf{x}^{\rm out}_t, \\
\!\!\!\!\!{\text{NLD}:}& \; \bf{s}_{t+1}^{\rm{out}} \! =\! \phi_t \left( \bf{s}^{\rm in}_t \right)\!=\!  \tfrac{1}{\epsilon^\phi_{t+1}} \left(  \hat\phi_t(\bf{s}^{\rm in}_t) \!+\! (\epsilon^\phi_{t+1}\!-\!1) \bf{s}^{\rm in}_t \right),\label{Eqn:OAMP/VAMP_NLE} \\
\!\!\!\!\!\!\!{\text{RT}:}& \;\bf{x}_{t+1}^{\rm{in}}=\bf{\Xi}\bf{s}^{\rm out}_{t+1}, 
\end{align}
where $\{\epsilon^\gamma_t, \epsilon^\phi_{t+1}\}$ denote the orthogonalization parameters, i.e.,
\begin{align}
 \epsilon^{\gamma}_t &= \tfrac{v_t^\phi}{v_t^\phi+v_t^\gamma}= \tfrac{1}{N}\mr{tr}\left\{\bf{A}^{\mr{H}}\left[ \tfrac{\sigma^2}{v_{t}^{\phi}}\bf{I}+\bf{A}\bf{A}^{\mr{H}}\right]^{-1}\bf{A}\right\}, \\
 \epsilon^{\phi}_{t+1} &= \tfrac{v_t^\gamma}{v_t^\gamma+v_{t+1}^\phi} = 1-\tfrac{1}{Nv_{t}^{\gamma}}\big\|\hat{\phi}_t\big(\bf{s} + \sqrt{v_t^\gamma}\bf{z}\big)-\bf{s}\big\|^2,
\end{align}
and 
\BE\label{Eqn:LMMSE}
\hat{\gamma}_t \left(\bf{x}_t^{\rm{in}}\right) \equiv \bf{A}^{\rm H} \big[(\sigma^2/v_{t}^{\phi})\bf{I} +   \bf{A}\bf{A}^{\mr H}\big]^{-1}(\bf{y}-\bf{A} \bf{x}_t^{\rm{in}} ).
\EE\ES

Based on \eqref{Eqn:MAMP_SE0} and \eqref{Eqn:OAMP/VAMP}, the SE of CD-OAMP/VAMP is single-input single-output based on scalar variances, i.e.,
\BS\label{Eqn:OAMPSE}
\begin{align}
\text{LD:}& \; v^\gamma_t =\gamma_{\mr{SE}}(v_t^\phi) = {v}^\phi_t  [({\epsilon}_t^\gamma) ^{-1} -1],\\
\text{NLD:}& \; {v}^\phi_{t+1} =  \phi_{\mr{SE}}(v^\gamma_t) = v^\gamma_t [(\epsilon^\phi_{t+1}) ^{-1} -1].
\end{align}
\ES
This scalar SE plays a critical role in the theoretical analysis of random multiplexing systems. Nevertheless, the sparsity of the time-domain channel matrix $\bf{H}$ is not effectively used in CD-OAMP/VAMP due to the matrix inversion in \eqref{Eqn:LMMSE}, resulting in high complexity and restricting their application in large-scale systems. The low-complexity CD-MAMP detector provides an effective solution to this limitation. 

\begin{figure}[t!]\vspace{-0.2cm}
    \centering
    \includegraphics[width = 0.45\textwidth]{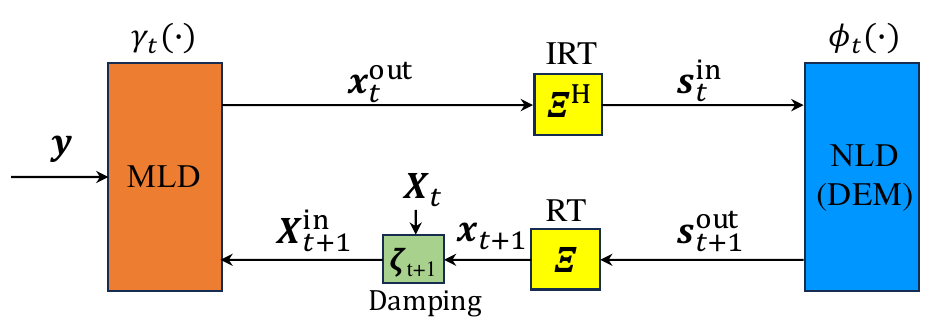}\vspace{-0.15cm}
    \caption{The CD-MAMP detector.}
    \label{fig:BO-MAMP}
\end{figure} 
\subsection{Cross-Domain Bayes-Optimal MAMP Detector}
Following the CD-MAMP framework in \eqref{Eqn:MIP}, a CD-MAMP detector is proposed as illustrated in Fig.~\ref{fig:BO-MAMP}. To avoid the prohibitive complexity of matrix inversion in \eqref{Eqn:LMMSE}, particularly in large-scale systems, we adopt a low-complexity memory matched filter ${\hat{\bf{r}}}_{t}$ as an efficient alternative for estimating the time-domain signal $\bf{x}$. The detailed process is as follows: Starting with $t=1$, $\bf{x}_1^{\rm in}=\bf{0}_N$ and $\hat{\bf{r}}_0 = \bf{0}_M$, 
\BS\label{Eqn:MAMP}
\begin{align}
\!\!\!{\text {MLD}:}&\;\; \bf{x}^{\rm out}_t  = \gamma_t(\bf{X}_t^{\rm{in}}) = \tfrac{1}{{\epsilon}^\gamma_t} \left( \bf{A}^{\rm H}{\hat{\bf{r}}}_{t} - \textstyle\sum\limits_{i=1}^{t}p_{t,i}\bf{x}_i^{\rm{in}} \right),\label{Eqn:MLE}\\%
\!\!\!{\text {IRT}:}& \;\; \bf{s}^{\rm in}_t =\bf{\Xi}^{\rm H}\bf{x}^{\rm out}_t, \label{Eqn:IRUT} \\
\!\!\!\!\!\!\!\!\!{\text {NLD}:}& \;\; \bf{s}^{\rm out}_{t+1}=\phi_t(\bf{s}^{\rm in}_t),\label{Eqn:NLE}\\ 
\!\!\!{\text {RT}:}& \;\; \bf{x}_{t+1} =\bf{\Xi}\bf{s}^{\rm out}_{t+1}, \label{Eqn:RUT}\\
\!\!\!{\text {Damp}:}&\;\; \bf{x}^{\rm in}_{t+1}  = [\bf{x}_1^{\rm{in}},\cdots\!,\bf{x}^{\rm{in}}_{t}, \bf{x}_{t+1}]\cdot \bf{\zeta}_{t+1},   \label{Eqn:damp}
\end{align}
\ES
where $\hat{\bf{r}}_t = \bf{B}_t \hat{\bf{r}}_{t-1} + \xi_t(\bf{y}-\bf{Ax}_t^{\rm in})$, $\bf{B}_t = \theta_t({\lambda^\dagger\bf{I}-\bf{AA}^{\rm H}})$ with ${\lambda}^\dag=[\lambda_{\max}+\lambda_{\min}]/2$, $\lambda_{\min}$ and $\lambda_{\max}$ denote the minimal and maximal eigenvalues of $\bf{A}\bf{A}^{\rm H}$, respectively. The parameters $\{\epsilon_t^{\gamma}, p_{t,i}\}$ ensure the orthogonality of input and output errors according to \eqref{Eqn:QP} and \eqref{Eqn:diverg}, while $\{\bf{\zeta}_{t+1}, \theta_t, \xi_t\}$ are optimized to ensure the convergence or enhance the convergence rate of CD-MAMP. For more details, refer to \cite{LeiMAMP}.

\begin{lemma}[Asymptotic IID Gaussianity \cite{Rishabh2024}]\label{The:IIDG_BOMAMP}
Suppose that Assumption \ref{ASS:Model} holds. The asymptotic IID Gaussianity in \eqref{Eqn:IIDG} holds for the CD-MAMP detector in \eqref{Eqn:MAMP} when applied to random multiplexing systems.
\end{lemma}

\emph{Remark:} Ref. \cite{Rishabh2024} extends the asymptotic IID Gaussianity of OAMP/VAMP to matrices in a general universality class, while \cite{LeiMAMP} demonstrates that MAMP satisfies the same asymptotic IID Gaussianity conditions as OAMP/VAMP. As a result, for matrices in the universality class, both OAMP/VAMP and MAMP exhibit rigorous state evolution and are replica optimal. Furthermore, CD-MAMP and MAMP are mathematically equivalent, differing only in implementation: CD-MAMP performs memory linear estimation in the time domain to leverage channel sparsity, whereas MAMP operates in the random transform domain. Consequently, the asymptotic IID Gaussianity, state evolution and replica optimality that hold for MAMP also apply for CD-MAMP.

By Lemma \ref{The:IIDG_BOMAMP}, the MSE performance of CD-MAMP can be tracked using the SE in \eqref{Eqn:MAMP_SE0}. However, the high-dimensional covariance matrices $\bf{V}_t^{\gamma}$ and $\bf{V}_t^{\phi}$ in \eqref{Eqn:MAMP_SE0} complicate direct application in performance analysis and optimization. This challenge is addressed by the following lemma, which establishes the fixed-point consistency between CD-MAMP and CD-OAMP/VAMP. This lemma simplifies the complex covariance-based SE analysis of CD-MAMP by using the scalar variance-based SE of CD-OAMP/VAMP in \eqref{Eqn:OAMPSE}.

\begin{lemma}[Fixed-Point Consistency\cite{LeiMAMP}]\label{Lem:same_fp}
Let the SE fixed point of CD-MAMP in \eqref{Eqn:MAMP} be $(v_{*}^{\gamma}, v_{*}^{\phi})$, where $v_{*}^{\gamma}=\lim\limits_{t \to \infty } v_{t,t}^{\gamma}$ and $v_{*}^{\phi}=\lim\limits_{t \to \infty } v_{t,t}^{\phi}$. The CD-MAMP in \eqref{Eqn:MAMP} has the same SE fixed point $(v_{*}^{\gamma}, v_{*}^{\phi})$ as that of CD-OAMP/VAMP in \eqref{Eqn:OAMP/VAMP}.
\end{lemma}

Following Lemma \ref{Lem:same_fp}, the lower-complexity CD-MAMP can be employed for practical signal detection, while the scalar SE of CD-OAMP/VAMP in \eqref{Eqn:OAMPSE} can be utilized for the performance analysis and optimization. Since CD-OAMP/VAMP (or CD-MAMP) aligns with that of OAMP/VAMP (or MAMP), the following corollary, derived from Lemma \ref{Prop:opti}, establishes the replica MAP-BER optimality of CD-MAMP in systems with random multiplexing.

\begin{corollary}[Replica MAP-BER Optimality] 
Suppose that Assumptions \ref{ASS:Model} and \ref{Asp:SEfixed} hold. For the linear systems with random multiplexing, i.e., $\bf{A\Xi} \in \mathscr{U}$, both the CD-OAMP/VAMP detector in \eqref{Eqn:OAMP/VAMP} and the CD-MAMP detector in \eqref{Eqn:MAMP} can achieve the replica MAP BER in \eqref{Eqn:replicaBER}.
\end{corollary}

To demonstrate the advantages of CD-MAMP in complexity, we present a comparison with existing state-of-the-art detectors, in which the measurement matrix $\bf{A}$ is assumed to be sparse and the number of non-zero elements per row in $\bf{A}$ is $\mathcal{K}$ ($\mathcal{K}\ll {\rm{min}}\{M,N\}$),  e.g., time-varying multipath channels\cite{tse2005fundamentals}, etc. For CD-MAMP, $\bf{AA}^{\rm H}\hat{\bf{r}}_t$ and $\bf{A}^{\rm H}\hat{\bf{r}}_t$ in \eqref{Eqn:MLE} are dominated and their time complexity is $\mathcal{O}(\mathcal{K}M\mathcal{T})$, where the maximum iteration number $\mathcal{T} \ll N$. In addition, $\bf{\Xi}$ can generally be implemented using a random interleaver and fast transformation matrices, such as DFT or Hadamard-Walsh transform matrices, with a time complexity of $\mathcal{O}(N{\rm{log}}N)$. 
The space complexity of CD-MAMP is $\mathcal{O}(\mathcal{K}M)$ to store the non-zero elements of $\bf{A}$,  ${\cal O}\big((M+N)\mathcal{T}\big)$ for $\{\bf{x}_t, \bf{x}_t^{\rm in}, \bf{x}_t^{\rm out}\}$ and $\{\bf{s}_t^{\rm in}, \bf{s}_t^{\rm out}\}$, and ${\cal O}\big(\mathcal{T}^2\big)$ for associated covariance matrix (see details in \cite{LeiMAMP}). Since $\bf{\Xi}$ can be constructed from an interleaver and a structured transform matrix (e.g., DFT or DCT), its fast algorithm avoids explicit storage. Table~\ref{Tab:complexity} presents the comparisons in time and space complexity of CD-MAMP, CD-OAMP/VAMP~\cite{OTFS-OAMP,TurboCS,EST-EQ}, symbol domain (SD) MAMP~\cite{MAMPOTFSconf}, SD Gaussian message passing (GMP)~\cite{OTFS_GMP}. Hence,  CD-OAMP/VAMP, SD-GMP, and SD-MAMP have higher complexity than CD-MAMP for $\mathcal{T}\ll N$ and $\mathcal{K}\ll {\rm{min}}\{M,N\}$.

\begin{table}[h!] \tiny
\renewcommand{\arraystretch}{1.3} 
\caption{Complexity comparisons of advanced detectors, where $\bf{\Xi}$ consists of an interleaver and a fast transform matrix (e.g., DFT or DCT), enabling fast computation without storage. $\bf{A}$ is an $M\times N$ matrix, $\mathcal{K}$ denotes the the number of non-zero elements per row in $\bf{A}$, and $\mathcal{T}$ denotes the maximum iteration number.)
}
\label{Tab:complexity} 
\centering \scriptsize  \setlength{\tabcolsep}{0.8mm}{
\begin{tabular}{c||c|c}
\hline
Algorithms & Time complexity & Space complexity   \\
\hline 
\hline 
\tabincell{c}{SD-GMP\!\!\vspace{-0.1cm}\\ \cite{OTFS_GMP}} &  $\mathcal{O}(N^2\mathcal{T})$ &$\mathcal{O}(MN)$  \\ 
\hline 
\tabincell{c}{CD-OAMP/VAMP\!\!\vspace{-0.1cm}\\ \cite{OTFS-OAMP,TurboCS,EST-EQ}}  &  $\mathcal{O}((M^2N+M^3)\mathcal{T}+2N\mathcal{T}{\mr{log}}N)$ &  $\mathcal{O}(MN)$ \\
\hline 
\tabincell{c}{SD-MAMP\!\!\vspace{-0.1cm}\\ \cite{MAMPOTFSconf}}  &  $\mathcal{O}(N^2\mathcal{T})$ &    $\mathcal{O}(MN\!+\!M\mathcal{T}\!+\!\mathcal{T}^2)$\\
\hline 
\tabincell{c}{CD-MAMP}   & $\mathcal{O}(\mathcal{K}N\mathcal{T}+2N\mathcal{T}{\mr{log}}N)$ &  $\mathcal{O}(\mathcal{K}M\!+\!M\mathcal{T}\!+\!\mathcal{T}^2)$ \\
\hline 
\end{tabular}}
\end{table}

\section{Power Allocation}\label{Sec:PA}
Generally, the base station can acquire channel state information (CSI), while low-cost transmitters face difficulties in obtaining perfect CSI directly. To address this challenge, CSI feedback has been widely studied as a means to transmit accurate CSI from the base station to the transmitter \cite{CSIFB_TCOM}. In this section, we assume that CSI is available at both the transmitter and receiver. We investigate optimal power allocation strategies for the linear system with random multiplexing, aiming to minimize the MAP BER and maximize the constrained channel capacity, respectively.

\subsection{Random Transform (RT) Domain Power Allocation}
In this section, we study the RT-domain power allocation in linear systems with random multiplexing, given by
\begin{align}
    \bm{y}=\bm{HP\Xi}\bm{s} + \bm{n}, \label{Eqn:GPA}
\end{align}
where $\bm{P} \in \mathbb{C}^{N \times N}$ is a power allocation matrix subject to ${\rm tr}\{\bm{P}\bm{P}^{\rm H}\}=P_{\rm sum}$ (total transmit power). 
\begin{theorem}[Optimal Power Allocation Matrix]\label{The:Opt_P}
    Let $\bf{H}=\bm{U}_H \bm{\Sigma}_{H} \bm{V}^{\rm H}_{H}$ be the singular value decomposition of $\bm{H}$. Then, regardless of whether the objective is to minimize the MAP BER or to maximize the constrained capacity of the asymptotic system in \eqref{Eqn:GPA}, the optimal form of $\bm{P}$ is
    \begin{align}
        \bm{P} = \bm{V}_H\bm{\Sigma}_P \bm{V}_P^{\rm H} , \label{Eqn:P_VS}
    \end{align}
    where $\bm{\Sigma}_P = \diag\{\sqrt{p_1}, \cdots\!, \sqrt{p_N}\}$ is a diagonal matrix to be optimized, and $\bm{V}_P$ is an arbitrary unitary matrix. For simplicity, and without loss of generality, we can set $\bm{V}_P = \bm{I}$ so that 
    \begin{align}
        \bm{P} = \bm{V}_H\bm{\Sigma}_P.
    \end{align}
\end{theorem}
\begin{IEEEproof}
    In \cite{EST-EQ}, it was proven that \eqref{Eqn:P_VS} is optimal for minimizing $P_{\rm sum}$ under a given MAP BER. We show that \eqref{Eqn:P_VS} is optimal for both minimizing the MAP BER and maximizing the constrained capacity under a given $P_{\rm sum}$. Our proof employs an intermediate result in their proof. See Appendix \ref{App:Opt_P} for details.
\end{IEEEproof}

Following Theorem \ref{The:Opt_P}, we reformulate \eqref{Eqn:GPA} as
\begin{align} 
    \bm{y} &= \bm{H}\bm{V}_H\bm{\Sigma}_P\bm{\Xi}\bm{s} + \bm{n}  \nonumber\\
    &=\bm{U}_H \bm{\Sigma}_{H} \bm{\Sigma}_P \bm{\Xi}\bm{s} + \bm{n}. \label{Eqn:pa2}
\end{align}
where $\bm{\Sigma}_{H}={\rm diag}\{\sigma_1,\dots, \sigma_{{\rm{min}}(M,N)}\}$ is an $M \times N$ rectangular diagonal matrix. The power allocation reduces to optimizing $\bm{p} \equiv [p_1, \cdots\!, p_N]$ subject to $\sum_{i=1}^N p_i = P_{\rm sum}$. The intuitions are as follows.
\begin{itemize}
    \item The RT matrix $\bm{\Xi}$ enhances the input isotropy of the equivalent channel matrix.
    \item The unitary matrix $\bm{V}_H$ enables us to diagonalize the channel matrix $\bf{H}$. 
    \item The power allocation matrix $\bm{\Sigma}_P$ optimizes the singular values of the equivalent channel matrix, thereby significantly improving the MAP BER or the constrained capacity of the systems.   
\end{itemize}

According to Theorem~\ref{The:PIM}, $\bm{HV}_H\bm{\Sigma}_P\bm{\Xi}$ belongs to the universality class $\mathscr{U}$. This leads to the following corollary.

\begin{corollary}[Replica MAP BER Optimality]\label{The:MAMP_opt_prob}
Suppose that Assumptions \ref{ASS:Model}  and \ref{Asp:SEfixed} hold. The replica MAP BER of the random multiplexing system in \eqref{Eqn:pa2} can be achieved by both the CD-OAMP/VAMP and CD-MAMP detectors in Section~\ref{Sec:MAMP} when substituting $\bm A$ with $\bm{HV}_H\bm{\Sigma}_P$ under the optimal power allocation vector $\bm{p}^*$. 
\end{corollary}

\textbf{Remark}: One may consider the symbol-domain power allocation, where the signal undergoes power allocation before being transmitted through multiplexing with $\bm{\Xi}$, given by
\BE\label{Eqn:pa3}
    \bm{y}=\bm{H}\bm{\Xi}\bm{\Sigma}_P\bm{s} + \bm{n}.
\EE
In this case, unequal power allocated coefficients amplify the asymmetry among the signal elements in $\bf{s}$, leading to the severe performance degradation, similar to the orthogonal case shown in Fig.~\ref{fig:ML_2D}. This is why we focus solely on the RT-domain power allocation in \eqref{Eqn:pa2}.

\subsection{Sate Evolution of CD-OAMP/VAMP and Key Properties}
For simplicity of discussion, we rewrite \eqref{Eqn:pa2} as:
\begin{align} \label{Eqn:pa0}
    \tilde{\bm y} = \tilde{\bm{\Sigma}}_{H}  \bm{\Sigma}_P \bm {\Xi} {\bm s}  + \tilde{\bm n},
\end{align}
where $\tilde{\bm y}=\bf{U}_H^{\rm{H}}\bf{y}$, $\tilde{\bm{\Sigma}}_{H}  ={\rm diag}\{\sigma_1,\cdots\!, \sigma_N\}$ with $\sigma_i = 0$ for $\min\{M,N\}<i\leq N$, and $\tilde{\bm n} \sim \mathcal{CN}(\bm{0},\sigma^2\bm{I})$. Then, the sate evolution (SE) of CD-OAMP/VAMP for the power allocated linear system is 
\BS\label{Eqn:iterSEb}\begin{align} 
    \rho_t^{\gamma} &= {\gamma}_{\mr{SE}}(v_t^{\phi},\bm{p})=\big[\hat{\gamma}_{\rm SE}(v_t^{\phi},\bm{p})\big]^{-1} - [v_t^{\phi}]^{-1}, \label{Eqn:iterSEa}\\
    v_{t+1}^{\phi}&={\phi}_{\mr{SE}}(\rho_t^{\gamma})=\big(\big[{\rm mmse}(\rho_t^{\gamma})\big]^{-1} - \rho_t^{\gamma}\big)^{-1},\label{Eqn:phi_se}
\end{align}
\ES
where
\BS\label{Eqn:iterSEc_total}
\begin{align}
    \hat{\gamma}_{\rm SE}(v_t^{\phi},\bm{p})&=\tfrac{1}{N}\mr{tr}\big\{\big([v_t^{\phi}]^{-1}\bf{I}\!+\!\sigma^{-2}\bm{\Sigma}_P^{\mr{H}}\tilde{\bm{\Sigma}}_H^{\rm H}\tilde{\bm{\Sigma}}_H\bm{\Sigma}_P\big)^{-1}\big\}\label{Eqn:iterSEc} \\
    & =  \tfrac{1}{N} \textstyle\sum_{i=1}^N ([v_t^{\phi}]^{-1} + \varrho_i p_i)^{-1}\label{Eqn:iterSEc2} \\
     & \mathop = \limits^{(a)}\tfrac{1}{N} \textstyle\sum_{i=1}^N ([v_t^{\phi}]^{-1} + \tilde{p}_i)^{-1}, 
\end{align}
\ES 
where  $\tilde{\bm{\Sigma}}_H={\text{diag}}\{\sigma_1, \cdots\!, \sigma_N\}$, $\varrho_i=\sigma_i^2/\sigma^{2}$, and (a) follows $\tilde{p}_i\equiv \varrho_i p_i$. 

Here, we provide two crucial properties essential for presenting our main results of optimum power allocation in Subsections \ref{Sec:PA_BER} and \ref{Sec:PA_Cap}.
\begin{lemma}[Concavity of $1/\hat{\gamma}_{\rm SE}$]\label{Lem:hat_gamma_Cav}
  The function $[\hat{\gamma}_{\rm SE}(v,\bm{p})]^{-1}$, where $\hat{\gamma}_{\rm SE}(\cdot)$ is given in \eqref{Eqn:iterSEc_total}, is concave w.r.t. (with respect to) $\bf p$.
\end{lemma}

\begin{IEEEproof}
See Appendix \ref{App:hat_gamma_Cav}. 
\end{IEEEproof}  

\begin{lemma}[Convexity of $\hat{\gamma}^{-1}_{\rm SE}$]\label{Lem:hat_gamma_inv_Cvex}
   Let $\hat{\gamma}^{-1}_{\rm SE}(\tilde{v},\bm{p})$ be the inverse function of $\tilde{v}=\hat{\gamma}_{\rm SE}(v,\bm{p})$ w.r.t. $v$. Then, $[\hat{\gamma}^{-1}_{\rm SE}(\tilde{v},\bm{p})]^{-1}$ is convex w.r.t. $\bf p$.
\end{lemma}

\begin{IEEEproof}
   See Appendix \ref{App:hat_gamma_inv_Cvex}. 
\end{IEEEproof} 
 
\subsection{Power Allocation to Minimize MAP BER}\label{Sec:PA_BER}
Power allocation to minimize MAP BER is tailored for the receivers employing a MAP detector cascaded by a decoder, without any iteration between the two. In this context, effective SISO-AWGN codes are sufficient for linear systems, rendering channel code optimization unnecessary. As a result, maximizing the achievable rate is equivalent to minimizing the MAP BER at the detection stage.

\subsubsection{Problem Formulation} Following Lemma \ref{Lem:replic-MMSE}, the MAP BER of the power allocated linear system in \eqref{Eqn:pa2} can be evaluated by the following lemma.

\vspace{-0.2cm}
\begin{lemma}[Replica MMSE and MAP BER]\label{Lem:MMSE_PA}
    The replica MMSE $v^*$ of the power allocated linear system in \eqref{Eqn:pa2} is the unique solution of  
\BE\label{Eqn:replicaMMSE_PA}
     {\rm mmse}^{-1}(v^*) = {\sigma^{-2}} \cdot \mathcal{R}_{\bf{R}}\left( -\sigma^{-2}v^* \right), 
\EE
where $\mathcal{R}_{\bf{R}}(\cdot)$ is the {R-transform} with $\bf{R}=\bm{\Sigma}_P^{\rm H}\tilde{\bm{\Sigma}}_H^{\rm H}\tilde{\bm{\Sigma}}_H \bm{\Sigma}_P$. Let ${\rho^*}={\rm mmse}^{-1}(v^*)$ . The replica MAP BER of the power allocated linear system in \eqref{Eqn:pa2} is 
\BE\label{Eqn:replicaBER_PA}
    {\rm BER}^*(\bm p) =Q_{\mathcal{S}}\big(\rho^*(\bm p)\big),
\EE
where $Q_{\mathcal{S}}(\rho)$ denotes the BER function of MAP demodulation for $\sqrt{\rho}\bf{x}+\bf{z}$ given signal constellation $\mathcal{S}$, i.e., $x_i \in \mathcal{S}, \forall i$, as defined in Appendix~\ref{App:MAP_BER}.
\end{lemma} 

\begin{framed}
    Following \eqref{Eqn:replicaBER_PA}, to minimize the replica MAP BER of the system in \eqref{Eqn:pa2}, the power allocation is formulated as the following optimization problem:
\BS\begin{align}
     {\mathcal P}_{1}:  \;\;   &\mathop{\rm{min}}_{\bm{p}}  \;\;  Q_{\mathcal{S}}\big(\rho^*(\bm p)\big), \\
     & {\rm s.t.}\quad  \textstyle\sum\limits_{i=1}^N p_i = P_{\rm sum}, \\ 
     & \;\;\quad\quad p_i\geq0,\ i \in [N]. 
\end{align}\ES
\vspace{-5mm}
\end{framed}

\subsubsection{Problem Transformation} 
Since $Q_{\mathcal{S}}(\cdot)$ is a monotonically decreasing function, Problem ${\mathcal P}_{1}$ reduces to
\BS\label{Eqn:MAP_BER_11}\begin{align}
     {\mathcal P}_{1.1}:  \;\;  &\mathop{\rm{max}}_{\bm{p}}  \;\;  \rho^*(\bm p), \\
     & {\rm s.t.}\quad  \textstyle\sum\limits_{i=1}^N p_i = P_{\rm sum}, \\ 
     & \;\;\quad\quad p_i\geq0,\ i\in [N].
\end{align} \ES
In general, obtaining $\rho^*$ by directly solving \eqref{Eqn:MAP_BER_11} is challenging. Fortunately, as shown in Corollary~\ref{The:MAMP_opt_prob}, CD-OAMP/VAMP can achieve the replica MAP-BER optimality in \eqref{Eqn:pa2}, and its convergence has been established in \cite{LeiMAMP, LMOAMP, liu2022sufficient}. Hence, we address this issue by analyzing the first fixed point $(\rho^*, v^*)$ of the SE of CD-OAMP/VAMP:
\BS\label{Eqn:iterSE}
\begin{align}
    \rho^{*} &= {\gamma}_{\mr{SE}}(v^*,\bm{p}), \\
    v^* &= {\phi}_{\mr{SE}}(\rho^{*}),
\end{align}
\ES
where ${\gamma}_{\mr{SE}}(\cdot)$ and ${\phi}_{\mr{SE}}(\cdot)$ are given in \eqref{Eqn:iterSEb}. Then, we have
\begin{align}
    \gamma_{\mr{SE}}(v,\bf{p})-\phi_{\mr{SE}}^{-1}(v)
    \begin{cases}
        > 0 & \quad \text{if}\ v \in (v^*, 1) \\[1mm]
        = 0 & \quad \text{if}\ v = v^*
    \end{cases}.
\end{align}
Based on this analysis, Problem $\mathcal{P}_{1.1}$ can equivalently transformed into a bi-level optimization problem: find $v_{\mr{goal}} \in (0, 1)$ such that ${\mathcal{P}}_{1.2}(v_{\mr{goal}}) = 0$, where
\BS \label{Eqn:P2_a} 
\begin{align} 
{\mathcal{P}}_{1.2}(v_{\mr{goal}}):\ \max_{\bm{p}} \; \min_{v \in [v_{\rm goal}, 1)} &\;  \gamma_{\mr{SE}}(v,\bf{p})-\phi_{\mr{SE}}^{-1}(v),\\
{\rm s.t.}\;&     \;\;  \textstyle\sum\limits_{i=1}^N p_i =P_{\rm sum},\\
& \;\; p_i\geq0,\ i \in [N].
\end{align}\ES
For any fixed $P_{\rm sum}$, $\mathcal{P}_{1.2}(v_{\mr{goal}})$ is monotonically increasing w.r.t $v_{\rm goal}$, which can be proven by contradiction. Hence, the unique $v_{\rm goal}$ satisfying ${\mathcal{P}}_{1.2}(v_{\mr{goal}}) = 0$ can be obtained via a bisection search. However, it is hard to compute the minimum of $\gamma_{\mr{SE}}(v,\bf{p})-\phi_{\mr{SE}}^{-1}(v)$ over the continuous interval ${v \in [v_{\rm goal}, 1)}$. To address this issue, we replace it with the minimum over $\mathcal{V}_{\rm goal}$, a discrete set of \emph{log-uniformly sampled points within $[v_{\rm goal},1)$.} In practice, $100$ sampling points are typically used, i.e., $|\mathcal{V}_{\rm goal}| = 100$.
\begin{framed}
In summary, we can solve $\mathcal{P}_{1}$ numerically by: finding $v_{\rm goal} \in (0, 1)$ via a bisection search, such that the corresponding $\mathcal{V}_{\rm goal}$ satisfies $\mathcal{P}_{1.3}(\mathcal{V}_{\rm goal}) = 0$, where
\BS \label{Eqn:P2_b} \begin{align} 
{\mathcal P}_{1.3}(\mathcal{V}_{\rm goal}): \ \mathop{\rm{max}}_{\bm{p}} \; \mathop{\rm{min}}_{v \in \mathcal{V}_{\rm goal}} &\;  \gamma_{\mr{SE}}(v,\bf{p})-\phi_{\mr{SE}}^{-1}(v),\\
{\rm s.t.}\;&     \;\;  \textstyle\sum\limits_{i=1}^N p_i =P_{\rm sum},\\
& \;\; p_i\geq0,\ i \in [N].
\end{align}\ES
\vspace{-5mm}
\end{framed}

\subsubsection{Concavity} 
The following lemma shows that ${\mathcal P}_{1.3}$ is a concave maximization problem over a convex set, which is well-known to be equivalent to a convex problem, enabling its solution via standard convex optimization solvers.
\begin{lemma}\label{Lem:convexity_P}
    ${\mathcal P}_{1.3}$ is a concave maximization problem over a convex set. 
\end{lemma}
\begin{IEEEproof}
   It is easy to verify that the constraints of ${\mathcal P}_{1.3}$ are convex. Hence, we only need to show the concavity of $\mathop{\rm{min}}_{v \in \mathcal{V}^*} \{\gamma_{\mr{SE}}(v,\bf{p})-\phi_{\mr{SE}}^{-1}(v)\}$ w.r.t. $\bm{p}$. Since $\phi_{\mr{SE}}^{-1}(v)$ is independent of $\bm{p}$ and the minimum operation preserves concavity, proving that $\gamma_{\mr{SE}}(v,\bm{p})$ is concave w.r.t. $\bm p$ is sufficient. Given $\gamma_{\mr{SE}}(v,\bf{p}) = [\hat{\gamma}_{\rm SE}(v_t^{\phi},\bm{p})]^{-1} - [v_t^{\phi}]^{-1}$, the problem reduces to showing the concavity of $[\hat{\gamma}_{\mr{SE}}(v,\bf{p})]^{-1}$ w.r.t. $\bm p$, which was established in Lemma \ref{Lem:hat_gamma_Cav}. Thus, we finish the proof.  
\end{IEEEproof}

\subsubsection{Numerical Solution} The search interval is initialized to $(0, v^{\rm up})$. Setting $v^{\rm up} = 1$ is always a natural choice. The iterative process is: Let $v_1^{\rm low} = 0$ and $v_1^{\rm up} = v^{\rm up}$. For $i \geq 1$, 
\begin{itemize}
    \item Step 1: Compute $v_{i} = (v_i^{\rm low}+v_i^{\rm up}) / 2$, and obtain the set $\mathcal{V}_i$ by uniformly sampling on $[v_i, 1)$ in log domain.
    \item Step 2: Find the solution $\bm{p}^{(i)}$ and the corresponding objective function value $c_i$ of ${\mathcal P}_{1.3}$ with $\mathcal{V}_{\rm goal} =\mathcal{V}_i$ by convex optimization solvers. 
    \item Step 3: 
    \begin{itemize} 
        \item If $c_i > 0$, $v_{i+1}^{\rm up} = v_i$ and $v_{i+1}^{\rm low} = v_{i}^{\rm low}$;
        \item If $c_i = 0$, stop iteration and return $\bm{p}^{(i)}$;
        \item If $c_i < 0$, $v_{i+1}^{\rm low} = v_i$ and $v_{i+1}^{\rm up} = v_{i}^{\rm up}$.
    \end{itemize}
\end{itemize}

In practice, the SE of CD-OAMP/VAMP is not perfectly accurate since the system size is finite. To ensure that CD-OAMP/VAMP converges to its SE fixed point, we revise step 1 and step 3 by introducing a small threshold $\epsilon > 0$:
\begin{itemize}
    \item Step 1: Compute $v_{i} = (v_i^{\rm low}+v_i^{\rm up}) / 2$, and obtain the set $\mathcal{V}_i$ by uniformly sampling on $(v_i, 1)$ in log domain.
    \item Step 3: Compute $d_i = \gamma_{\mr{SE}}(v_i,\bf{p}^{(i)})\!-\!\phi_{\mr{SE}}^{-1}(v_i)$.
    \begin{itemize} 
        \item If $c_i > \epsilon$ and $d_i > \epsilon$, $v_{i+1}^{\rm up} = v_i$ and $v_{i+1}^{\rm low} = v_{i}^{\rm low}$;
        \item If $c_i > \epsilon$ and $d_i \leq \epsilon$, stop searching and return $\bf{p}^{(i)}$;
        \item Otherwise, $v_{i+1}^{\rm low} = v_i$ and $v_{i+1}^{\rm up} = v_{i}^{\rm up}$.
    \end{itemize}
\end{itemize}
The pseudocode is given in Algorithm \ref{Alg:PA_MAP}. The stopping threshold $\Delta_v$ controls the precision of the output $v$, and the threshold $\epsilon$ serves as a safety margin to account for potential mismatch between the SE and the actual performance of CD-OAMP/VAMP in finite systems. In our simulations, we set $\Delta v = 10^{-6}$ to ensure $\Delta v/v \ll 1$. For systems with $N=2048$, we set $\epsilon=0.45$ empirically.
\begin{algorithm}[htb]
    \caption{Power allocation to minimize MAP BER}\label{Alg:PA_MAP}
    \begin{algorithmic}[1]
        \renewcommand{\algorithmicrequire}{\textbf{Input:}}
        \renewcommand{\algorithmicensure}{\textbf{Output:}}
        \Require $\sigma^2$, $N$, $P_{\rm sum}$, $\sigma_1, \cdots\!, \sigma_N$ 
        \Ensure $\bf{p}$, $v$
        \State Set $\epsilon$, $\Delta v$, $v_{\rm up}$ and $v_{\rm low}=0$
        \While{$v_{\rm up} - v_{\rm low} > \Delta v$}
            \State $v = (v_{\rm low} + v_{\rm up}) / 2$
            \State Get $\mathcal{V}_{\rm goal}$ by log-uniformly sampling in $(v, 1)$
            \State Get $\bm{p}$ and $c$ by solving ${\mathcal P}_{1.3}$
            \State $c = \mathop{\rm{min}}_{v \in\mathcal{V}_{\rm goal}}\{\gamma_{\mr{SE}}(v,\bm{p})-\phi_{\mr{SE}}^{-1}(v)\}$
            \If{$c > \epsilon$}
                \State $d = \gamma_{\mr{SE}}(v,\bm{p})-\phi_{\mr{SE}}^{-1}(v)$
                \If{$d \leq \epsilon$}
                    \State \textbf{break}
                \EndIf
                \State $v_{\rm up} = v$
            \Else
                \State $v_{\rm low} = v$
            \EndIf
        \EndWhile
    \end{algorithmic}
\end{algorithm}

 \emph{Note:} In \cite{EST-EQ}, power allocation is addressed by minimizing the total power under a target MAP BER, formulated as optimizing $\bm{p}$ to minimize $P_{\rm sum}= \sum_{i=1}^N p_i$ subject to $\gamma_{\mr{SE}}(v,\bf{p})-\phi_{\mr{SE}}^{-1}(v)\geq 0$ for $v \in [v^*, 1]$. This problem is convex if $1/\hat{\gamma}_{\rm SE}$ is concave w.r.t $\bm{p}$, with the proof, omitted in \cite{EST-EQ}, derived in Lemma \ref{Lem:hat_gamma_Cav}. Consequently, the problem can be solved by convex optimization solvers. Nevertheless, directly solving $\mathcal{P}_{1.1}$ in \eqref{Eqn:MAP_BER_11} is challenging. To overcome this, we reformulate $\mathcal{P}_{1.1}$ into a bisection search framework based on $\mathcal{P}_{1.3}$ in \eqref{Eqn:P2_b}. Furthermore, we offer practical implementation details, such as discretizing the continuous intervals in the logarithmic domain and introducing a threshold $\epsilon$ to account for the inaccuracy of the SE. These measures are essential for solving both the power allocation problem in this paper and the one in \cite{EST-EQ}.

\subsection{Power Allocation to Maximize Constrained Capacity}\label{Sec:PA_Cap}
In this subsection, we study the power allocation to maximize the constrained capacity of the power allocated linear system in \eqref{Eqn:pa2}, an issue not addressed in \cite{EST-EQ}.
\subsubsection{Problem Formulation} 
The constrained capacity of the system in \eqref{Eqn:pa2} can be evaluated by the following lemma.

\begin{lemma}[Replica Constrained Capacity]\label{Pro:Cap_PA}
     The replica constrained capacity per transmit symbol of the linear system in \eqref{Eqn:pa2} is given by
\BE\label{Eqn:C_MIMO_PA}
C_{\rm MIMO}(\bm{p}) =  \int_0^{v^*{\rm snr}} \!\!\!\!\!\!\!\!\!\! \mathcal{R}_{\bf{R}}(-z)dz + C_{\rm SISO}(\rho^*) - \rho^*v^*,
\EE
where $\bf{R}=\bm{\Sigma}_P^{\rm H}\tilde{\bm{\Sigma}}_H^{\rm H}\tilde{\bm{\Sigma}}_H\bm{\Sigma}_P$, $\rho^*={\rm mmse}^{-1}(v^*)$, and $v^*$ is the replica MMSE given by the unique solution of \eqref{Eqn:replicaMMSE_PA}. 
\end{lemma}     

\begin{framed}
To maximize the constrained capacity of the system in \eqref{Eqn:pa2}, the power allocation is formulated as the following optimization problem:
\BS\label{Eqn:optP_rate}
\begin{align}
      {\mathcal P}_{2}:  \;\;    &\mathop{\rm{max}}_{\bm{p}}  \;\;  C_{\rm MIMO}(\bm{p}),\\
     & {\rm s.t.}\quad  \textstyle\sum\limits_{i=1}^N p_i = P_{\rm sum}, \\ 
     & \;\;\quad\quad p_i\geq 0,\; i=1,\cdots\!, N,
\end{align}\ES 
where $C_{\rm MIMO}(\bm{p})$ is given in \eqref{Eqn:C_MIMO_PA}. 
\end{framed}

\begin{figure}[t!]
    \centering
    \includegraphics[width = 0.9\columnwidth]{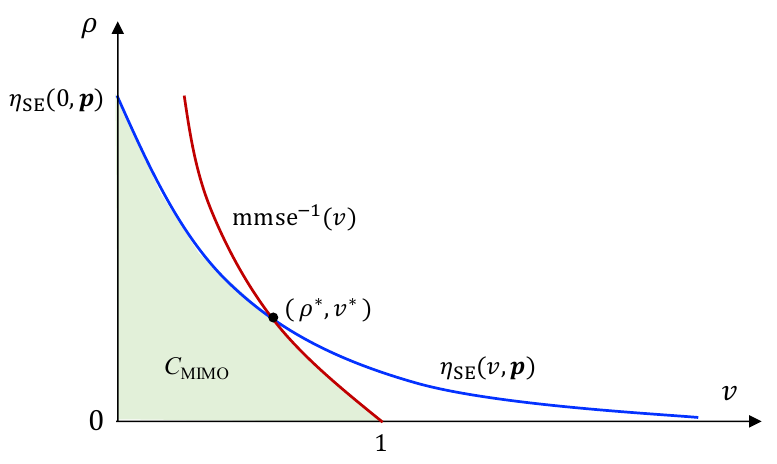}
    \caption{Constrained capacity per transmit symbol of the random multiplexing system.}
    \label{fig:const_cap}
\end{figure}

For Gaussian signaling $\bf{s}$, $C_{\rm MIMO}$ simplifies to the Gaussian MIMO capacity $C_{\rm Gau-MIMO}(\bm{p}) = \frac{1}{N} \sum_{i=1}^N \log\left(1 + p_i \varrho_i\right)$, where $\varrho_i=\sigma_i^2/\sigma^2$. In this case, the optimal power allocation corresponds to the \emph{waterfilling} solution\cite{Lozano2006}. Conversely, for non-Gaussian signaling $\bf{s}$, the power allocation becomes more complex. Following the capacity-area theorem in \cite[Theorem 1]{LeiOptOAMP}, the linear capacity $C_{\rm MIMO}$ in \eqref{Eqn:C_MIMO_PA} can be reformulated to
\BS\label{Eqn:TF}\BE
    C_{\rm MIMO} (\bf{p}) = \int_{0}^1 \min\big\{{\eta}_{\rm SE}(v,{\bm p}), {\rm mmse}^{-1}(v)\big\} d v,
\EE
where ${\eta}_{\rm SE}(\cdot)$ is the variational transform function given by
\begin{align}
    {\eta}_{\rm SE}(v,\bf{p})\equiv v^{-1}-[\hat{\gamma}_{\rm SE}^{-1}(v,{\bm p})]^{-1} \label{Eqn:eta_p}
\end{align}\ES
with $\hat{\gamma}^{-1}_{\rm SE}(v,\bm{p})$ being the inverse function of $v=\hat{\gamma}_{\rm SE}(\tilde{v},\bm{p})$ w.r.t. $\tilde{v}$.
Fig. \ref{fig:const_cap} provides a graphic illustration of \eqref{Eqn:TF}. Note that the ${\phi}_{\mr{SE}}(\cdot)$ in \eqref{Eqn:phi_se} involves an orthogonalization operation in the SE of CD-OAMP/VAMP, rendering it no longer locally MMSE optimal. This complicates the analysis of constrained capacity and achievable rates using the I-MMSE lemma \cite{Guo2005}. This difficulty can be overcame by utilizing the variational transform functions of CD-OAMP/VAMP, as outlined in \cite[Equation (37)]{LeiOptOAMP}, preserving the same fixed point as the SE described in \eqref{Eqn:iterSEb}.

\subsubsection{Convexity} 
The following lemma shows that ${\mathcal P}_{2}$ is
a concave maximization problem over a convex set, thus
enabling its solution via standard convex optimization solvers.
\begin{lemma}\label{Lem:P2_convex}
    $ {\mathcal P}_{2}$ is a concave maximization problem over a convex set.
\end{lemma}
\begin{IEEEproof}
It is straightforward to verify that the constraints of Problem ${\mathcal P}_{2}$ are convex. Hence, we only need to show that $C_{\rm MIMO} (\bf{p})$ is concave w.r.t. $\bm{p}$. Following Lemma \ref{Lem:hat_gamma_inv_Cvex}, $[\hat{\gamma}^{-1}_{\mr{SE}}(\tilde{v},\bf{p})]^{-1}$ is convex w.r.t. $\bm p$. Therefore, ${\eta}_{\rm SE}(\tilde{v},\bf{p})$ is concave w.r.t. $\bm p$. Furthermore, since ${\rm mmse}^{-1}(\tilde{v})$ is independent of $\bm{p}$, and the integral operation and the minimum operation both preserve concavity, we obtain that $C_{\rm MIMO} (\bf{p})$ is concave w.r.t. $\bm{p}$. Hence, we complete the proof of Lemma \ref{Lem:P2_convex}.
\end{IEEEproof}\vspace{3mm}

\begin{algorithm}[b!]
    \caption{Power allocation to maximize capacity}\label{Alg:PA_Cap}
    \begin{algorithmic}[1]
        \renewcommand{\algorithmicrequire}{\textbf{Input:}}
        \renewcommand{\algorithmicensure}{\textbf{Output:}}
        \renewcommand{\algorithmicprocedure}{\textbf{Function}}
        \Require $\sigma^2$, $N$, $P_{\rm sum}$
        \Ensure $\bf{p}$, $C(\bf{p})$
        \State Set $\epsilon$, $v^{\rm low}$ 
        \State Get $\bf{p}$ and $C(\bf{p})$ by solving ${\mathcal P}_{2}$ with the target function $C(\bf{p}) = \int_{v^{\rm low}}^1 {\rm fun}(v, \bf{p}) d v + v^{\rm low} {\rm fun}(v^{\rm low},\bf{p})$.
        \Procedure{${\rm fun}$}{$v$, $\bf{p}$}
        \State $\tilde{v}^{\rm up} = 1$, $\tilde{v}^{\rm low} = 0$
        \While{$\tilde{v}^{\rm up} - \tilde{v}^{\rm low} > \epsilon$}
            \State $\tilde{v} = (\tilde{v}^{\rm up} + \tilde{v}^{\rm low})/2$
            \If{$\big|\hat{\gamma}_{\rm SE}(\tilde{v}, \bf{p}) - v\big| < \epsilon$}
            \State \textbf{break}
            \ElsIf{$\hat{\gamma}_{\rm SE}(\tilde{v}, \bf{p}) - v > \epsilon$}
            \State $\tilde{v}^{\rm up} = \tilde{v}$
            \Else
            \State $\tilde{v}^{\rm low} = \tilde{v}$
            \EndIf
        \EndWhile
        \State $\eta = v^{-1} - \tilde{v}^{-1}$
        \State \textbf{return} $\min\{\eta, {\rm mmse}^{-1}(v)\}$
        \EndProcedure
    \end{algorithmic}
\end{algorithm}

\subsubsection{Numerical Solution} It is intractable to find the closed-form expressions of ${\rm mmse}^{-1}(v)$ and $\eta_{\rm SE}(v,\bf{p})$. Therefore, we compute them as follows:
\begin{itemize}
    \item Since $v = {\rm mmse}(\rho)$ is independent of $\bf{p}$, we precompute a lookup table of ${\rm mmse}(\rho)$, where $\rho$ takes values from a uniformly spaced discrete set $\{\rho_i\}$ over the interval $[0, \hat{\rho}]$. Typically, we ensure $|\{\rho_i\}| \geq 10^4$ and choose $\hat{\rho}$ such that ${\rm mmse}(\hat{\rho})$ is small enough. This allows us to obtain ${\rm mmse}^{-1}(v)$ via the lookup table and interpolation.
    \item Following \eqref{Eqn:iterSEc2}, $\hat{\gamma}_{\rm SE}(\tilde{v}, \bf{p})$ is monotonically increasing w.r.t $\tilde{v}$, which implies that $\hat{\gamma}_{\rm SE}^{-1}(v,{\bm p})$ is increasing w.r.t $v$. Therefore, we obtain $\hat{\gamma}_{\rm SE}^{-1}(v,{\bm p})$ using a bisection search. We initialize $\tilde{v}^{\rm up}_1 = 1$, $\tilde{v}^{\rm low}_1 = 0$, and set a small threshold $\epsilon > 0$. For each iteration $i \geq 1$, $\tilde{v}_i = (\tilde{v}^{\rm up}_i + \tilde{v}^{\rm low}_i) / 2$, and 
    \begin{itemize}
        \item If $\hat{\gamma}_{\rm SE}(\tilde{v}_i, \bf{p}) > v + \epsilon$, $\tilde{v}^{\rm up}_{i+1} = \tilde{v}_i$ and $\tilde{v}^{\rm low}_{i+1} = \tilde{v}^{\rm low}_i$.
        \item If $\hat{\gamma}_{\rm SE}(\tilde{v}_i, \bf{p}) < v - \epsilon$, $\tilde{v}^{\rm up}_{i+1} = \tilde{v}^{\rm up}_{i}$ and $\tilde{v}^{\rm low}_{i+1} = \tilde{v}_i$.
        \item Otherwise, stop searching and return $\hat{\gamma}_{\rm SE}^{-1}(v,{\bm p}) = \tilde{v}_i$.
    \end{itemize}
    
\end{itemize}
Furthermore, we approximate $C_{\rm MIMO} (\bf{p})$ in (\ref{Eqn:TF}) by
\begin{align}\label{Eqn:Cap_app}
    C_{\rm MIMO} (\bf{p}) \approx \int_{v^{\rm low}}^1 g(v, \bf{p}) d v + v^{\rm low} g(v^{\rm low},\bf{p}),
\end{align}
where $g(v, \bf{p}) \equiv \min\big\{{\eta}_{\rm SE}(v, \bf{p}), {\rm mmse}^{-1}(v)\big\}$, and $v^{\rm low}>0$ is a small threshold. The term $v^{\rm low} g(v^{\rm low},\bf{p})$ approximates $\int_{0}^{v^{\rm low}} g(v, \bf{p}) d v$ using a rectangular rule. Typically, setting $v^{\rm low} \in [10^{-4}, 10^{-3}]$ is suitable. The pseudocode is given in Algorithm \ref{Alg:PA_Cap}. The stopping threshold $\epsilon$ controls the precision of $\mr{fun}(v, \bm{p})$, thereby determining the accuracy of the integral in $C(\bm{p})$. Therefore, $\epsilon$ is required to be sufficiently small. In our simulations, we set $v^{\rm low} = 10^{-4}$ and $\epsilon=10^{-15}$.

\section{Channel Coding}\label{Sec:code}
Channel coding is crucial for achieving high-speed, reliable information transmission. Most current studies focus on well-designed codes that approach the SISO channel capacity. However, these schemes often neglect inter-symbol interference introduced by the measurement matrix $\bf{A}$. Previous research has optimized codes and demonstrated replica constrained capacity for linear systems like \eqref{Eqn:linear_sys} under the assumption that the channel $\bf{A}$ is IID Gaussian \cite{liu2021capacity} or right unitarily invariant \cite{LeiOptOAMP, Code_MAMP}. Unfortunately, real-world wireless channels rarely meet this assumption, which limits the applicability of these conclusions. This assumption is relaxed in this paper by random multiplexing, allowing both code optimization and replica constrained capacity optimality hold for arbitrary norm-bounded and spectrally convergent channel matrices. This section introduces the optimal channel coding for the linear system with random multiplexing.

\subsection{Problem Formulation}
Based on \eqref{Eqn:pa2}, a coded random multiplexing system with RT-domain power allocation is given by
\BS\label{Eqn:coded_RUP}
\begin{align}
    \text{Linear constraint}\; \Gamma: \; & \bf{y}=\bf{H}\bm{V}_H\bf{\Sigma}_P\bf{x}+\bf{n}, \\ 
    \text{Random Transform}\; T:\; &\bf{x}=\bf{\Xi}\bf{s}, \\
    \text{Code  constraint}\; \Phi_{\mathcal{C}}: \; & \bf{s}\in \bf{\mathcal{C}}, \;\; \bf{s}\sim P_S(\bf{s}), \label{Eqn:codec}
\end{align}
\ES
where $\bf{\mathcal{C}}$ is a codebook, and the power allocation matrix $\bf{\Sigma}_P$ is optimized to maximize the constrained capacity, as detailed in Section \ref{Sec:PA_Cap}. Under the constellation constraint $P_S(\bf{s})$, the random transform matrix $\bf{\Xi}$, and the optimized power allocation $\bf{\Sigma}_P$, the primary objective of code design is to achieve the constrained capacity. Thanks to the input isotropy of $\bf{\Xi}$, we can analyse the achievable rates and derive the optimal coding principle for coded random multiplexing systems by means of the SE of CD-OAMP/VAMP. Since $\bm{V}_H$ does not affect the input isotropy of $\bf{\Xi}$, when $\bf{\Sigma}_P = \bm{I}$, $\bm{V}_H\bf{\Sigma}_P\bf{\Xi}$ can be considered an equivalent RT matrix. 

\subsection{Achievable Rate Analysis}
\begin{figure}[t!]
    \centering
    \includegraphics[width = 0.8\columnwidth]{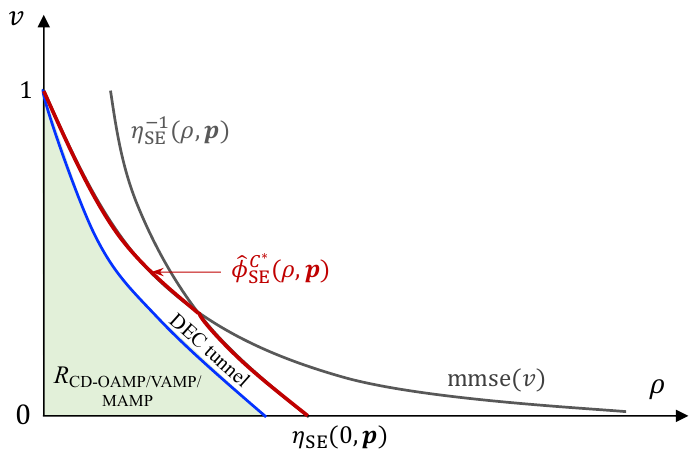}
    \caption{The variational transform curves of CD-OAMP/VAMP receiver, and achievable rates of CD-OAMP/VAMP and CD-MAMP receivers.}
    \label{fig:VSE_oamp}
\end{figure}

As shown in Fig.~\ref{fig:VSE_oamp}, we assume that there is a unique fixed point between ${\eta}_{\mr{SE}}^{-1}(\cdot)$ (see \eqref{Eqn:eta_p}) and ${\rm mmse}(\cdot)$  of CD-OAMP/VAMP, i.e., $(\rho_*, v_*)$, where ${\eta}_{\mr{SE}}^{-1}(\cdot)$ is the inverse function of ${\eta}_{\mr{SE}}(\cdot)$. Therefore, to achieve error-free signal transmission, a proper channel coding scheme is necessary to enable an available decoding tunnel between the decoder transfer function $\hat{\phi}_{\rm{SE}}^{\mathcal{C}}(\cdot)$ and linear transfer function ${\eta}_{\mr{SE}}^{-1}(\cdot)$, ensuring the error-free signal recovery, i.e., estimate variance $\lim_{t\to\infty} v_t^{\phi} \rightarrow 0$, where $\hat{\phi}_{\text{SE}}^{\mathcal{C}}({\rho}_t)=\frac{1}{N}{\mr{E}}\{||\hat{\phi}_t^{\mathcal{C}}(\bf{s}_t^{\rm{in}})-\bf{s}||^2\}$, $\hat{\phi}_t^{\mathcal{C}}(\bf{s}_t^{\rm{in}})\equiv{\mr{E}}\{\bf{s}|\bf{s}_t^{\rm{in}},\Phi_{\mathcal{C}}\}$, $\bf{s}_t^{\rm{in}}$ is given in \eqref{Eqn:NLE}, and $\Phi_{\mathcal{C}}$ is given in \eqref{Eqn:codec}. 

\begin{assumption}\label{Asp:fec}
 The a-posteriori probability (APP) decoder $\hat{\phi}_t^{\mathcal{C}}(\cdot)$ used in this paper is  uniformly Lipschitz-continuous.
\end{assumption}

Assumption~\ref{Asp:fec} is a necessary condition for the accuracy of the SE of CD-OAMP/VAMP and CD-MAMP in coded random multiplexing systems. Similar to \cite[Proposition 1]{LeiOptOAMP}, the following lemma presents the necessary and sufficient condition for error-free signal recovery in the random multiplexing system.
\begin{lemma}[Error-Free Condition]\label{Lem:err-free}
Suppose that Assumptions~\ref{ASS:Model} and \ref{Asp:fec} hold. The necessary and sufficient condition of error-free signal recovery in the coded random multiplexing system with the CD-OAMP/VAMP and CD-MAMP receiver is
\BE\label{Eqn:err-free}
\hat{\phi}_{\mr{SE}}^{\mathcal{C}}({\rho})< {\eta}_{\mr{SE}}^{-1}({\rho}, \bf{p}),\quad {\mr{for}} \;\; 0\le {\rho} \le \eta_{\text{SE}}(0,\bf{p}),
\EE
where ${\eta}_{\mr{SE}}^{-1}(\cdot)$ denotes the inverse function of ${\eta}_{\mr{SE}}(\cdot)$, and ${\eta}_{\mr{SE}}(\cdot)$ is given in \eqref{Eqn:eta_p}.
\end{lemma}

Due to the coding gain, an upper bound for the decoder can be obtained as 
\BE\label{Eqn:upperboud}
\hat{\phi}_{\mr{SE}}^{\mathcal{C}}({\rho})< {\rm mmse}({\rho}), \quad {\mr{for}} \;\;  {\rho} >0.
\EE

Based on the I-MMSE lemma in \cite{Guo2005} and \cite[Lemma 7]{LeiOptOAMP}, the achievable rate of the CD-OAMP/VAMP and CD-MAMP receiver  is given in the following lemma.
\begin{lemma}[Achievable Rate]\label{Lem:ach_rate}
Suppose that Assumptions~\ref{ASS:Model} and \ref{Asp:fec} hold. Based on Lemma~\ref{Lem:same_fp} and \cite[Theorem 3]{Code_MAMP}, given the decoder $\hat{\phi}_{\mr{SE}}^{\mathcal{C}}(\cdot)$,  the achievable rate of the CD-OAMP/VAMP and CD-MAMP per transmit symbol is given by 
\BE\label{Eqn:ach_rate}
R_{\rm{CD-OAMP/VAMP}}=R_{\rm{CD-MAMP}}=\int_{0}^{\rho_{\mathcal{C}}}\hat{\phi}_{\mr{SE}}^{\mathcal{C}}(\rho) d \rho,
\EE
where $\hat{\phi}_{\mr{SE}}^{\mathcal{C}}(\rho)<{\mr{min}}\{{\rm mmse}(\rho), {\eta}_{\mr{SE}}^{-1}(\rho, \bf{p})\}$ according to \eqref{Eqn:err-free} and \eqref{Eqn:upperboud}.
\end{lemma}

\subsection{Optimal Coding Principle and Capacity Optimality}

Based on Lemmas~\ref{Lem:err-free} and \ref{Lem:ach_rate}, the following lemma presents the optimal coding principle to maximize the achievable rate while ensuring error-free signal recovery.

\begin{lemma}[Optimal Coding Principle]\label{Lem:opt_code}
Suppose that Assumptions~\ref{ASS:Model} and \ref{Asp:fec} hold. In random multiplexing systems, the optimal decoding transform function is given by 
\BE\label{Eqn:opt_code}
\hat{\phi}_{\mr{SE}}^{\mathcal{C}}(\rho) \rightarrow \hat{\phi}_{\mr{SE}}^{\mathcal{C}^*}(\rho)={\mr{min}}\{{\rm mmse}(\rho), {\eta}_{\mr{SE}}^{-1}(\rho, \bf{p})\}.
\EE
\end{lemma}

\textbf{Remark:} The optimal coding principle relies on the availability of  $\hat{\phi}_{\mr{SE}}^{\mathcal{C}^*}(\rho)$ at the transmitter. To meet this requirement, we assume that the transmitter knows the spectral distribution of $\bm{H}$, which is more practical than requiring the full CSI.

Building on Lemmas~\ref{Prop:opti} and \ref{Lem:opt_code}, the following theorem demonstrates that CD-OAMP/VAMP and the low-complexity CD-MAMP achieve the replica-constrained capacity in random multiplexing systems.

\begin{theorem}[Capacity Optimality]\label{The:maxrate}
Suppose that Assumptions~\ref{ASS:Model}-\ref{Asp:fec} hold. When $\hat{\phi}_{\mr{SE}}^{\mathcal{C}}(\rho) \to  \hat{\phi}_{\mr{SE}}^{\mathcal{C}^*}(\rho)\!=\!{\mr{min}}\{{\rm mmse}(\rho), {\eta}_{\mr{SE}}^{-1}(\rho, \bf{p})\}$, the achievable rate of the CD-OAMP/VAMP and CD-MAMP receiver per transmit symbol is maximized and expressed as 
\BE\label{Eqn:rateMAMP}
  	R_{\rm{CD-OAMP/VAMP}}^{\rm{max}}=R_{\rm{CD-MAMP}}^{\rm{max}} \to  \int_0^{\eta_{\text{SE}}(0, \bf{p})}\hat{\phi}_{\mr{SE}}^{\mathcal{C}^*}(\rho) d \rho,
\EE
which corresponds to the replica constrained capacity of the random multiplexing system, i.e., 
\begin{align}\label{Eqn:cap_opt}
     R_{\rm{CD-OAMP/VAMP}}^{\rm{max}}=R_{\rm{CD-MAMP}}^{\rm{max}} \to  C_{\rm MIMO}.
\end{align}
\end{theorem}

The optimal coding principle in Lemma~\ref{Lem:opt_code} and the capacity optimality of the CD-MAMP in Theorem~\ref{The:maxrate} are established based on the SE fixed-point consistency in Lemma~\ref{Lem:same_fp}, rather than the consistency of the SE trajectory. This has been proven in \cite[Theorem 3]{Code_MAMP}, namely, when the channel matrices belong to the universality class, iterative algorithms with the same SE fixed point, such as OAMP, VAMP, and MAMP, share identical rate analysis, optimal coding principle, and capacity optimality. As a result, the codes optimized for CD-OAMP/VAMP are also optimal for CD-MAMP.

\section{Discussions}\label{Sec:dis}
In this section, we explore the generality of random multiplexing by highlighting its potential in various common wireless application scenarios.

\subsection{Conjecture: Beyond the Universality Class}
In practice, it is typical to set the random transform matrix $\bm{\Xi} = \bm{\Pi}\bm{F}\bm{D}$, where $\bm{\Pi}$ is a random permutation, $\bm{F}$  a normalized fast transform matrix, and $\bm{D}$  a random phase matrix independent of $\bm{\Pi}$. To guarantee that $\bm{A}\bm{\Xi} \in \mathscr{U}$, the condition in \eqref{Eqn:off_diag} for Theorem \ref{The:PIM} is necessary. However, it is hard to claim that practical channel matrices always satisfy this condition. We can even consider an extreme example as:
\begin{align}\label{Eqn:ZR1}
    \bm{A} = \frac{1}{\sqrt{N}}\bm{Z} + \frac{1}{N}\bm{1}_N\bm{1}_N^T,
\end{align}
where $\bm{Z}$ is an IID Gaussian matrix with $Z_{i, j} \overset{i.i.d.}{\sim} \mathcal{CN}(0, 1)$. We have
\begin{itemize}
    \item $\|\bm{A}\|_2 \leq \|\bm{Z}/\sqrt{N}\|_2 + \|\bm{1}\bm{1}^{\rm T}/N\|_2$, where the upper bound almost surely converges to $3$. Hence, $\|\bm{A}\|_2 \lesssim 1$.
    \item The empirical spectral distribution of $\bm{A}^{\rm H}\bm{A}$ converges to a Marchenko–Pastur distribution supported on $[0, 4]$. Hence, $\bm{A}$ is spectrally convergent.
    \item $\textstyle\sum_{i, j \in [N], i \neq j}[\bm{A}^{\mr{H}}\bm{A}]_{i, j} = (N-1) + \mathcal{O}(N^{1/2})$, meaning \\[0.5mm] that it does not satisfy the condition in \eqref{Eqn:off_diag}.
\end{itemize}
Let $\bm{A}\bm{\Xi} = \bm{J}\bm{D}$. Then, $\|\bm{J}^{\rm H} \bm{J} - \tfrac{{\rm tr}[\bm{J}^{\rm H} \bm{J}]\bm{I}}{N}\|_{\max} = \Theta(1)$. As a result, $\bm{A}\bm{\Xi} \notin \mathscr{U}$.

However,  for any spectrally convergent and norm-bounded matrix $\bm{A}$, including that in \eqref{Eqn:ZR1}, we observe that the performance of OAMP/VAMP with $\bm{A\bm{\Xi}}$ always matches the SE well in large-scale systems (e.g. $N>1000$). This motivates the following conjecture.
\begin{conjecture}\label{Conj:MC}
    There exists a broader matrix class $\mathscr{Q}$, which contains the universality class $\mathscr{U}$ (i.e., $\mathscr{U} \subset \mathscr{Q}$), such that the error vectors in OAMP/VAMP are asymptotically IID Gaussian and thus the SE is accurate. Moreover, for any spectrally convergent $\bm{A}$ with $\|\bm{A}\|_2 \lesssim 1$, and any $\bm{\Xi}_{\rm PI}$ defined in Theorem \ref{The:PIM}, $\bm{A}\bm{\Xi}_{\rm PI} \in \mathscr{Q}$.
\end{conjecture}
\begin{figure}[t!]
    \centering
    \includegraphics[width = 0.85\columnwidth]{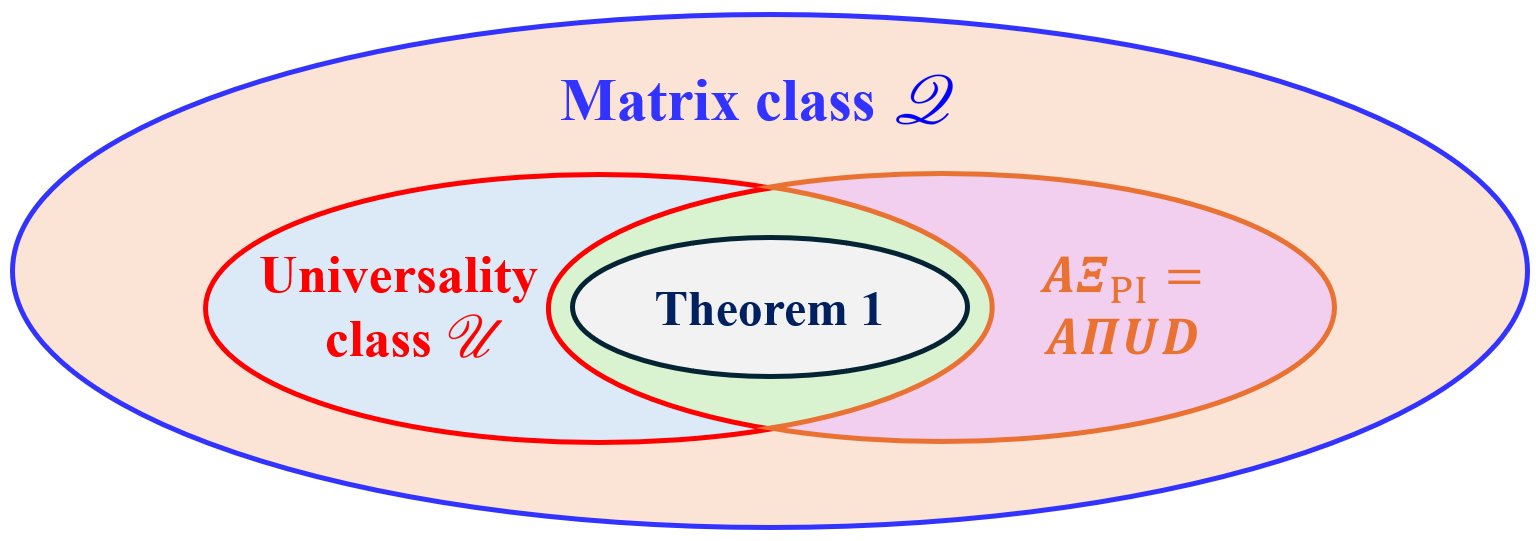}
    \caption{Conjecture \ref{Conj:MC}: SE of OAMP/VAMP/MAMP is accurate under measurement matrix class $\mathscr{Q}$.}
    \label{fig:matrix_class}
\end{figure}

\subsection{Low-Complexity Random Transform}
To reduce computational complexity and storage requirements, we typically set the random transform matrix as $\bm{\Xi} = \bf{\Pi T}$, where $\bm{\Pi}$ is a random permutation matrix and $\bm T$ is a fast transform matrix, representing a row-permuted fast transform matrix. In large-scale linear systems, practical implementations often struggle to support the ultra-high-dimensional fast transforms due to high complexity and hardware limitations. This creates significant challenges for the low-complexity detection, e.g., CD-MAMP. Processing the message sequence segment by segment with low-dimensional transforms fails to ensure adequate input isotropy, resulting in considerable performance degradation. To tackle this issue, \cite{IBST} proposed the interleaved block-sparse transform (IBST):
\BE
     {\bf \Xi}_{\rm IBST} = \bf{\Pi} {\rm diag} \{\bf{T}, \cdots\!, \bf{T} \},
\EE
which provides an effective trade-off between complexity and performance in random multiplexing systems. 
Besides, the design of a low-complexity interleaver ${\bf{\Pi}}$ is essential in practical communication systems. Traditional methods rely on generating pseudo-random numbers for interleaving, which requires significant storage—especially in large-scale systems with long sequences. In practical implementations, low-complexity interleaver designs, such as those employed in Turbo codes, serve as valuable references. These designs typically leverage modular arithmetic to efficiently generate interleaving patterns, allowing the system to store only the polynomial coefficients and greatly reducing memory requirements.

\subsection{Higher Spectral Efficiency}
Traditional multiplexing schemes, such as OFDM, OTFS, and AFDM, require inserting guard intervals into the transmitted sequence to send the cyclic prefix (CP). The key idea behind this is to use CP to construct an equivalent channel with a special structure, such as cyclic shifts. This ensures that the channel exhibits diagonal or sparse characteristics in the transform (frequency, delay Doppler, or affine frequency) domain, reducing inter-symbol interference and enabling signal recovery through simple equalization algorithms. However, the extra CP overhead reduces spectral efficiency. In contrast, random multiplexing eliminate the need for CP, as they don’t require a diagonal or sparse channel matrix in the transform domain. Therefore, compared to OFDM, OTFS, and AFDM, random multiplexing offers higher spectral efficiency by eliminating the need for CP overhead, without compromising performance.

\subsection{Adaptability to Other Multiplexing Schemes}
The random multiplexing system in \eqref{Eqn:RUP_sys} is compatible with current multiplexing schemes. Specifically, at the transmitter, the signal $\bf{s}$ is first processed by the RT matrix
$\bm{\Xi}$, followed by a arbitrarily given multiplexing through the unitary
matrix $\bf{\Upsilon}^{\rm H}$, as given by
\begin{align}\label{Eqn:RUP_M} 
     \bm{y}   = \bm{H}\underbrace{\bm{\Upsilon}^{\rm H} \bm{\Xi s}}_{\bf{x}} + \bm{n}. 
\end{align}
For example, $\bf{\Upsilon} = {\bf{F}}_{N}$ in OFDM, $\bf{\Upsilon} = {\bf{F}}_{K} \otimes{\bf{I}}_{L}$ in Zak transform-based OTFS using rectangular windows with $N=KL$, and $\bf{\Upsilon} = \bf{\Lambda}_{c_2}\bf{F}_{{N}}\bf{\Lambda}_{c_1}$ in AFDM with $\bf{\Lambda}_{c_i}\triangleq\mr{diag}(e^{-j2\pi c_in^2}, n=0, \cdots\!, N-1)$, $i=1,2$, where $c_1$ and $c_2$ are determined by Doppler shift in time-varying multipath channels~\cite{AFDM}, and in OCDM with $c_1=c_2=\tfrac{1}{2N}$. Notably, the system in \eqref{Eqn:RUP_M} is equivalent to that in \eqref{Eqn:RUP_sys}, as the unitary transformed RT matrix $\bf{\Upsilon}^{\rm H}\bf{\Xi}$ remains an RT matrix. Defining $\bf{\bar{\Xi}}=\bm{\Upsilon}^{\rm H} \bm{\Xi}$, \eqref{Eqn:RUP_M} is rewritten as
\BE\label{Eqn:otmo}
   \bm{y} = \bm{H}\bf{\bar{\Xi}} \bm{s} + \bm{n}.
\EE
Since $\bm{\Upsilon}$ is a unitary matrix, it does not change the system performance, and $\bm{H}\bf{\bar{\Xi}}$ still belongs to the universality class. Therefore, employing the CD-MAMP detector can achieve the same performance as in random multiplexing systems. Moreover, we have the following equivalent system for given multiplexing with $\boldsymbol{\Upsilon}$, i.e.,
\BE\label{Eqn:RUP_M2} 
   \tilde{\bm{y}} = \underbrace{ \bm{\Upsilon}\bm{H}\bm{\Upsilon}^{\rm H}}_{\tilde{\bf{H}}} \bm{\Xi s} + \tilde{\bm{n}},
\EE
where $\tilde{\bm{y}}=\bm{\Upsilon}\bm{y}$, $\tilde{\bm{n}}=\bm{\Upsilon}\bm{n}$, and $\tilde{\bf{H}}$ denotes the effective channel matrix at the receiver, which generally exhibits unique features. For example, $\tilde{\bf{H}}$ is diagonal in OFDM for SISO static multipath channels, and is sparse in OTFS and AFDM for time-varying multipath channels. By leveraging the diagonal or sparse characteristic of $\tilde{\bf{H}}$, one can further reduce the complexity of the CD-MAMP detection and channel estimation. Taking $\tilde{\bf{H}}={\bm{\Sigma}}_{H}$ in OFDM as an example, the complexity of the linear estimator in CD-OAMP/VAMP and CD-MAMP is reduced to $\mathcal{O}(M)$ per iteration.

\subsection{Compressed/Spread Random Multiplexing}
This paper primarily focuses on square random multiplexing, where $\bf{\Xi}$ is square RT matrix. However, the results in this paper can be directly extended to cover both compressed and spread random multiplexing, allowing for fine-tuned data rate adjustment by modifying the compression ratio without compromising performance.

\begin{itemize}
    \item \emph{Compressed Random Multiplexing}: A compressed random multiplexing system can be described as
        \BE
           \bm{y}_{M\times 1}   = \bm{A}_{M\times N} \bm{\Xi}_{N\times L} \bf{s}_{L\times 1} + \bm{n}_{M\times 1},
        \EE
        where $N<L$, meaning $\bm{\Xi}_{N\times L}$ is a random compression matrix, which can be constructed as a partial RT matrix by selecting $N$ rows from the square RT matrix $\bm{\Xi}_{L\times L}$. This can be reformulated into an equivalent square random multiplexing system as:
         \BE
           \!\!\!\! \bm{y}_{M\times 1}   = \underbrace{[
            \bm{A}_{M\times N} \; 
            \bf{0}_{M\times (L-N)}]}_{\tilde{\bf{A}}_{M\times L}} \bm{\Xi}_{L\times L} \bf{s}_{L\times 1} + \bm{n}_{M\times 1},
        \EE
        where $\tilde{\bf{A}}_{M\times L}= [
            \bm{A}_{M\times N} \; 
            \bf{0}_{M\times (L-N)}]$ is treated as an equivalent measurement matrix. In practice, $\bm{\Xi}_{N\times L}$ can be constructed using IBST \cite{IBST}, providing a trade-off between complexity and performance:
            \BE
                 {\bf \Xi}_{\rm IBS}={\bf{\hat\Pi}}\mathrm{diag}\{[{\bf{{\Pi}}}_{1}{\bf{T}}_{1}]_{1:N/J},\cdots\!,[{\bf{{\Pi}}}_{J}{\bf{T}}_{J}]_{1:N/J}\},
            \EE
           where $\bf{\hat\Pi}$ is an $N\times N$ uniformly random permutation matrix, $[\bf{\hat\Pi}_j{\bf{T}}_{j}]_{1:N/J}$ represents the extraction of the first $N/J$ rows of $\bf{\hat\Pi}_j{\bf{T}}_{j}$, and $\{\bf{\hat\Pi}_j\}$ and $\{{\bf{T}}_{j}\}$) are $L/J\times L/J$ permutation and transform matrices, respectively.  
            
    \item \emph{Spread Random Multiplexing}: A spread random multiplexing system can be written as
        \BE
           \bm{y}_{M\times 1}   = \bm{A}_{M\times N} \bm{\Xi}_{N\times S} \bf{s}_{S\times 1} + \bm{n}_{M\times 1},
        \EE
        where $N>S$, meaning $\bm{\Xi}_{N\times S}$ is a random spread matrix, which can be constructed as a partial RT matrix by selecting $S$ columns from the square RT matrix $\bm{\Xi}_{S\times S}$. This can be reformulated into a square random multiplexing system as:
         \BE
           \bm{y}_{M\times 1}   = 
            \bm{A}_{M\times N} \bm{\Xi}_{N\times N} \underbrace{\left[\begin{array}{c}
        \bf{s}_{S\times 1}  \\
        \bf{0}_{(N-S)\times 1}
    \end{array} \right]}_{\tilde{\bf{s}}_{N\times 1}} + \bm{n}_{M\times 1},
        \EE
        where $\tilde{\bf{s}}_{N\times 1}= \left[\begin{array}{c}
        \bf{s}_{S\times 1}  \\
        \bf{0}_{(N-S)\times 1}
    \end{array} \right]$ is treated as an equivalent signal vector.  
\end{itemize}

\subsection{Integration with Artificial Intelligence (AI) Technology}
Note that the replica MAP BER and capacity optimality of RM rely on certain assumptions, such as large-scale systems and unique SE fixed points (Assumptions 1 and 2). When these conditions are not met, the combination of RM and CD-MAMP may experience performance degradation. For instance, the optimality and coding principles of CD-MAMP and CD-OAMP/VAMP may not hold in small- to medium-dimensional systems. A promising research avenue, inspired by \cite{DL_MIMO-OFDM}, is to leverage AI techniques to facilitate the design of finite-length coding and detection algorithms. Meanwhile, under extremely harsh wireless channels, the equivalent channel matrices may still belong to the universality class $\mathscr{U}$ but with large condition numbers, causing severe instability—an open problem for message passing algorithms. Thus, integrating AI techniques holds promise for overcoming this bottleneck. In summary, RM offers a practical framework for optimal transceiver design, and its integration with AI is expected to enable broader applicability across diverse scenarios.

\section{Numerical Results}\label{Sec:sim}
In this section, we provide numerical results to demonstrate the achievable rates and finite-length BER/BLER performance of the random multiplexing systems with CD-MAMP receiver, optimal power allocation, and channel coding.

\subsection{Correlated Time-Varying Multipath Channel}
\begin{figure}[t!]
    \centering
    \includegraphics[width =1\columnwidth]{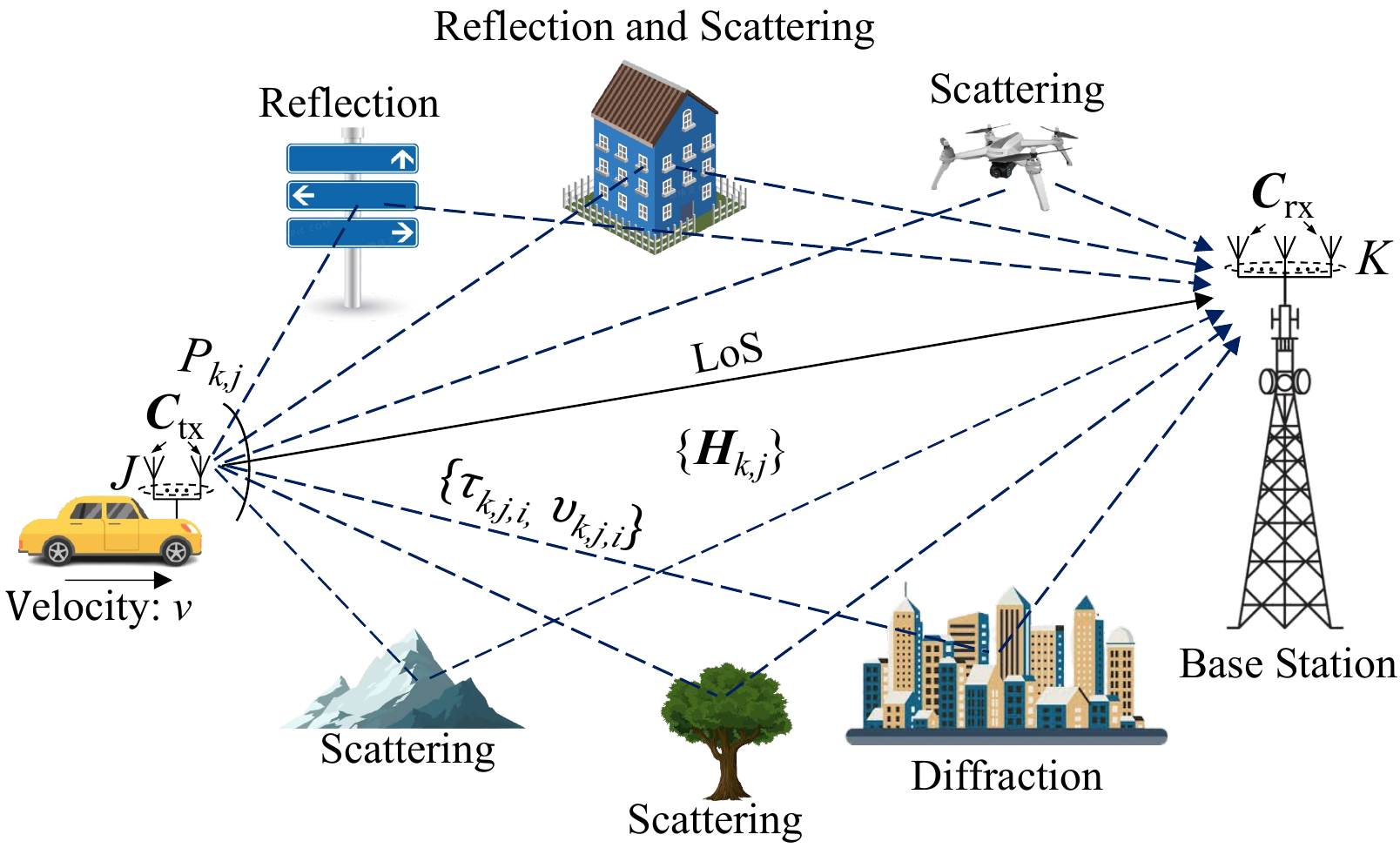}
    \caption{Illustration of the correlated time-varying multipath channels $\{\bf{H}_{ k,j}\}$ in MIMO RM systems with a mobile transmitter with $J$ antennas and a receiver with $K$ antennas,  where $P_{k,j}$ denotes the maximal number of multipaths between the $j$-th transmit antenna and the $k$-th receive antenna, $\tau_{k,j,i}$  channel delay and $\nu_{k,j,i}$ Doppler shift in the $i$-th path, and $\bf{C}_{\rm{tx}}$ and $\bf{C}_{\rm{rx}}$ correlation-shaping matrices at the transmitter and receiver, respectively, $1\le i\le P_{k,j}, 1\le j\le J$ and $1\le k \le K$.}
    \label{fig:channel_model}
\end{figure}
Consider a time-varying multipath MIMO channel, as shown in Fig.~\ref{fig:channel_model}, i.e., $\bf{H}=[\bf{H}^{\rm{T}}_{1}, \cdots\!, \bf{H}^{\rm{T}}_{K}]\in \mathbb{C}^{K\bar{M}\times J\bar{N}}$ with $M=K\bar{M}$ and $N=J\bar{N}$, $\bf{H}_{k}=[\bf{H}_{k,1},\cdots\!, \bf{H}_{k,J}]\in \mathbb{C}^{\bar{M}\times J\bar{N}}$, and channel impulse response ${H}_{k,j}[u,\ell]$ between $j$-th transmit antenna and $k$-th receive antenna, $1\le j\le J$, $1\le k \le K$, is given by
    \BE\label{Eqn:chl}
    \!\!{H}_{k,j}[u, \ell] = \sum\limits_{i=1}^{P_{k, j}} {H}_{k,j,i} e^{j2\pi \nu_{k,j,i}(uT_s -\ell T_s)}\mr{P_{rc}}(\ell T_s-\tau_{k,j,i}),
    \EE
    where $u=1, \cdots\!, \bar{M}$, $\ell=1, \cdots\!, \mathcal{L}_{k,j}$, $\mathcal{L}_{k,j}$ denotes the maximal number of channel taps, $P_{k,j}$ is the maximal number of multipaths, $T_s$ denotes the system sampling interval, $H_{k,j,i}$, $\tau_{k,j,i}$, and $\nu_{k,j,i}$ denote the channel gain, delay, and Doppler shift in the $i$-th path, respectively. $\mr{P_{rc}}(\cdot)$ is the overall raised-cosine rolloff filter when the practical root raised-cosine (RRC) pulse shaping filters are employed at the transceiver to control signal bandwidth and reject out-of-band emissions. Meanwhile, $H_{k,j,i}$ also contains the spatial correlations among the antennas \cite[Equation (6)]{MIMO-OTFS-WCL}, i.e., $H_{k,j,i}={{{\mathcal{\bar{H}}}}}_{i}[k,j]$, where 
    \BE\label{Eqn:correlatedChannelMatrix}
    {\bf{{\mathcal{\bar{H}}}}}_{i}=\bf{C}_{\rm{rx}} \cdot \bf{\mathcal{H}}_{i} \cdot \bf{C}_{\rm{tx}}^{\rm H},
    \EE
    with IID complex Gaussian random matrix~$\bf{\mathcal{H}}_{i}$, correlation-shaping matrices $\bf{C}_{\rm{rx}}$ and $\bf{C}_{\rm{tx}}^{\rm H}$ obtained by the Cholesky decomposition of $\bf{R}_{\rm{tx}}$ and $\bf{R}_{\rm{rx}}$, i.e. $\bf{R}_{\rm{tx}}=\bf{C}_{\rm{tx}} \cdot \bf{C}_{\rm{tx}}^{\rm{H}}$ and $\bf{R}_{\rm{rx}}=\bf{C}_{\rm{rx}} \cdot \bf{C}_{\rm{rx}}^{\rm{H}}$. The elements of $\bf{R}_{\rm tx}$ and $\bf{R}_{\rm rx}$ are 
\BE\label{Eqn:TRx}
\begin{aligned}
&\bf{R}_{\rm{tx}}[p, q]=\left\{\begin{array}{ll}
\rho_{\rm{tx}}^{p-q}, & q \leq p \\
\left(\rho_{\rm{tx}}^{q-p}\right)^*, & q>p
\end{array}, \quad p, q \in\left\{1, \ldots, J\right\}, \right.  \\
&\bf{R}_{\rm{rx}}[p, q]=\left\{\begin{array}{ll}
\rho_{\rm{rx}}^{p-q}, & q \leq p \\
\left(\rho_{\rm{rx}}^{q-p}\right)^*, & q>p
\end{array}, \quad p, q \in\left\{1, \ldots, K\right\}, \right.
\end{aligned}
\EE
where $\rho_{\rm{tx}}$, $\rho_{\rm{rx}} \in[0,1)$ denote the correlation level at $\bf{R}_{\rm tx}$ and $\bf{R}_{\rm rx}$, respectively. Here, we assume that $\rho_{\rm{tx}}=\rho_{\rm{rx}}=\rho$.

We consider that the carrier frequency is $4$~GHz with $\Delta f = 15$ kHz, the velocity of the device is $v \in \{150, 300\}$ km/h with a maximum Doppler frequency shift $\nu_{\text{max}} \in \{555.5, 1111\}$~Hz, $P_{k,j}=5$, and the channel Doppler shift is generated by using Jakes information \cite{OTFS_GMP}. The RRC rolloff factor in the transceiver is set at $0.4$. Here, SISO and MIMO random multiplexing systems are considered with $(J, K)=(1,1)$, $(2,2)$, $(4,4)$, and $(8, 4)$, where $\rho \in \{0.3, 0.6\}$ and $\bar{M}=\bar{N} \in \{2048, 1024, 512, 256\}$, respectively. In addition, the common QPSK, 8PSK, 16QAM, and Gaussian signaling are employed. Here, the random transform matrix is set as $\bm{\Xi}=\bf{\Pi}\bf{T}_{\rm HW}$ for Figs.~\ref{fig:RM_withoutmd} and \ref{fig:RM_diffm} and  $\bm{\Xi}=\bf{\Pi}\bf{F}^{\rm{H}}$ for Figs.~\ref{fig:RM_PA}-\ref{fig:MIMO_optLDPC_RM}, where $\bf{T}_{\rm HW}$ denotes the normalized Hadamard-Walsh transform matrix and $\bf{F}^{\rm{H}}$ the normalized IDFT matrix.

\subsection{Uncoded Random Multiplexing Systems}
\begin{figure}[t!]
    \centering
    \includegraphics[width = 0.85\columnwidth]{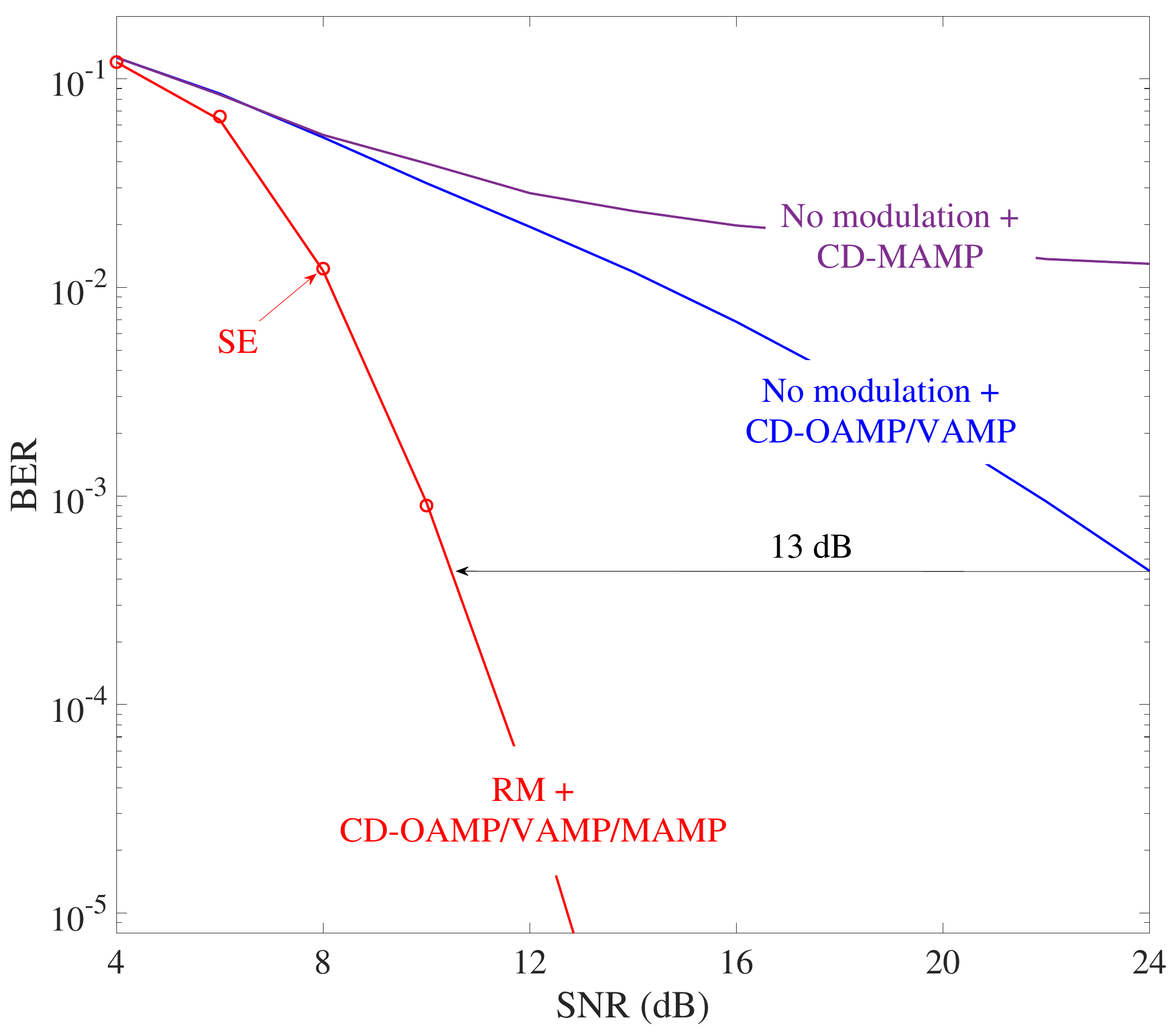}
    \caption{BER comparisons of MIMO systems with and without random multiplexing (RM), where CD-OAMP/VAMP and CD-MAMP detectors are employed with $(J, K)=(2,2)$, $\rho=0.3$, QPSK signaling, $\bar{M}=\bar{N}=1024$, and $v=150$~km/h.}
    \label{fig:RM_withoutmd}
\end{figure}

Fig.~\ref{fig:RM_withoutmd} shows BER comparisons of MIMO systems with and without random multiplexing, where CD-OAMP/VAMP and CD-MAMP detectors are employed, $(J, K)=(2,2)$, $\bar{M}=\bar{N}=1024$, and $\rho=0.3$.  As can be seen, thanks to the random multiplexing, both CD-MAMP and CD-OAMP/VAMP achieve the same BER performance consistent with the predictions of SE analysis in \eqref{Eqn:MAMP_SE0} and \eqref{Eqn:OAMPSE}. In contrast, when no multiplexing is employed and the BER curves are at $10^{-4}$, the BER performance of CD-OAMP/VAMP deteriorates significantly by approximately $13$ dB, while CD-MAMP converges to a BER of around $10^{-2}$. This indicates that channel matrices in practical communication systems rarely satisfy the input isotropy condition, resulting in degraded detection performance. Random multiplexing is therefore required to ensure that the equivalent channel matrix meets the input isotropy condition for CD-MAMP and CD-OAMP/VAMP detectors, enabling replica-MAP BER performance. These results are presented for the first time in this paper.

\begin{figure}[t!]
    \centering
    \includegraphics[width = 0.85\columnwidth]{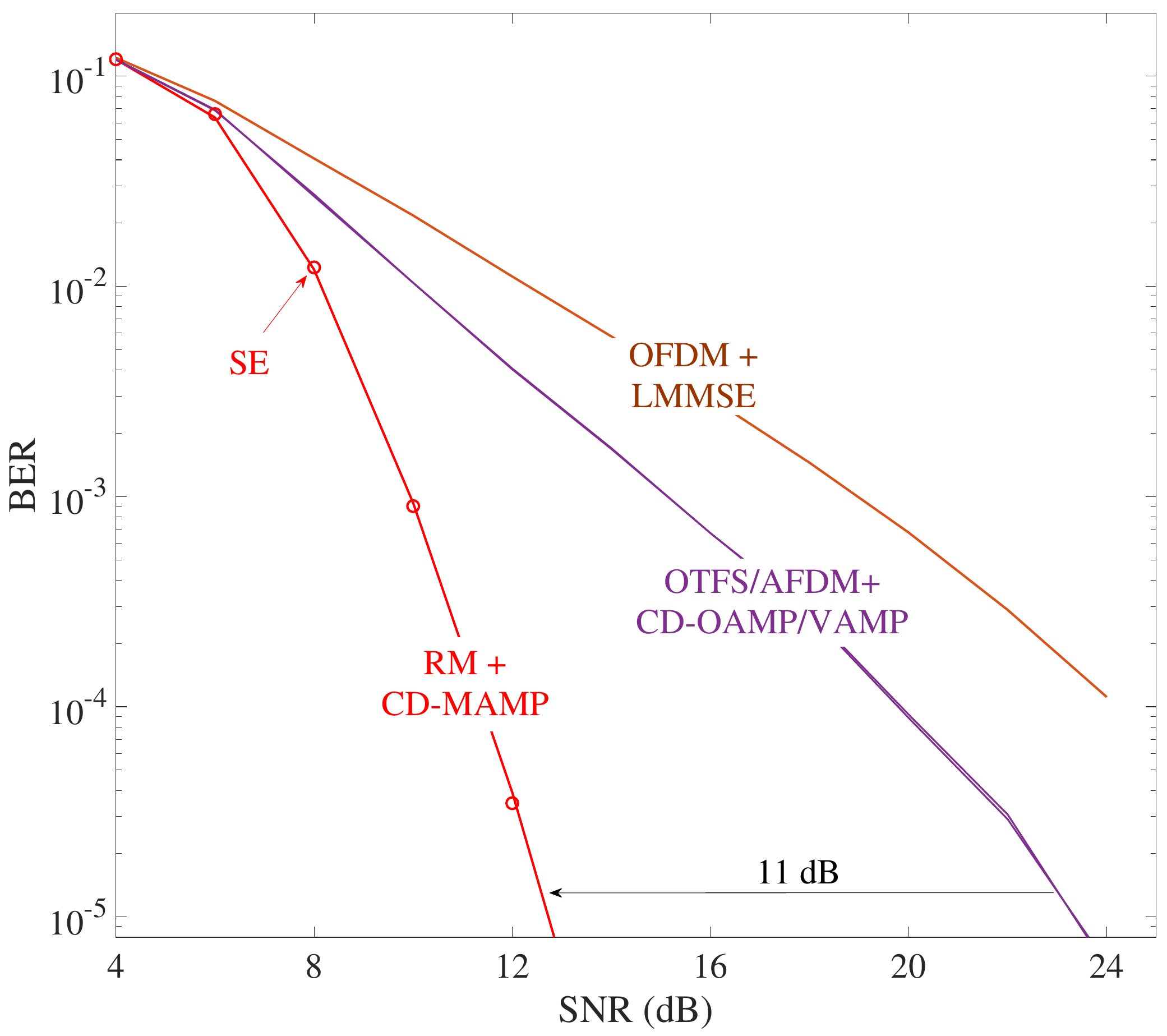}
    \caption{BER comparisons of random multiplexing (RM), OFDM, OTFS, and AFDM in MIMO systems with CD-OAMP/VAMP and CD-MAMP detectors, QPSK signaling, $\bar{M}=\bar{N}=1024$, $(J, K)=(2,2)$, $\rho=0.3$, and $v=150$ km/h.}
    \label{fig:RM_diffm}
\end{figure}

Fig.~\ref{fig:RM_diffm} illustrates the BER comparisons of random multiplexing, OFDM, OTFS, and AFDM in MIMO systems with CD-OAMP/VAMP and CD-MAMP detectors, where QPSK signaling is employed, $(J, K)=(2,2)$, and $\bar{M}=\bar{N}=1024$. To be fair, we ensure that the multiplexing systems use the same bandwidth, i.e., the subcarrier spacing is $\Delta f$ kHz for OTFS, and $\tfrac{\Delta f}{Q}$ kHz for IFDM, AFDM, and OFDM, where the OTFS multiplexing matrix is $\bf{F}_{Q}^{\mr{H}}\otimes \bf{I}_{P}$ with $Q=32$ and $P=32$. For BER curves of $10^{-5}$,  it can be clearly seen that random multiplexing with CD-MAMP can achieve about a $11$ dB gain with much lower complexity than OTFS/AFDM with CD-OAMP/VAMP, which also significantly outperforms OFDM with LMMSE. These results are similar to that reported in \cite{IBST}. This also indicates that CD-OAMP/VAMP is no longer replica MAP optimal for existing OTFS and AFDM. The design of low-complexity, near-optimal detectors for OTFS and AFDM remains an open challenge.

\begin{figure}[t!]
    \centering
    \includegraphics[width = 0.85\columnwidth]{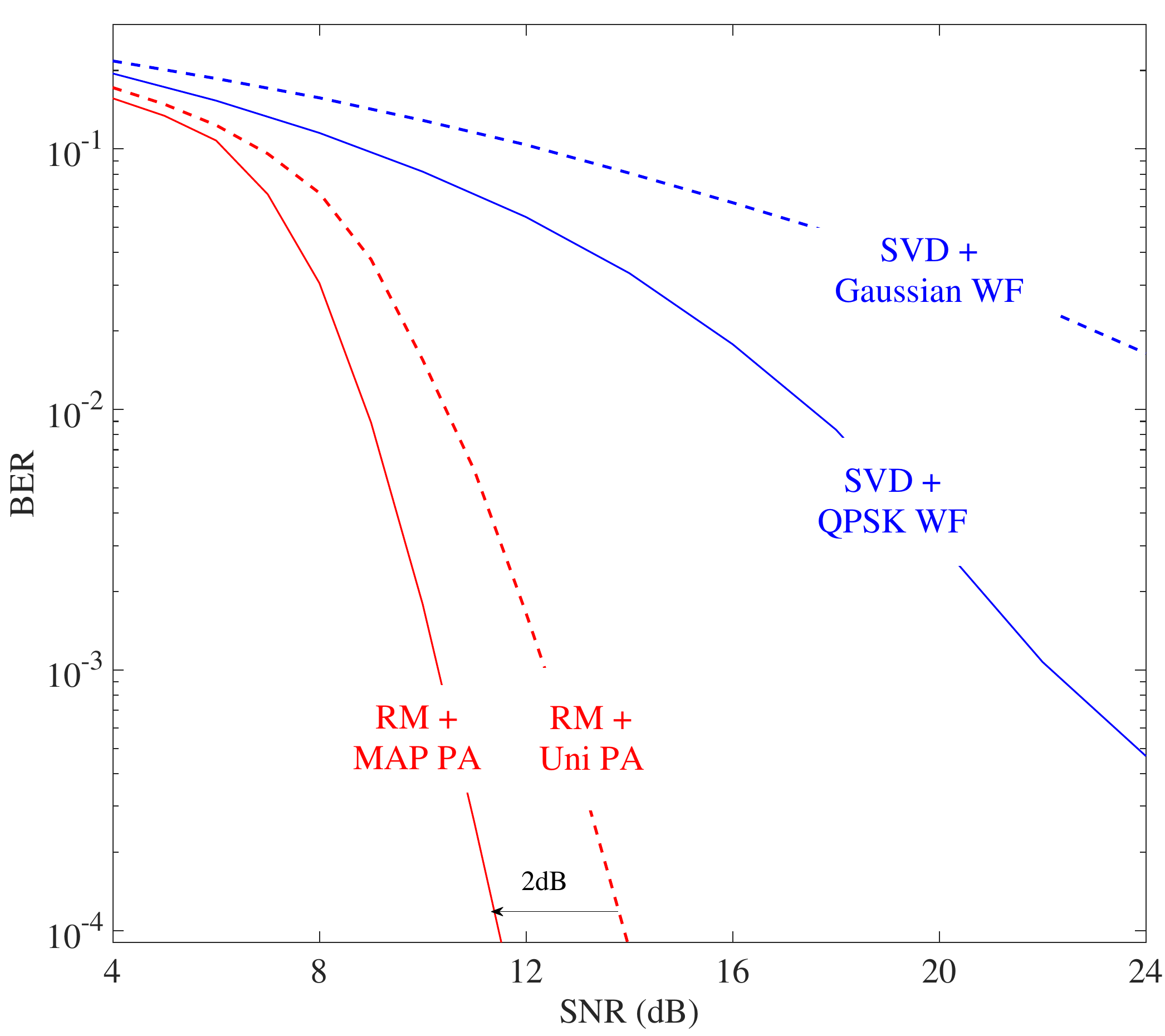}
    \caption{BER comparisons of random multiplexing (RM) with CD-MAMP detector in SISO systems with QSPK signaling, $M=N=2048$, and $v=300$ km/h, where optimal-MAP power allocation (PA) in \eqref{Eqn:P2_a}, uniform (Uni) PA, and water filling (WF) under Gaussian and QPSK signaling based on channel parallelization via SVD are employed.}
    \label{fig:RM_PA}
\end{figure}

Fig.~\ref{fig:RM_PA} shows the BER comparisons of random multiplexing with optimal-MAP power allocation (PA) in \eqref{Eqn:P2_b}, uniform PA, and channel parallelization via SVD with Gaussian and QPSK water filling, where QSPK signaling is employed. In random multiplexing, the optimal power allocation, aimed at achieving MAP performance, can achieve a $2$ dB gain over uniform power allocation. In contrast, waterfilling for Gaussian and QPSK signals using channel parallelization via SVD decomposition results in a significant performance loss of more than $12$ dB. This confirms the optimality of power allocation in random multiplexing.

\subsection{Achievable Rates of Coded Random Multiplexing Systems}
\begin{figure}[t!]
    \centering
    \includegraphics[width = 0.85\columnwidth]{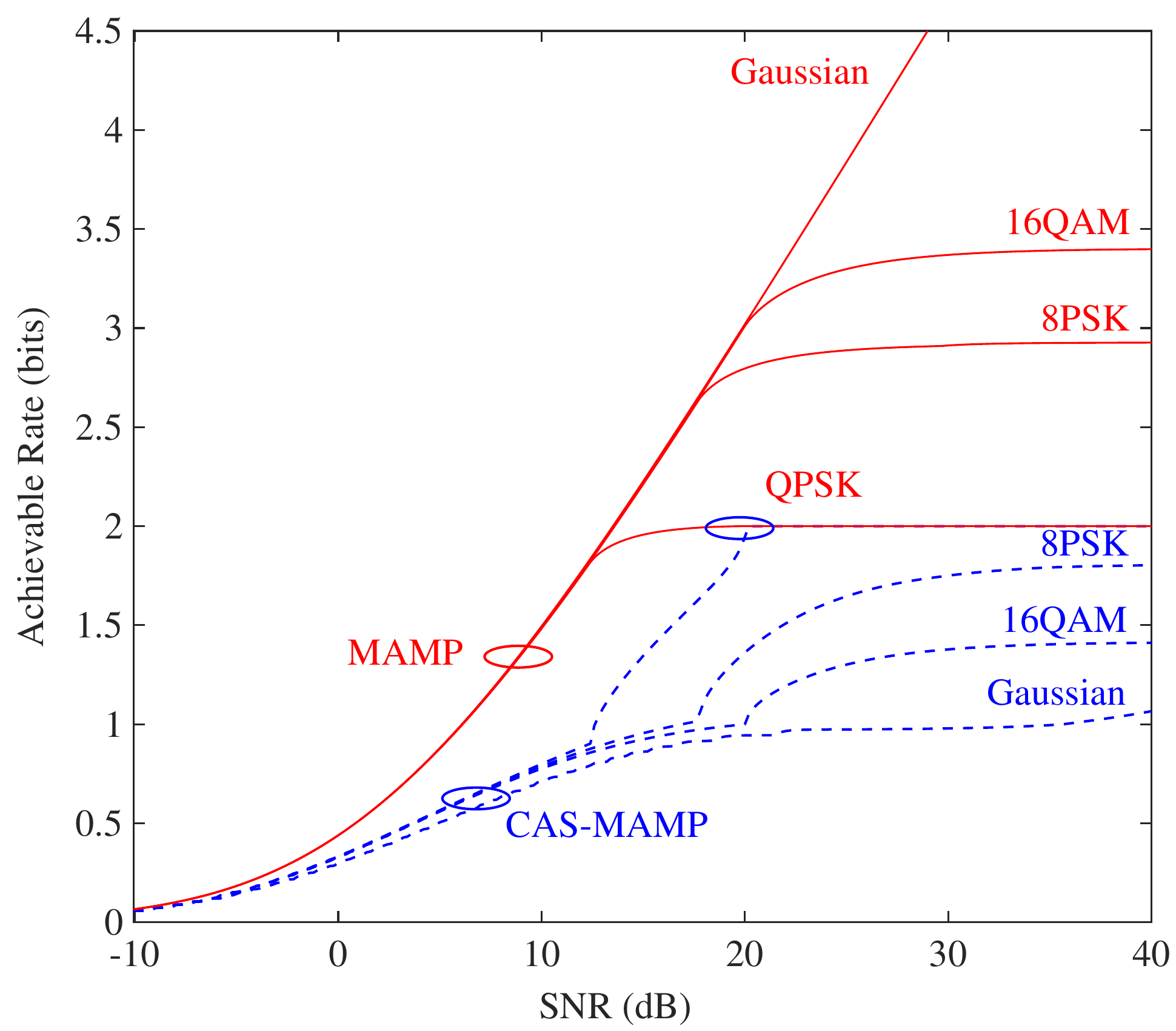}
    \caption{Maximum achievable rate comparison of CD-MAMP and CAS-CD-MAMP (abbr. MAMP and CAS-MAMP) in MIMO random multiplexing systems, where $\bar{M}=\bar{N}=256$, $(J, K) = (8, 4)$, $P_{k.j}=5$, $\rho=0.6$, $v=300$km/h, and \{QPSK, 8PSK, 16QAM, Gaussian\} signaling.}
    \label{fig:MIMO_8_4}
\end{figure}
\begin{figure}[t!]
    \centering
    \includegraphics[width = 0.85\columnwidth]{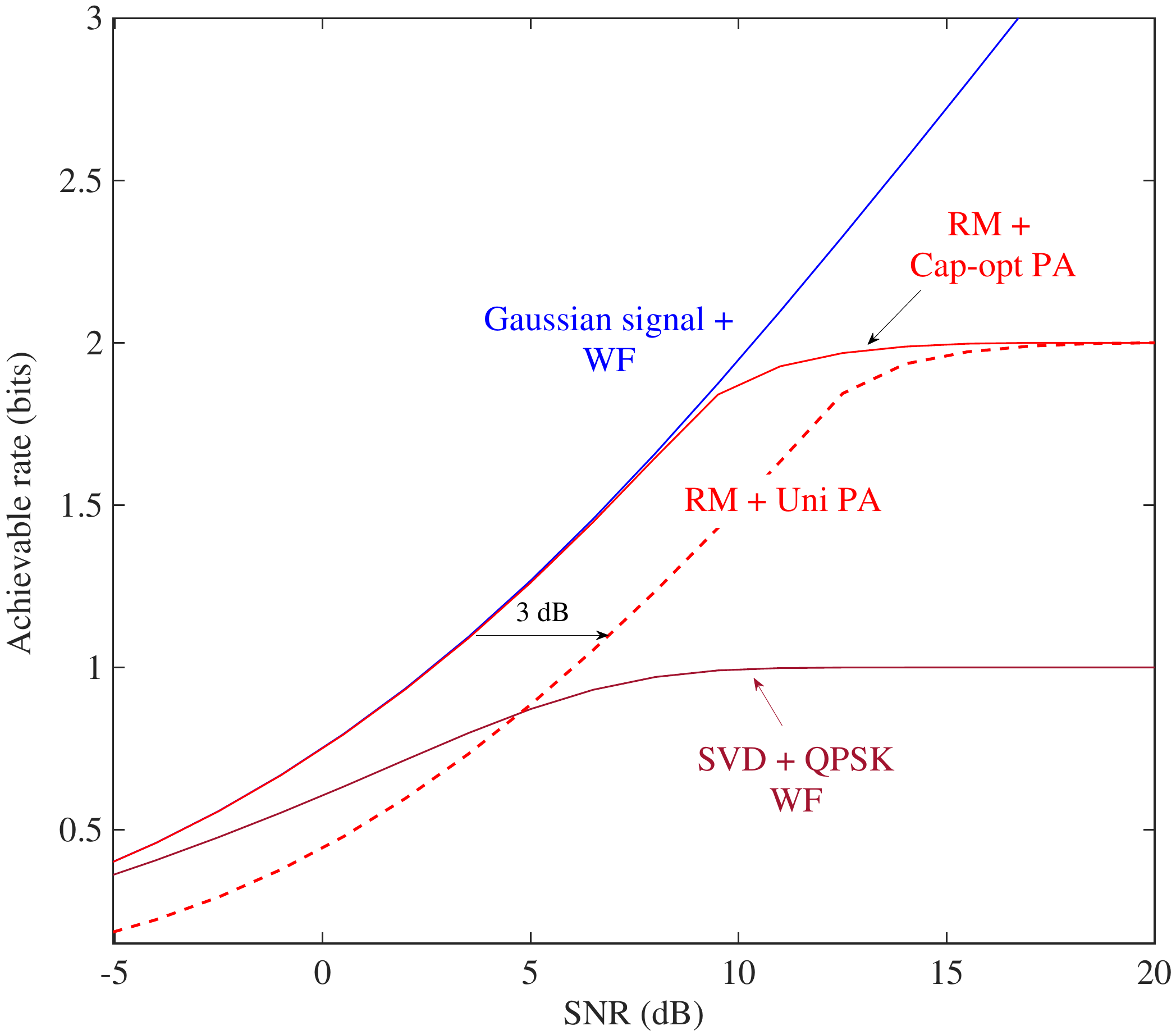}
    \caption{Maximum achievable rate comparison of CD-MAMP with 
    random multiplexing (RM) and channel parallelization via SVD in MIMO linear systems, where QPSK signaling, and different power allocation (PA) (i.e. optimization in \eqref{Eqn:optP_rate} and waterfilling (WF) in \cite[Sec. 10.4]{Cover1990} and \cite{Lozano2006}) are employed with $\bar{M}=\bar{N}=256$, $(J, K)=(8, 4)$, $P_{k.j}=5$, $\rho=0.6$, and $v=300$km/h.}
    \label{fig:MIMO_qpsk}
\end{figure}

\begin{figure}[t!]
    \centering
    \includegraphics[width = 0.85\columnwidth]{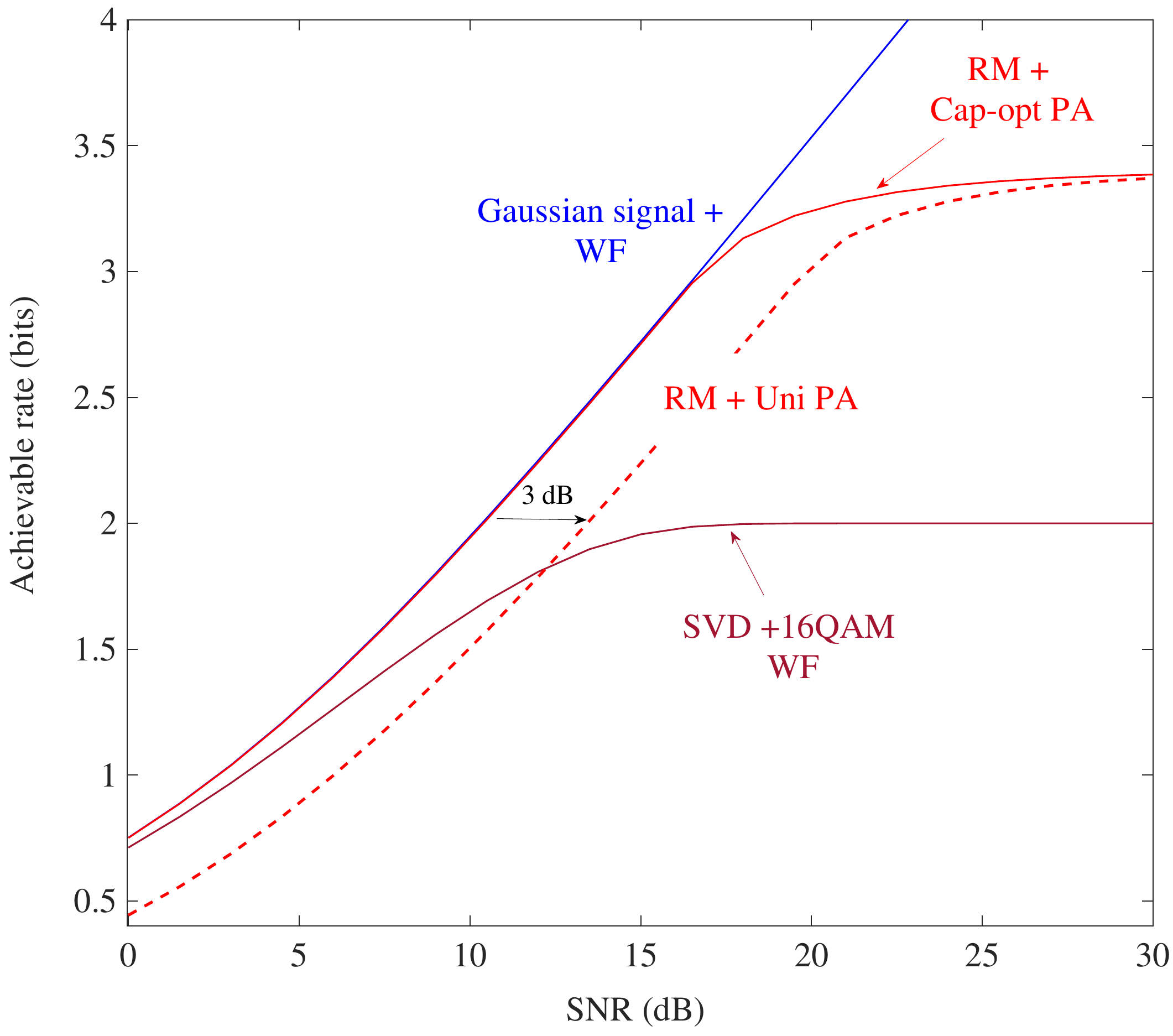}
    \caption{Maximum achievable rate comparison of CD-MAMP with 
    random multiplexing (RM) and channel parallelization via SVD in MIMO linear systems, where 16QAM signaling, and different power allocation (PA) (i.e. optimization in \eqref{Eqn:optP_rate} and waterfilling (WF) in \cite[Sec. 10.4]{Cover1990} and \cite{Lozano2006}) are employed with $\bar{M}=\bar{N}=256$, $(J, K) =(8, 4)$, $P_{k,j}=5$, $\rho=0.6$, and $v=300$km/h.}
    \label{fig:MIMO_16qam}
\end{figure}
Based on Theorem~\ref{The:maxrate}, we provide the achievable rate analysis of CD-MAMP in MIMO random multiplexing systems.
\begin{itemize}
    \item \emph{Uniform Power Allocation:} Fig.~\ref{fig:MIMO_8_4} shows the maximum achievable rates of CD-MAMP in MIMO random multiplexing systems with $(J=8, K=4)$, $(\bar{M}=\bar{N}=256)$, uniform power allocation, and \{QSPK, 8PSK, 16QAM, Gaussian\} signaling when CSI is unknown at the transmitter and available only at the receiver. Note that the achievable rate of CD-MAMP increases monotonically with the signal constellation order, where it converges to a constant value for discrete signaling and increases monotonically for Gaussian signaling as the SNR increases. Although Fig.~\ref{fig:MIMO_8_4} resembles our previous work\cite{LeiOptOAMP,Code_MAMP}, the earlier study relied on the channel matrix being right-unitarily invariant, whereas this paper leverages random multiplexing to remove the need for such idealized channel assumptions. In contrast, a conventional cascading CD-MAMP (CAS-CD-MAMP) receiver is used as a baseline method, as defined in~\cite{DongningTIT2005,Tanaka}, which operates by first using CD-MAMP for detection and then utilizing its result for decoding, without any iteration between the two stages. The achievable rate of CAS-CD-MAMP is $R_{\rm{CAS-CD-MAMP}}=\int_{0}^{\rho^*} {\rm mmse}(\rho)d\rho$, where $\rho^*$ is the solution of ${\rm mmse}(\rho)=\eta^{-1}_{\rm{SE}}(\rho,\bf{p})$ as shown in Fig.~\ref{fig:VSE_oamp}. As a result, the rate loss of CAS-CD-MAMP is $\int_{\rho^*}^{\eta_{\text{SE}}(0, \bf{p})}{\eta}_{\mr{SE}}^{-1}(\rho, \bf{p}) d \rho$. Note that in this scenario, its achievable rate of CAS-CD-MAMP decreases instead with increasing signal constellation order.
    \item \emph{Optimal Power Allocation:} Fig.~\ref{fig:MIMO_qpsk} shows the  achievable rate comparison of CD-MAMP with random multiplexing and channel parallelization via SVD in MIMO linear systems, where QPSK signaling is employed. When CSI is available in the transmitter, power allocation is employed to enhance the system capacity. For MIMO random multiplexing linear systems, the optimal power allocation in \eqref{Eqn:optP_rate} has a $3$ dB gain over the uniform power allocation at the same achievable rate and achieves the optimal rate for Gaussian signaling with Gaussian waterfilling in \cite[Sec. 10.4]{Cover1990} when SNR $<10$ dB. Compared to conventional waterfilling for QPSK signaling based on channel parallelization, random multiplexing with optimal power allocation can achieve nearly a twofold rate improvement in the high SNR region (i.e., SNR $> 10$ dB). Fig.~\ref{fig:MIMO_16qam} shows the achievable rate comparison of CD-MAMP with random multiplexing and channel parallelization via SVD in MIMO linear systems, where 16QAM signaling is employed. Similar to the QPSK case, the random multiplexing with optimal power allocation can achieve the optimal rate for Gaussian signaling, a $3$~dB gain over the random multiplexing with uniform power allocation for SNR $\le 15$~dB, and a $1.5$ bits improvement per symbol for SNR $> 15$~dB.
\end{itemize}

\begin{figure}[t!]
    \centering
    \includegraphics[width = 0.85\columnwidth]{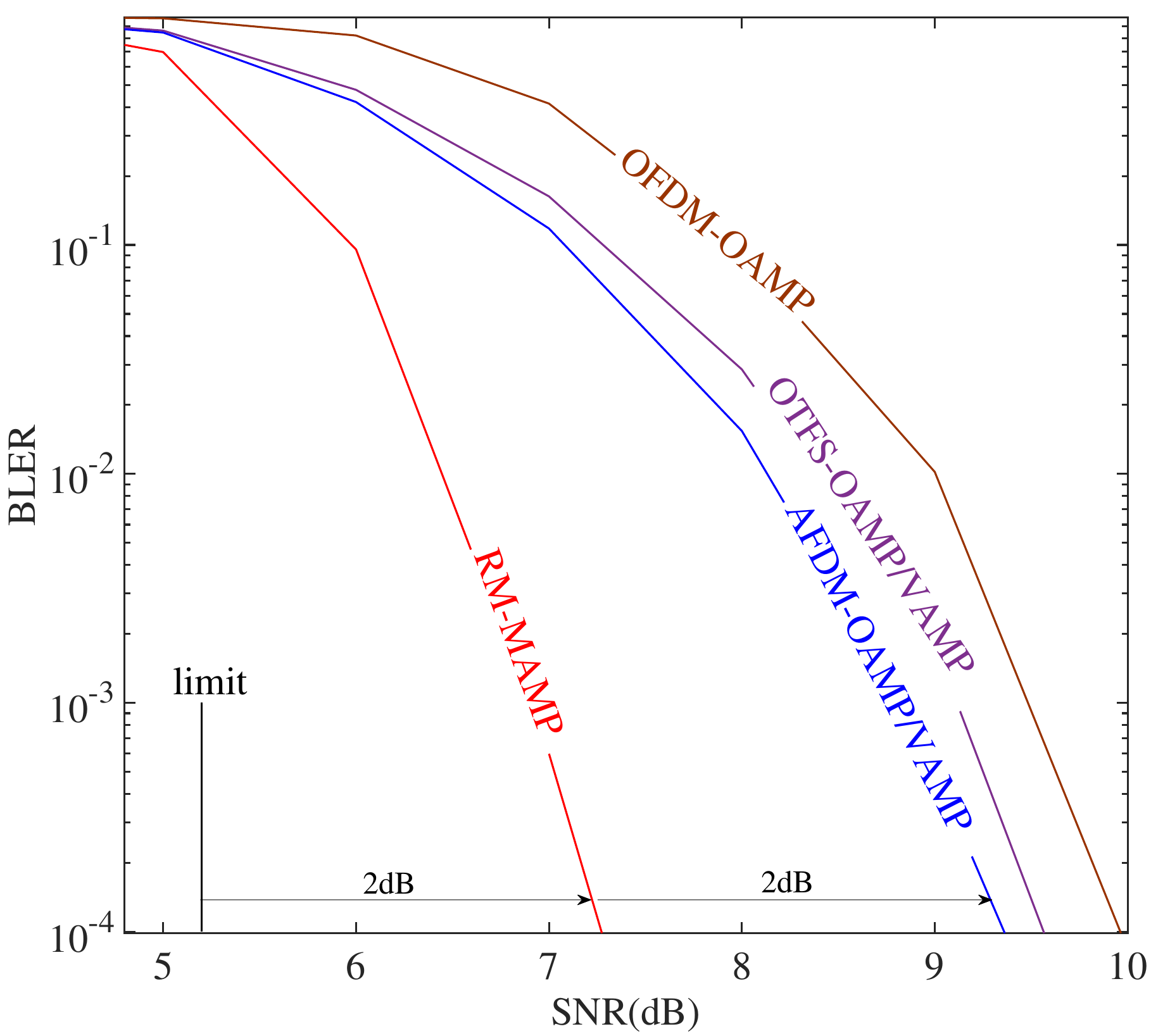}
    \caption{BLER performances of CD-MAMP with 5G-NR LDPC codes in MIMO systems with coding rate is $0.625$, $(J, K)=(4,4)$, $\bar{M}=\bar{N}=512$, $\rho=0.6$, $P_{k,j}=5$, and $v=300$~km/h, where random multiplexing (RM), OFDM, OTFS, and AFDM are employed. The CAS-MAMP and CAS-OAMP/VAMP are abbreviated as MAMP and OAMP/VAMP, respectively.}
    \label{fig:MIMO_SIC}
\end{figure}
\subsection{BERs of Coded Random Multiplexing Systems}

Fig.~\ref{fig:MIMO_SIC} shows the BLER comparisons of the different multiplexing with 5G-NR LDPC codes in MIMO linear systems, where the coding rate is $0.625$ and coding length is $2048$,  $(J, K)=(4,4)$, $\bar{M}=\bar{N}=512$, $\rho=0.6$, $P_{k,j}=5$, $v=300$~km/h. To be consistent with practical commercial communication systems, we employ CAS-MAMP for random multiplexing and CAS-OAMP/VAMP for OFDM, OTFS, and AFDM. Note that the gap between the BLER curve of random multiplexing with CAS-MAMP at $10^{-4}$ and the corresponding performance limit is $2$ dB, which achieves up to a $2$ dB gain with lower complexity compared to OTFS, AFDM, and OFDM. This also validates the advantages of random multiplexing in 5G-NR LDPC coded linear systems.

\begin{figure}[t!]
    \centering
    \includegraphics[width = 0.85\columnwidth]{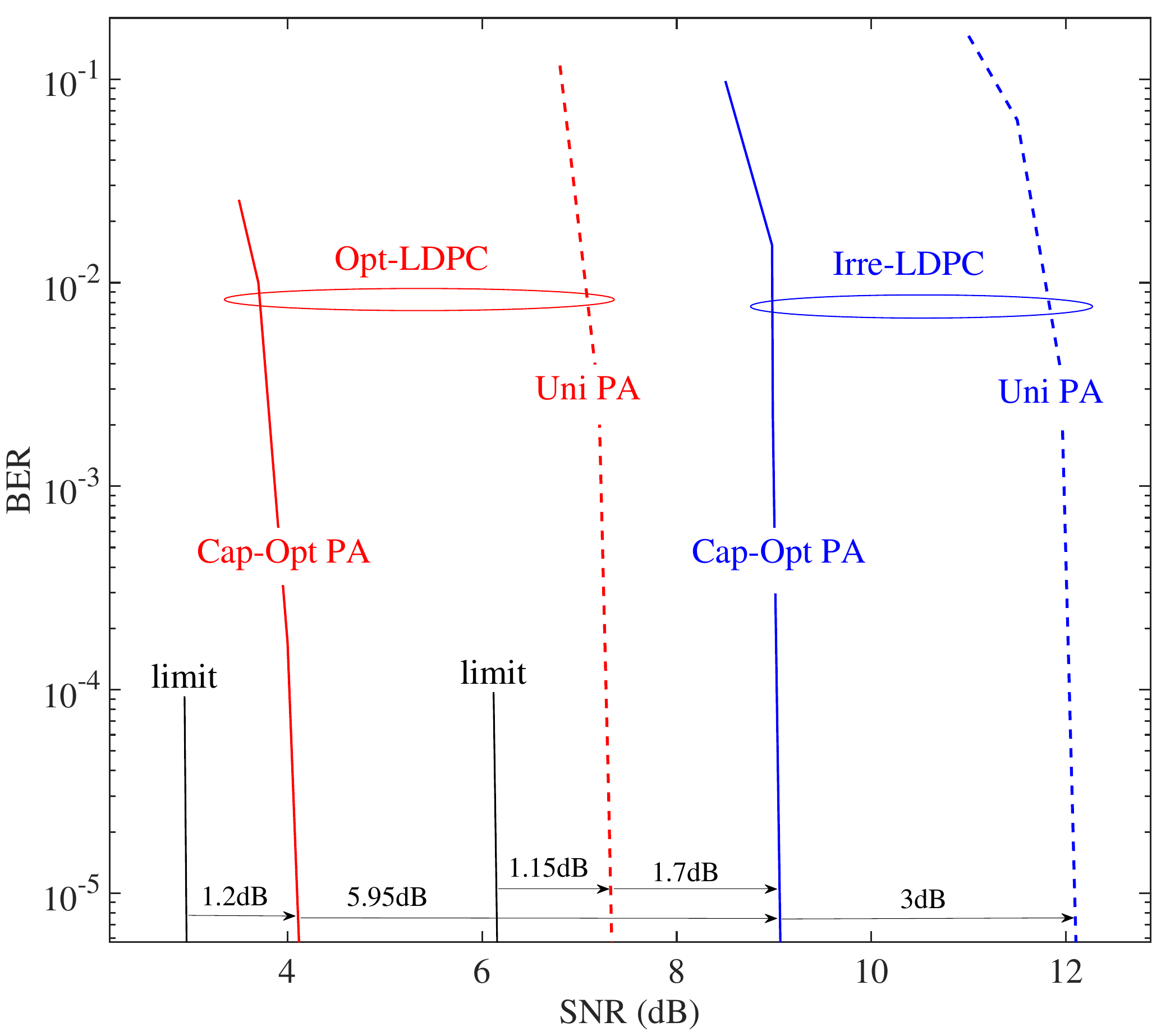}
    \caption{BER performances of CD-MAMP with optimized LDPC (Opt-LDPC) codes in Table~\ref{Opt_degree1} and well-designed P2P irregular LDPC (Irre-LDPC) codes in \cite{Richardson2001} in MIMO random multiplexing systems, where $(J, K)=(8,4)$, $\bar{M}=\bar{N}=256$, $\rho=0.6$, $P_{k,j}=5$, and $v=300$~km/h.}
    \label{fig:MIMO_optLDPC_RM}
\end{figure}

Fig.~\ref{fig:MIMO_optLDPC_RM} provides the BER comparisons of the optimized irregular LDPC codes in Table~\ref{Opt_degree1} and well-designed SISO irregular LDPC codes in \cite{Richardson2001} for MIMO random multiplexing systems, where capacity-optimal power allocation (PA) in \eqref{Eqn:optP_rate} and uniform PA are employed for QPSK signaling with $\bar{M}=\bar{N}=256$, $(J, K) = (8, 4)$, $P_{k.j}=5$, $\rho=0.6$, and $v=300$km/h. The degree distributions of P2P irregular LDPC codes are $\lambda(X)=0.24426x+0.25907x^2+0.01054x^3+0.05510x^4+0.01455^7+0.01275x^9+0.40373x^{11}$ and $\mu(X)=0.25475x^6+0.73438x^7+0.01087x^8$, whose rate $R_{\rm{LDPC}}$ is $0.5$ and the decoding threshold is $0.18$~dB away from the P2P-AWGN capacity. Notably, with optimized and irregular LDPC codes, the capacity-optimal PA can achieve about $4.25$ dB and $3$ dB gains over the uniform PA, respectively. Furthermore, for fixed capacity-optimal and uniform PA, the optimized LDPC codes outperform the P2P irregular LDPC codes by about $5.95$ dB and $4.7$ dB gains. This indicates that the well-designed P2P irregular LDPC codes are not optimal anymore with significant performance losses in MIMO random multiplexing systems.

\begin{table}[!t] \small
\caption{Optimized Irregular LDPC Codes with CD-MAMP and Capacity Optimal and Uniform Power Allocation (PA) in $8\times 4$ MIMO Random Multiplexing Systems.}\label{Opt_degree1}
\centering\setlength{\tabcolsep}{0.8mm}
\begin{tabular}{|c||c|c|}
\hline
{\multirow {2}{*}{\makecell[c]{System\\[0.5mm] parameters}}} & $\bar{M}=\bar{N}=256$ & $(P_{k,j}=5, v=300{\text{km/h}})$\\
\cline{2-3}
 & \multicolumn{1}{c|}{Cap-Opt PA}  & Uni PA \\ 
\hline
Power allocation & \multicolumn{1}{c|}{Capacity Optimal}  & Uniform \\ 
\hline
Code length & \multicolumn{2}{c|}{102400}  \\ 
\hline
$R_\text{LDPC}$& \multicolumn{1}{c|}{0.5014}  & 0.4993 \\ 
\hline
$\mu(x)$ & \multicolumn{1}{c|}{${\it{\mu}}_{\text{6}}=0.1$, ${\it{\mu}}_{\text{9}}=0.9$}& ${\it{\mu}}_{\text{6}}=1$ \\ 
\hline
$\lambda (x)$& \multicolumn{1}{c|}{\begin{tabular}[c]{@{}c@{}} $\lambda_{2}=0.4353$\\$\lambda_{22}=0.2640$\\$\lambda_{23}=0.0519$\\$\lambda_{250}=0.1646$\\ $\lambda_{300}=0.0842$\end{tabular}} &
\begin{tabular}[c]{@{}c@{}}$\lambda_{2}=0.6318$\\ $\lambda_{19}=0.0357$\\ $\lambda_{20}=0.2848$\\ $\lambda_{60}=0.0476$\end{tabular} \\ 
\hline
${\text{(SNR)}^{\it{\ast}}_{\text{dB}}}$ & \multicolumn{1}{c|}{2.86} & 6.20 \\ 
\hline
${\text{(Capacity)}_{\text{dB}}}$ & \multicolumn{1}{c|}{2.82}  & 6.12  \\ 
\hline
\end{tabular}
\end{table}

\section{Conclusion}
This paper presents a random multiplexing framework that decouples the multiplexing process from channel matrix dependencies, thereby enabling its application to arbitrary norm-bounded and spectrally convergent channel environments. By employing a random transformation, the random multiplexing builds an equivalent input-isotropic channel matrix in the random transform domain, facilitating the asymptotic replica MAP-BER optimality of AMP-type detectors. To further enhance efficiency, a CD-MAMP detector is considered, which effectively leverages the sparsity of time-domain channels and the input isotropy of the transform-domain channels, significantly reducing computational complexity without compromising detection performance. Furthermore, optimal power allocation strategies are developed to minimize the MAP BER and maximize the constrained channel capacity, complemented by the establishment of optimal coding principles and the constrained capacity optimality of CD-MAMP. The random multiplexing framework demonstrates remarkable versatility across diverse wireless applications, delivering advantages in spectral efficiency, and compatibility with established multiplexing techniques. Numerical results show that in correlated time-varying multipath MIMO channels, random multiplexing achieves BER and BLER performance improvements of up to $2\sim10$ dB compared to OFDM/OTFS/AFDM schemes, including conventional multiplexing and SISO channel coding, under both uniform and optimized power allocation, as well as with optimal channel coding strategies.

Future work will explore: 1) Robustness against burst noise in practical channels, leveraging random multiplexing’s ability to convert such noise into equivalent white Gaussian noise; 2) Enhanced physical layer security, where the random transform matrix acts as an encryption key to thwart eavesdroppers; and 3) Applications to sparse regression coded linear systems, extending the theoretical framework of compressed random multiplexing to systems with structured sparsity. These avenues will further validate the versatility and advantages of random multiplexing in real-world communication scenarios.

\section*{Acknowledgment}  
The authors would like to thank Junjie Ma, Burak \c{C}akmak, Weihua Liu, Hao Yan, Ming Wang, Jiazhen Dong, Wenyi Zhang (Associate Editor), and the anonymous reviewers for their comments or discussions that have improved the quality of the manuscript greatly.

\appendices

\section{MAP Demodulation BER} \label{App:MAP_BER}
Given an arbitrary signal constellation $\mathcal{S}$, we consider that the received signal in scalar AWGN channels is given by
\begin{align}
    y = \sqrt{\rho}{x} + {z},
\end{align}
where the transmitted signal ${x}\in \mathcal{S}$, $\rho>0$ is the SNR parameter, and ${z} \sim \mathcal{CN}(0, 1)$ is an AWGN noise. As a result, when MAP detection is employed at the receiver, the estimated signal is $\hat{{x}}={\rm{arg max}}_{{x}}P({x} | {y})$, which is equivalent to ML detection when the signal prior probabilities are uniform, i.e. $\hat{{x}}={\rm{arg min}}_{{x}}||{y}-\sqrt{\rho}{x}||^2$. The union bound on the BER $Q_{\mathcal{S}}(\rho)$ is given by \cite{Goldsmith2005}:
\begin{align}
    Q_{\mathcal{S}}(\rho) \leq \frac{1}{|\mathcal{S}{\rm{log}}_2\mathcal{S}|}\sum_{i=1}^{|\mathcal{S}|}\sum_{j \neq i}N(x_i,x_j)Q\left(\frac{\sqrt{\rho}d_{ij}}{\sqrt{2}}\right),
\end{align}
where $N(x_i,x_j)$ denotes the number of bit errors that occur when $x_i$ is mistakenly detected as $x_j$, $d_{ij}$ denotes the Euclidean distance between the constellation points $x_i$ and $x_j$, and $Q(\cdot)$ denotes the right-tail probability of the standard Gaussian distribution. At high SNR, where the error probability is dominated by the minimum-distance error events, this union bound provides a tight approximation to the true MAP BER $Q_{\mathcal{S}}(\rho)$.

\section{Proof of Theorem \ref{The:PIM}} \label{App:PIM}
We can perform singular value decomposition on $\bm{A}\bm{\Xi}_{\mathrm{PI}}$. Then, Theorem \ref{The:PIM} reduces to the following lemma.
\begin{lemma}\label{The:PIM_2}
Consider a matrix $\bm{W} \in \mathbb{C}^{M \times N}$ with singular value decomposition $\bm{W} = \bm{U} \bm{\Sigma} \bm{V}^{\mr{H}}$, where:
\begin{itemize}
    \item  
    $\bm{U}$ is an $M \times M$ deterministic unitary matrix,
    \item 
    $\bm{\Sigma}$ is an $M \times N$ deterministic diagonal matrix satisfying $\|\bm{\Sigma}\|_{2} \lesssim 1$ and 
    \begin{align}
        \frac{\mr{tr}[(\bm{\Sigma}^{\mr{H}} \bm{\Sigma})^k]}{N} \rightarrow  \int \lambda^k \mu(\mr{d} \lambda),
    \end{align}
    for any fixed $k \in \mathbb{N}^*$ and a compactly supported probability distribution $\mu$ on $[0, \infty)$,
    \item $\bm{V}= \bm{D}^{\mr{H}}\bm{OPQ}$ is a unitary matrix, where $\bm{D}$, $\bm{O}$, $\bm{P}$, and $\bm{Q}$ are all $N \times N$ matrices defined as follows:
\begin{enumerate}
    \item $\bm{D} = \diag\big\{[\mr{e}^{\mr{i}\theta_1}, \cdots\!, \mr{e}^{\mr{i}\theta_N}]\big\}$ is a uniformly random phase matrix with $\theta_{1:N} \overset{\mr{i.i.d.}}{\sim} \mathsf{Unif}\big\{[0, 2\pi)\big\}$. 
    \item $\bm{O}$ is a deterministic unitary matrix satisfying $\|\bm{O}\|_{\max}\lesssim N^{-1/2+\epsilon}$ for any $\epsilon > 0$. 
    \item $\bm{P}$ is a uniformly random permutation matrix independent of $\bm{D}$.
    \item $\bm{Q}$ is a deterministic unitary matrix satisfying
    \begin{align} 
        \Big|\textstyle\sum_{i,j\in [N], i\neq j} \big[\bm{Q}(\bm{\Sigma}^{\mr{H}}\bm{\Sigma})^k\bm{Q}^{\mr{H}}\big]_{i, j}\Big| \lesssim N^{1/2+\epsilon},  \label{Eqn:con_Q}
    \end{align}
    for any fixed $k \in \mathbb{N}^*, \epsilon > 0$.
\end{enumerate}
\end{itemize}
Then, $\bm{W}$ lies in the universality class $\mathscr{U}$.
\end{lemma}
\begin{IEEEproof}
    See Appendix \ref{App:PIM_a}.
\end{IEEEproof}

\subsection{Proof of Lemma \ref{The:PIM_2}}\label{App:PIM_a}
We can write $\bm{W} = \bm{J}\bm{D}$, where $\bm{J}=\bm{U}\bm{\Sigma}\bm{Q}^{\mr{H}}\bm{P}^{\mr{H}}\bm{O}^{\mr{H}}$. Clearly, $\|\bm{J}\|_2 = \|\bm{\Sigma}\|_2 \lesssim 1$ and for any fixed $k \in \mathbb{N}^*$,
\begin{align}
    \frac{\mr{tr}[(\bm{J}^{\mr{H}}\bm{J})^k]}{N} = \frac{\mr{tr}[(\bm{\Sigma}^{\mr{H}} \bm{\Sigma})^k]}{N} \rightarrow  \int \lambda^k \mu(\mr{d} \lambda).
\end{align}
Hence, we only need to show that for any $k \in \mathbb{N}^*, \epsilon > 0$,
\begin{align}
    \Big\| (\bm{J}^{\mr{H}} \bm{J})^k -  \frac{ \mr{tr}[(\bm{J}^{\mr{H}} \bm{J})^k]}{N} \bm{I}_N \Big\|_{\max} \lesssim N^{-1/2+\epsilon},
\end{align}
where $(\bm{J}^{\mr{H}}\bm{J})^k = \bm{O}\bm{P}\bm{Q}(\bm{\Sigma}^{\mr{H}}\bm{\Sigma})^k\bm{Q}^{\mr{H}}\bm{P}^{\mr{H}}\bm{O}^{\mr{H}}$. Next, we introduce a crucial lemma as follows.
\begin{lemma}\label{Lem:con_ineq}
Let $\bm{M} \in \mathbb{C}^{N \times N}$ be a random matrix satisfying $\bm{M} = \bm{O}\bm{P}\bm{Q}\bm{\Lambda}\bm{Q}^{\mr{H}}\bm{P}^{\mr{H}}\bm{O}^{\mr{H}}$, where
\begin{itemize}
    \item $\bm{O}, \bm{Q} \in \mathbb{C}^{N \times N}$ are deterministic matrices.
    \item $\bm{P}$ is an $N \times N$ uniformly random permutation matrix.
    \item $\bm{\Lambda} \in \mathbb{C}^{N \times N}$ is a deterministic diagonal matrix.
\end{itemize}
Then, there exists universal constants $C$ such that 
\begin{align}
\begin{split}
    \mr{Pr}\bigg(&\big\|\bm{M}-\mathbb{E}[\bm{M}]\big\|_{\max} 
    > C \|\bm{Q}\|_2^2 \|\bm{\Lambda}\|_2 \|\bm{O}\|_2 \\
    &\Big( \|\bm{O}\|_{\max} \ln{N}
    + \|\bm{O}\|_2\sqrt{\ln{N}/N} \Big) \bigg) \leq \frac{8}{N^2}.
\end{split}
\end{align}
In particular, if $\bm{O}$ and $\bm{Q}$ are unitary matrices, then
\begin{align}
\begin{split}
    \mr{Pr}\bigg(&\big\|\bm{M}-\mathbb{E}[\bm{M}]\big\|_{\max} 
    > C \|\bm{\Lambda}\|_{2} \\
    &\Big(\|\bm{O}\|_{\max} \ln{N}
    + \sqrt{\ln{N}/N} \Big) \bigg) \leq \frac{8}{N^2}.
\end{split}
\end{align}
\end{lemma}
\begin{IEEEproof}
    See Appendix \ref{App:PIM_b}.
\end{IEEEproof}
By Lemma \ref{Lem:con_ineq}, for any fixed $k \in \mathbb{N}^*, \epsilon > 0$, 
\begin{align}
\begin{split}
     \mr{Pr}\bigg(&\big\|(\bm{J}^{\mr{H}} \bm{J})^k-\mathbb{E}[(\bm{J}^{\mr{H}} \bm{J})^k]\big\|_{\max} > C \|\bm{\Sigma}^{\mr{H}} \bm{\Sigma}\|_{2}^k \\
     &\Big(\|\bm{O}\|_{\max} \ln{N} + \sqrt{\ln{N}/N}\Big)\bigg) \leq \frac{8}{N^2}.
\end{split}
\end{align}
From the Borel-Cantelli lemma and the fact that $\|\bm{O}\|_{\max}\lesssim N^{-1/2+\epsilon}$,
\begin{align}\label{Eqn:con_6}
    \mr{Pr}\Big(\big\|(\bm{J}^{\mr{H}} \bm{J})^k-\mathbb{E}[(\bm{J}^{\mr{H}} \bm{J})^k]\big\|_{\max} \lesssim N^{-1/2+\epsilon}\Big) = 1.
\end{align}
Subsequently, we introduce the following lemma.
\begin{lemma}\label{Pro:mean_p}
Let $\bm{B} \in \mathbb{C}^{N \times N}$ be a deterministic matrix, $\bm{P}$ be an $N \times N$ uniformly random permutation matrix. Then,
\begin{align}
    \mathbb{E}[\bm{P}\bm{B}\bm{P}^{\mr{H}}] = \beta \bm{1}_N\bm{1}^{\mr{T}}_N + (\alpha - \beta) \bm{I}_N,
\end{align}
where $\bm{1}_N$ denotes the $N$-dimensional all-ones vector, $\alpha$ and $\beta$ are the averages of diagonal and off-diagonal entries of $\bm{B}$ respectively, i.e., 
\begin{align}
    & \alpha = \frac{\mr{tr}[\bm{B}]}{N},\ \beta = \frac{\textstyle\sum_{i,j \in [N], i \neq j} B_{i,j}}{N(N-1)}.
\end{align}    
\end{lemma}
\begin{IEEEproof}
    See Appendix \ref{App:PIM_c}.
\end{IEEEproof}
By Lemma \ref{Pro:mean_p}, we have
\begin{align}\label{Eqn:con_7}
    \mathbb{E}\big[(\bm{J}^{\mr{H}}\bm{J})^k\big] &= \bm{O}\,\mathbb{E}\big[\bm{P}\bm{Q}(\bm{\Sigma}^{\mr{H}}\bm{\Sigma})^k\bm{Q}^{\mr{H}}\bm{P}^{\mr{H}}\big]\,\bm{O}^{\mr{H}} \nonumber \\
    &= (\tilde{\alpha} - \tilde{\beta}) \bm{I}_N + \tilde{\beta}(\bm{O} \bm{1}_N)(\bm{O} \bm{1}_N)^{\mr{H}}, 
\end{align}
where
\begin{align}
    \tilde{\alpha} &= \frac{\mr{tr}[(\bm{J}^{\mr{H}}\bm{J})^k]}{N}, \\
    \tilde{\beta} &= \frac{\sum_{i,j \in [N], i \neq j} \big[\bm{Q}(\bm{\Sigma}^{\mr{H}}\bm{\Sigma})^k\bm{Q}^{\mr{H}}\big]_{i,j}}{N(N-1)}.
\end{align}
From \eqref{Eqn:con_Q}, we have
\begin{align}
    |\tilde{\beta}| \lesssim N^{-3/2+\epsilon} \quad \forall k \in \mathbb{N}^*,\ \epsilon > 0. \label{Eqn:con_8}
\end{align}
In addition,
\begin{align}
    \|\tilde{\beta}(\bm{O} \bm{1}_N)(\bm{O} \bm{1}_N)^{\mr{H}}\|_{\max} &= |\tilde{\beta}| \cdot \|\bm{O} \bm{1}_N\|_\infty^2 \nonumber \\
    &\leq |\tilde{\beta}| (N \|\bm{O}\|_{\max})^2 \nonumber \\
    &\lesssim N^{-1/2+\epsilon}. \label{Eqn:con_9}
\end{align}
From \eqref{Eqn:con_6}, \eqref{Eqn:con_7}, \eqref{Eqn:con_8} and \eqref{Eqn:con_9}, we obtain that
\begin{align}
    \mr{Pr}\bigg(\Big\|(\bm{J}^{\mr{H}} \bm{J})^k-\frac{\mr{tr}[(\bm{J}^{\mr{H}}\bm{J})^k]}{N}\bm{I}_N\Big\|_{\max} \lesssim N^{-1/2+\epsilon}\bigg) = 1,
\end{align}
for any fixed $k \in \mathbb{N}^*, \epsilon > 0$.
Finally, we take a union bound over $k \in \mathbb{N}^*$ and $\epsilon \in \mathbb{Q}_{+}$ to finish the proof.

\subsection{Proof of Lemma \ref{Lem:con_ineq}}\label{App:PIM_b}
We first state the following lemma, a straightforward complex-valued extension of \cite[Theorem 4.3]{bercu2015concentration}.
\begin{lemma}\label{Lemma:con_ineq_pre}
    For any $\bm{B} \in \mathbb{C}^{N \times N}$ and $t > 0$, we have
    \begin{align}
        \begin{split}
            \mr{Pr}\Big(&\big|T(\bm{B})-\mathbb{E}[T(\bm{B})]\big| \geq t \Big) \leq \\ &8\,\mr{exp}\bigg(-\frac{t^2}{32(\eta b_N + t\|\bm{B}\|_{\max}/3\sqrt{2})}\bigg)
        \end{split}
    \end{align}
    where 
    \begin{align}
        T(\bm{B}) \equiv &\sum_{\ell = 1}^N B_{\ell, \pi(\ell)}, \\
         \eta = \frac{5}{2}\ln{3}-\frac{2}{3}&,\ 
        b_N = \frac{1}{N}\|\bm{B}\|_{\mr{F}}^2,
    \end{align}
    where $\pi:[N] \to [N]$ is a uniformly random permutation.
\end{lemma}
\begin{IEEEproof}
    Let $c = T(\bm{B})-\mathbb{E}[T(\bm{B})] \in \mathbb{C}$. For any $t > 0$,
    \begin{align}
        \big\{|c| \geq t \big\} \subset \big\{|\Re(c)| \geq t/\sqrt{2} \big\} \cup \big\{|\Im(c)| \geq t/\sqrt{2} \big\}.
    \end{align}
    Hence, we have
    \begin{align}
        \begin{split}
            \mr{Pr}\big(|c|\geq t\big) \leq \; &\mr{Pr}\big(|\Re(c)|\geq t/\sqrt{2}\big) \\ &+ \mr{Pr}\big(|\Im(c)|\geq t/\sqrt{2}\big).
        \end{split}
    \end{align}
    The proof is then completed by applying \cite[Theorem 4.3]{bercu2015concentration} to $\Re(\bm{B})$ and $\Im(\bm{B})$, noting that their Frobenius and max norms are bounded by those of $\bm{B}$.
\end{IEEEproof}
Following Lemma \ref{Lemma:con_ineq_pre} with
\begin{align}
    t = 16\eta s \|\bm{B}\|_{\max} + \sqrt{256(\eta s \|\bm{B}\|_{\max})^2 + 32\eta b_N s},
\end{align}
and bounding $\|\bm{B}\|_{\max}/3\sqrt{2}$ by $\eta\|\bm{B}\|_{\max}$, we obtain
\begin{align}
    \mr{Pr}\Big(\big|T(\bm{B})-\mathbb{E}[T(\bm{B})]\big| \geq t \Big) \leq 8\,\mr{e}^{-s} \quad \forall s \geq 0.
\end{align}
Since
\begin{align}
    t \leq 32\eta \big(s \|\bm{B}\|_{\max}+\sqrt{b_N s}\big),
\end{align}
there exists explicit universal constants $C$ such that 
\begin{align}\label{Eqn:con_1}
\begin{split}
    \mr{Pr}\Big(&\big|T(\bm{B})-\mathbb{E}[T(\bm{B})]\big| \geq \\ C &\big(s \|\bm{B}\|_{\max}+\sqrt{s/N}\|\bm{B}\|_\mr{F}\big) \Big) \leq 8\,\mr{e}^{-s} \quad \forall s \geq 0.
\end{split}
\end{align}
Let $\bm{P}$ be the permutation matrix corresponding to $\pi$, i.e., 
\begin{align}
    P_{i,j} =
    \begin{cases}
        1, & \quad \text{if} \ \pi(i)=j \\
        0, & \quad \text{otherwise}
    \end{cases}\quad \forall i, j \in [N],
\end{align}
the entries of $\bm{M}$ can be expressed as 
\begin{align}
    M_{i, j} &= \sum_{\ell=1}^N O_{i, \ell} \sum_{v=1}^N P_{\ell, v} \big[\bm{Q}\bm{\Lambda}\bm{Q}^{\mr{H}}\bm{P}^{\mr{H}}\bm{O}^{\mr{H}}\big]_{v, j} \nonumber \\
    &= \sum_{\ell=1}^N O_{i, \ell} \big[\bm{Q}\bm{\Lambda}\bm{Q}^{\mr{H}}\bm{P}^{\mr{H}}\bm{O}^{\mr{H}}\big]_{\pi(\ell), j} \nonumber \\
    &= T(\bm{D}^{(i, j)}), \\
    \bm{D}^{(i, j)} &= \bm{u}^{(i)} (\bm{v}^{(j)})^{\mr{T}},
\end{align}
where $\bm{u}^{(i)}$ is the $i$-th row of $\bm{O}$ and $\bm{v}^{(j)}$ is the $j$-th column of $\bm{Q}\bm{\Lambda}\bm{Q}^{\mr{H}}\bm{P}^{\mr{H}}\bm{O}^{\mr{H}}$, i.e.,
\begin{align}
    \bm{u}^{(i)} = \bm{O}^{\mr{T}}\bm{e}_i,\  
    \bm{v}^{(j)} = \bm{Q}\bm{\Lambda}\bm{Q}^{\mr{H}}\bm{P}^{\mr{H}}\bm{O}^{\mr{H}}\bm{e}_j,
\end{align}
where $\bm{e}_i$ denotes the $i$-th standard basis vector. Next, we apply the concentration inequality in \eqref{Eqn:con_1} with $s = 4\ln{N}$ to obtain
\begin{align}
\begin{split}
    \mr{Pr}\bigg(&\Big|M_{i, j}-\mathbb{E}[M_{i, j}]\Big| \geq C \Big(4\|\bm{D}^{(i, j)}\|_{\max}\ln{N} + \\
    & 2\|\bm{D}^{(i, j)}\|_\mr{F}\sqrt{\ln{N}/N}\Big)\bigg) \leq 8N^{-4} \quad \forall i, j \in [N].
\end{split}
\end{align}
It can be verified that
\begin{align}
    \|\bm{D}^{(i, j)}\|_{\mr{F}} &= \|\bm{u}^{(i)}\|_2 \|\bm{v}^{(j)}\|_2, \\
    \|\bm{D}^{(i, j)}\|_{\max} &= \|\bm{u}^{(i)}\|_\infty 
    \|\bm{v}^{(j)}\|_\infty.
\end{align}
Utilizing the inequalities:
\begin{align}
    \|\bm{u}^{(i)}\|_2 &\leq \|\bm{O}\|_2,\  
    \|\bm{u}^{(i)}\|_\infty \leq \|\bm{O}\|_{\max}, \\
    \|\bm{v}^{(j)}\|_\infty &\leq \|\bm{v}^{(j)}\|_2 \leq \|\bm{Q}\|_2^2 \|\bm{\Lambda}\|_2  \|\bm{O}\|_2,
\end{align}
we obtain 
\begin{align}
\begin{split}
    \mr{Pr}\bigg(&\Big|M_{i, j}-\mathbb{E}[M_{i, j}]\Big| > C \|\bm{Q}\|_2^2 \|\bm{\Lambda}\|_2  \|\bm{O}\|_2  \\
    &\Big(\|\bm{O}\|_{\max} \ln{N} + \|\bm{O}\|_2 \sqrt{\ln{N}/N}\Big)\bigg) \leq 8N^{-4},
\end{split}
\end{align}
for any $i, j \in [N]$. Finally, we take a union bound over $i, j$ to finish the proof.

\subsection{Proof of Lemma \ref{Pro:mean_p}}\label{App:PIM_c}
Let $\pi:[N] \to [N]$ be the permutation corresponding to $\bm{P}$: 
\begin{align}
    P_{i,j} =
    \begin{cases}
        1, & \quad \text{if} \ \pi(i)=j \\
        0, & \quad \text{otherwise}
    \end{cases}\quad \forall i, j \in [N].
\end{align}
Hence, for any $i, j \in [N]$, 
\begin{align*}
    [\bm{P}\bm{B}\bm{P}^{\mr{H}}]_{i,j} = 
    \textstyle\sum_{k, \ell \in [N]} P_{i, k} B_{k, \ell} P_{\ell, j} = 
    B_{\pi(i), \pi(j)}.
\end{align*}
When $i = j$, the probability of $\pi(i)=k$ is $1/N$. As a result,
\begin{align*}
    \mathbb{E}[B_{\pi(i), \pi(i)}] = \frac{\textstyle\sum_{k=1}^N B_{k, k}}{N} = \frac{\mr{tr}[\bm{B}]}{N}.
\end{align*}
When $i \neq j$, for any $k \neq \ell$, the probability of $\pi(i)=k$ and $\pi(j)=\ell$ is $1/[N(N-1)]$. As a result, 
\begin{align*}
    \mathbb{E}[B_{\pi(i), \pi(j)}] = \frac{\textstyle\sum_{k, \ell \in [N], k \neq \ell}B_{k, \ell}}{N(N-1)}.
\end{align*}
Thus, we have completed the proof.

\section{Proof of Lemma \ref{Lem:hat_gamma_Cav}}\label{App:hat_gamma_Cav}
Our goal is to prove that
\begin{align}
    [\hat{\gamma}_{\mr{SE}}(v,\bm{p})]^{-1} = N\big[\textstyle\sum_{i=1}^N (v^{-1}+\sigma^{-2}\sigma_i^2 p_i)^{-1}\big]^{-1}
\end{align}
is concave w.r.t $\bm{p}$. Since $v > 0$ and $\sigma^{-2}\sigma_i^2 \geq 0$, it follows from the concavity of 
\begin{align}
    g_N(\hat{\bm{p}}_N) = \big[\textstyle\sum_{i=1}^N  \hat{p}_i^{-1}\big]^{-1},
\end{align}
where $\hat{\bm{p}}_N = [\hat{p}_i, \cdots\!, \hat{p}_N] $ with $\hat{p}_i \equiv v^{-1} + \sigma^{-2}\sigma_i^2 p_i > 0$. We prove the concavity of $g_N(\hat{\bm{p}}_N)$ by induction on $N$.
\begin{itemize}
    \item When $N=1$, $g_1(\hat{p}_1)=\hat{p}_1$ is concave.
        \item When $N=2$, the Hessian matrix of $g_2(\hat{p}_1, \hat{p}_2)$ is
            \begin{align}
            \bm{H}_{g_2} &= 
            \begin{bmatrix}
                \frac{\partial^2 g_2}{ \partial \hat{p}_1^2} & \frac{\partial^2 g_2}{ \partial \hat{p}_1 \partial \hat{p}_2} \\[2mm]
                \frac{\partial^2 g_2}{ \partial \hat{p}_2 \partial \hat{p}_1} &
                \frac{\partial^2 g_2}{ \partial \hat{p}_2^2}
            \end{bmatrix} \nonumber \\
            &= \frac{1}{(\hat{p}_1 + \hat{p}_2)^3}
            \begin{bmatrix}
                -2\hat{p}_2^2 & 2\hat{p}_1 \hat{p}_2 
                \\[2mm]
                2\hat{p}_1 \hat{p}_2 & -2\hat{p}_1^2
            \end{bmatrix}.
        \end{align}
    Since ${\rm tr}\{\bm{H}_{g_2}\} = -2(\hat{p}_1^2 + \hat{p}_2^2)/(\hat{p}_1 + \hat{p}_2)^3 < 0$ and ${\rm det}(\bm{H}_{g_2}) = 0$, the eigenvalues of $\bm{H}_{g_2}$ must be one zero and one non-positive number. Thus, $\bm{H}_{g_2}\preceq 0$, meaning that $g_2(\hat{p}_1, \hat{p}_2)$ is concave. 
    \item When $N \geq 2$, suppose that $g_N(\hat{\bf{p}}_N)$ is concave.
    Then, for any $0 \leq \alpha \leq 1, \bar{\alpha}=1-\alpha$, we have
    \BS
    \begin{align}
    & \alpha  g_{N+1}(\hat{\bf{p}}_{N+1}) + \bar{\alpha} g_{N+1}(\bf{q}_{N+1}) \nonumber\\
    &= \alpha  g_{2}\big(g_N(\hat{\bf{p}}_N), \hat{p}_{N+1}\big) + \bar{\alpha} g_{2}\big(g_N(\bf{q}_N), q_{N+1}\big)  \nonumber\\
    &\mathop \leq \limits^{(a)}  g_2\big(\alpha g_N(\hat{\bf{p}}_N) + \bar{\alpha} g_N(\bf{q}_N),\, \alpha \hat{p}_{N+1}+ \bar{\alpha}  q_{N+1}\big) \nonumber\\
    &\mathop  \leq \limits^{(b)}  g_2\big(g_N(\alpha \hat{\bf{p}}_N+ \bar{\alpha} \bf{q}_N\big), \alpha  \hat{p}_{N+1}+\bar{\alpha} q_{N+1})\big) \nonumber\\
    &\!\! = g_{N+1}( \alpha  \hat{\bf{p}}_{N+1} + \bar{\alpha} \bf{q}_{N+1}), \nonumber
    \end{align}
    \ES
    where $\hat{\bf{p}}_N = [\hat{p}_1, \cdots\!, \hat{p}_N]$, $\bf{q}_N = [q_1, \cdots\!, q_N]$, $(a)$ follows from the concavity of $g_2$, and $(b)$ follows from the monotonic increase of $g_2(\hat{p}_1, \hat{p}_2) = (\hat{p}_1^{-1}+\hat{p}_2^{-1})^{-1}$ w.r.t. $\hat{p}_1$. Thus, $g_{N+1}(\hat{\bf{p}}_{N+1})$ is concave. 
    \end{itemize}    
    Thus, we have completed the proof.

\section{Proof of Lemma \ref{Lem:hat_gamma_inv_Cvex}}\label{App:hat_gamma_inv_Cvex}

   Following \eqref{Eqn:iterSEb}, we have 
\begin{align}
    \tilde{v} = \hat{\gamma}_{\rm SE}(v,\bm{p}) = \tfrac{1}{N} \textstyle\sum_{i=1}^N (v^{-1} + \tilde{p}_i)^{-1}, 
\end{align}
Let $z=v^{-1}$ and $\hat{v}= N\tilde{v}$. Therefore, $[\hat{\gamma}^{-1}_{\rm SE}(\tilde{v},\bm{p})]^{-1}$ can be formulated as
\BE
    z = [\hat{\gamma}^{-1}_{\rm SE}(\tilde{v},\bm{p})]^{-1} = g^{-1}(\hat{v}, \tilde{\bf{p}}),
\EE
where $g^{-1}(\hat{v}, \tilde{\bf{p}})$ is the inverse function of $g(z, \tilde{\bf{p}})$ w.r.t. $z$, given by
\BE
    \hat{v} = g(z, \tilde{\bf{p}}) =  \textstyle\sum_{i=1}^N (z + \tilde{p}_i)^{-1}.
\EE
It then suffices to prove the convexity of $g^{-1}(\hat{v}, \tilde{\bf{p}})$.

The following equation is obtained by taking the partial derivative of both sides of $ g^{-1}(\hat{v}, \tilde{\bf{p}})-z = 0$ w.r.t. $\tilde{p}_i$, treating $z$ as a variable independent of $\tilde{\bm p}$.
    \begin{align}
        \frac{\partial \left[g^{-1}(\hat{v},\tilde{\bm{p}})\right]}{\partial \tilde{p}_i}=0,
    \end{align}
    i.e., 
    \begin{align}
        \frac{\partial g^{-1}(\hat{v}, \tilde{\bf{p}})}{\partial \hat{v}}\cdot \frac{\partial \hat{v}}{\partial \tilde{p}_i}+\frac{\partial g^{-1}(\hat{v}, \tilde{\bf{p}})}{\partial \tilde{p}_i}=0.\label{equ33}
    \end{align}
    The relationship between the first-order partial derivatives of an inverse function and its original function is given by the following equation: \begin{align}\label{Eqn:d_g_inv_v}
        \frac{\partial g^{-1}(\hat{v},\tilde{\bm{p}})}{\partial \hat{v}}=\frac{1}{\frac{\partial g(z,\tilde{\bm{p}})}{\partial z}}=-\frac{1}{\sum^N_{k=1}\frac{1}{(\tilde{p}_k+z)^2}},
    \end{align}
    where $z=g^{-1}(\hat{v},\tilde{\bm{p}})$ is a function w.r.t. $\hat{v}$ and $\tilde{\bm{p}}$. Following \eqref{Eqn:d_g_inv_v}, Equation (\ref{equ33}) can be written as
    \begin{align}
        \frac{1}{\frac{\partial g(z,\tilde{\bm{p}})}{\partial z}}\cdot \frac{\partial \hat{v}}{\partial \tilde{p}_i}+\frac{\partial g^{-1}(\hat{v}, \tilde{\bf{p}})}{\partial \tilde{p}_i}=0.\label{equ35}
    \end{align}
    Therefore, we have     \begin{align}\label{Eqn:pd_g_inv}
        \frac{\partial g^{-1}(\hat{v}, \tilde{\bf{p}})}{\partial \tilde{p}_i}=-\frac{1}{\frac{\partial g(z, \tilde{\bf{p}})}{\partial z}}\cdot \frac{\partial \hat{v}}{\partial \tilde{p}_i}  =-\frac{\frac{1}{(\tilde{p}_i+z)^2}}{\textstyle\sum^N_{k=1}\frac{1}{(\tilde{p}_k+z)^2}}.
    \end{align}
    The second order partial derivatives are expressed as follows:
    \begin{align}
       & \frac{\partial^2 g^{-1}(\hat{v},\bf{p})}{\partial \tilde{p}^2_i}\notag\\
       &=\tfrac{1}{\Omega^2_2}\Big\{2a^3_i(1\!+\!\tfrac{\partial z}{\partial \tilde{p}_i})\Omega_2\!-a^2_i\Big[\textstyle\sum\limits_{k\neq i}\!{(2a^3_k\frac{\partial z}{\partial \tilde{p}_i})}\!+2a^3_i(1\!+\!\frac{\partial z}{\partial \tilde{p}_i})\!\Big]\!\Big\}\notag\\
        &=\tfrac{1}{\Omega^2_2}\Big\{2a^3_i(\Omega_2-a^2_i)-a^2_i\Big[\textstyle\sum\limits_{k=1}^N{2a^3_k\tfrac{\partial z}{\partial \tilde{p}_i}}+2a^3_i\Big]\Big\}\notag\\
        &\overset{(a)}{=}\tfrac{1}{\Omega_2^2}\left[2a^3_i(\Omega_2-a^2_i)+a^2_i(\tfrac{2a^2_i\Omega_3}{\Omega_2}-2a^3_i)\right]\notag\\
        &=\tfrac{2}{\Omega^3_2}(a^4_i\Omega_3-2a^5_i\Omega_2+a^3_i\Omega^2_2),
    \end{align} 
    where $a_i=\frac{1}{\tilde{p}_i+z}$, $\Omega_2=\sum_{i=1}^{N}{a^2_i}$,  $\Omega_3=\sum_{i=1}^{N}{a^3_i}$, and (a) follows $z=g^{-1}(\hat{v},\bf{p})$ and \eqref{Eqn:pd_g_inv}. Similarly, we have
    \begin{align}
        \frac{\partial^2 g^{-1}(\hat{v},\bf{p})}{\partial \tilde{p}_i\partial \tilde{p}_j}=\tfrac{2}{\Omega_2^3}\left[a^2_ia^2_j\Omega_3-(a^3_ia^2_j+a^2_ia^3_j)\Omega_2\right].
    \end{align} 
    The convexity of $g^{-1}(\hat{v},\bf{p})$ can be established by proving that the Hessian matrix $\nabla_{\bf{p}}^2 g^{-1}(\hat{v},\bf{p})$ is semi-positive definite. Given that $\Omega_2=\sum_{i=1}^{N}{a^2_i}>0$, the task reduces to proving the semi-positive definiteness of $\frac{\Omega^3_2}{2}\nabla_{\bf{p}}^2 g^{-1}(\hat{v},\bf{p})$, which can be shown by verifying the non-negativity of its quadratic form, i.e., $\forall \bf{x}\in \mathbb{R}^N$,    
    \begin{align}
        &\bm{x}^T \big[\tfrac{\Omega^3_2}{2}\nabla_{\bf{p}}^2 g^{-1}(\hat{v},\bf{p})\big]\bm{x}\notag\\ 
        &=\textstyle\sum\limits_{i=1}^N{(a^3_ix^2_i\Omega^2_2-2a^5_ix^2_i\Omega_2+a^4_ix^2_i\Omega_3)}\notag\\ 
        &\hspace{2.4cm} +\textstyle\sum\limits_{i\neq j}\left[a^2_ia^2_jx_ix_j\Omega_3-(a^3_ia^2_j+a^2_ia^3_j)x_ix_j\Omega_2\right]\notag\\       &=\textstyle\sum\limits_{i=1}^N{a^3_ix^2_i\Omega^2_2}+\textstyle\sum\limits_{i,j\leq N}{\left[a^2_ia^2_jx_ix_j\Omega_3-(a^3_ia^2_j+a^2_ia^3_j)x_ix_j\Omega_2\right]}\notag\\
        &=\textstyle\sum\limits_{i=1}^N{a^3_ix^2_i}(\textstyle\sum\limits_{j=1}^N{a^2_j})^2+\textstyle\sum\limits_{i,j\leq N}{a^2_ia^2_jx_ix_j}\textstyle\sum\limits_{k=1}^N{a^3_k} \notag\\ 
        &\hspace{3.9cm} -\textstyle\sum\limits_{i,j\leq N}{(a^3_ia^2_j+a^2_ia^3_j)x_ix_j}\textstyle\sum\limits_{k=1}^N{a^2_k}\notag\\ &=\textstyle\sum\limits_{i=1}^N{a^3_ix^2_i}\textstyle\sum\limits_{j=1}^N{a^2_j}\textstyle\sum\limits_{k=1}^N{a^2_k}+\textstyle\sum\limits_{i,j,k\leq N}{a^3_ka^2_ia^2_jx_ix_j}\notag\\ 
        &\hspace{4.2cm} -\textstyle\sum\limits_{i,j,k\leq N}{a^2_k(a^3_ia^2_j+a^2_ia^3_j)x_ix_j}\notag\\
        &=\textstyle\sum\limits_{i,j,k\leq N}{a^3_ia^2_ja^2_kx^2_i}+\textstyle\sum\limits_{i,j,k\leq N}{a^3_ia^2_ja^2_kx_kx_j}\notag\\ 
        &\hspace{2cm} -\textstyle\sum\limits_{i,j,k\leq N}{(a^3_ia^2_ja^2_k)x_ix_j}-\textstyle\sum\limits_{i,j,k\leq N}{(a^2_ia^3_ja^2_k)x_ix_j}\notag\\
        &=\textstyle\sum\limits_{i,j,k\leq N}{a^3_ia^2_ja^2_kx^2_i}+\textstyle\sum\limits_{i,j,k\leq N}{a^3_ia^2_ja^2_kx_kx_j}\notag\\ 
        &\hspace{2cm} -\textstyle\sum\limits_{i,j,k\leq N}{(a^3_ia^2_ja^2_k)x_ix_j}-\textstyle\sum\limits_{i,j,k\leq N}{(a^3_ia^2_ja^2_k)x_ix_k}\notag\\
        &=\textstyle\sum\limits_{i,j,k\leq N}{(a^3_ia^2_ja^2_k)(x_j-x_i)(x_k-x_i)}\notag\\
        &=\textstyle\sum\limits_{i=1}^N{a^3_i}\cdot\textstyle\sum\limits_{j=1}^N{a^2_j(x_j-x_i)}\cdot\textstyle\sum\limits_{k=1}^N{a^2_k(x_k-x_i)}\notag\\
        &=\textstyle\sum\limits_{i=1}^N{a^3_i}\Big[\textstyle\sum\limits_{j=1}^N{a^2_j(x_j-x_i)}\Big]^2\geq 0.
    \end{align}
    Thus, we have completed the proof.

\section{Proof of Theorem \ref{The:Opt_P}} \label{App:Opt_P}
Let $\bf{H}=\bm{U}_H \bm{\Sigma}_{H} \bm{V}_H^{\rm H}$ and $\bm{P} = \bm{U}_P \bm{\Sigma}_P \bm{V}_P^{\rm H}$ denote the singular value decomposition of $\bm{H}$ and $\bm{P}$, respectively. The proof is completed by establishing the following two lemmas.
\begin{lemma}\label{Lem:Opt_P_a}
    Choosing $\bm{U}_P = \bm{V}_H$ with arbitrary unitary $\bm{V}_P$ minimizes the MAP BER of the power-allocated system with random multiplexing in \eqref{Eqn:GPA}.
\end{lemma}
\begin{IEEEproof}
    See Appendix \ref{App:Opt_P_a}.
\end{IEEEproof}
\begin{lemma}\label{Lem:Opt_P_b}
     Choosing $\bm{U}_P = \bm{V}_H$ with arbitrary unitary $\bm{V}_P$ maximizes the constrained capacity of the power-allocated system with random multiplexing in \eqref{Eqn:GPA}. 
\end{lemma}
\begin{IEEEproof}
    See Appendix \ref{App:Opt_P_b}.
\end{IEEEproof}

\subsection{Proof of Lemma \ref{Lem:Opt_P_a}} \label{App:Opt_P_a}
We formulate the problem as
\BS\begin{align}
     {\mathcal P}_{3.0}:  \;\;   &\mathop{\rm{min}}_{\bm{P}}  \;\;  Q_{\mathcal{S}}\big(\rho^*(\bm{P})\big), \\
     & {\rm s.t.}\quad {\rm tr}\{\bm{P}\bm{P}^{\rm H}\}=P_{\rm sum}.
\end{align}\ES
Since the MAP demodulation function $Q_{\mathcal{S}}(\cdot)$ is monotonically decreasing, it is reduced to
\BS\begin{align}
     {\mathcal P}_{3}:  \;\;  &\mathop{\rm{max}}_{\bm{P}}  \;\;  \rho^*(\bm{P}), \\
     & {\rm s.t.}\quad {\rm tr}\{\bm{P}\bm{P}^{\rm H}\}=P_{\rm sum},
\end{align}\ES
where $\rho^*$ can be obtained by the first fixed point of the SE of CD-OAMP/VAMP. The SE is presented as
\BS\begin{align} 
 \rho_t^{\gamma} &= {\gamma}_{\mr{SE}}(v_t^{\phi},\bm{P})=\big[\hat{\gamma}_{\rm SE}(v_t^{\phi},\bm{P})\big]^{-1} - [v_t^{\phi}]^{-1}, \\
v_{t+1}^{\phi}&={\phi}_{\mr{SE}}(\rho_t^{\gamma})=\big(\big[{\rm mmse}(\rho_t^{\gamma})\big]^{-1} - \rho_t^{\gamma}\big)^{-1},
\end{align}
where
\begin{align}
    & \hat{\gamma}_{\rm SE}(v_t^{\phi},\bm{P}) \nonumber \\
    &= \tfrac{1}{N}\mr{tr}\big\{\big([v_t^{\phi}]^{-1}\bf{I}+\sigma^{-2}\bf{P}^{\mr{H}}\bm{H}^{\rm H}\bm{H}\bf{P}\big)^{-1}\big\} \\
    &= \tfrac{1}{N}\mr{tr}\big\{\!\big([v_t^{\phi}]^{-1}\bm{I}+\sigma^{-2}\bm{H}\bm{P}\bm{P}^{\rm H}\bm{H}^{\rm H}\big)^{\!-1}\!\big\} + \tfrac{N-M}{N}v_t^{\phi}\\
    &= \underbrace{\tfrac{1}{N}\mr{tr}\big\{\big([v_t^{\phi}]^{-1}\bm{I}+\sigma^{-2}{\bm{\Sigma}}_{H}\bm{Q}{\bm{\Sigma}}_{H}^{\rm H}\big)^{-1}\big\}  + \tfrac{N-M}{N}v_t^{\phi}}_{\hat{\gamma}(v_t^{\phi}, \bm{Q})}, \label{Eqn:gamma_Q}
\end{align}\ES 
where $\hat{\gamma}(v_t^{\phi}, \bm{Q})$ denotes the posterior variance w.r.t $\bm{Q}$, and 
\begin{align}\label{Eqn:def_Q}
    \bm{Q} \equiv \bm{V}_H^{\rm H}\bm{P}\bm{P}^{\mr{H}}\bm{V}_H = \bm{V}_H^{\rm H}\bm{U}_P \bm{\Sigma}_P^2 \bm{U}_P^{\rm H}\bm{V}_H. 
\end{align}
As a result, Problem ${\mathcal P}_{3}$ can be rewritten as
\BS\begin{align}
     {\mathcal P}_{3.1}:  \;\;  &\mathop{\rm{max}}_{\bm{Q}}  \;\;  \rho^*(\bm{Q}), \\
     &{\rm s.t.}\quad {\rm tr}\{\bm{Q}\}=P_{\rm sum}.
\end{align}\ES
Arbitrary $\bm{V}_P$ is optimal since $\bm{Q}$ is independent to $\bm{V}_P$. Next, we introduce a crucial lemma proposed in \cite{EST-EQ}. 
\begin{lemma}[Posterior Variance \cite{EST-EQ}]\label{Lem:diag_Q}
    For $v > 0$,
    \begin{align}
        \hat{\gamma}(v, \bm{Q}_{\rm diag}) \leq \hat{\gamma}(v, \bm{Q}),
    \end{align}
    where $\hat{\gamma}(\cdot)$ is given in \eqref{Eqn:gamma_Q}, and $\bm{Q}_{\rm diag}$ denotes the diagonal of $\bm{Q}$ with zeros elsewhere.
\end{lemma}
\begin{IEEEproof}
    Building on the proof in \cite{EST-EQ}, we provide the omitted convexity proof of $\hat{\gamma} (v, \bm{Q})$ w.r.t $\bm{Q}$. See Appendix \ref{App:diag_Q} for details.
\end{IEEEproof}
From Lemma \ref{Lem:diag_Q} and the fact that ${\rm tr}\{\bm{Q}\} = {\rm tr}\{\bm{Q}_{\rm diag}\}$, it is clear that choosing $\bm{U}_P=\bm{V}_H$ and arbitrary $\bm{V}_P$ is optimal to minimize the MAP BER of the system in \eqref{Eqn:GPA}.
 
\subsection{Proof of Lemma \ref{Lem:Opt_P_b}}\label{App:Opt_P_b}
Following $\bm{Q} \equiv \bm{V}_H^{\rm H}\bm{P}\bm{P}^{\mr{H}}\bm{V}_H$ in \eqref{Eqn:def_Q}, we formulate the problem as
\BS\label{Eqn:C_MIMO_Q}
\begin{align}
      {\mathcal P}_{4}:  \;\;   &\mathop{\rm{max}}_{\bm{Q}}  \;\;  C_{\rm MIMO}(\bm{Q}),\\
     &{\rm s.t.}\quad {\rm tr}\{\bm{Q}\}=P_{\rm sum},
\end{align}
\ES
where 
\BS\label{Eqn:C_Q}
\begin{align}
    C_{\rm MIMO} (\bm{Q}) &= \int_{0}^1 \min\big\{\eta(v, {\bm Q}), {\rm mmse}^{-1}(v)\big\} d v, \\
    \eta(v,\bm{Q}) &\equiv v^{-1}-\big[\hat{\gamma}^{-1}(v, \bm{Q})\big]^{-1}.
\end{align}
\ES 
where $C_{\rm MIMO} (\bm{Q})$ is from \cite[Theorem 1]{LeiOptOAMP}, $\hat{\gamma}(v, \bm{Q})$ is given in \eqref{Eqn:gamma_Q}. Arbitrary $\bm{V}_P$ is optimal since $\bm{Q}$ is independent to $\bm{V}_P$. Next, we introduce an important lemma.
\begin{lemma}\label{Lem:diag_Q_inv}
    For $v > 0$,
    \begin{align}
        \hat{\gamma}^{-1}(v, \bm{Q}_{\rm diag}) \geq \hat{\gamma}^{-1}(v, \bm{Q}),
    \end{align}
    where $\hat{\gamma}^{-1}(v, \bm{Q})$ denotes the inverse function of $\hat{\gamma}(\tilde{v}, \bm{Q})$ in \eqref{Eqn:gamma_Q} w.r.t $\tilde{v}>0$.
\end{lemma}
\begin{IEEEproof}
    For $\tilde{v} > 0$, $\hat{\gamma}(\tilde{v}, \bm{Q})$ is always positive and strictly increasing w.r.t $\tilde{v}$, i.e.,
    \begin{align}
        \hat{\gamma}(\tilde{v}_1, \bm{Q}) > \hat{\gamma}(\tilde{v}_2, \bm{Q}) > 0 \ \text{if and only if} \  \tilde{v}_1 > \tilde{v}_2 > 0.
    \end{align}
    Furthermore, following Lemma \ref{Lem:diag_Q}, we have
    \begin{align}
        \hat{\gamma}(\tilde{v}, \bm{Q}_{\rm diag}) \leq \hat{\gamma}(\tilde{v}, \bm{Q}).
    \end{align}
    Thus, for any $v > 0$,
    \begin{align}
        \hat{\gamma}(\tilde{v}_1, \bm{Q}_{\rm diag}) = \hat{\gamma}(\tilde{v}_2, \bm{Q}) = v \quad \text{only if} \quad \tilde{v}_1 \geq \tilde{v}_2,
    \end{align}
    which means that $\hat{\gamma}^{-1}(v, \bm{Q}_{\rm diag}) \geq \hat{\gamma}^{-1}(v, \bm{Q})$. Thus, we completed the proof. 
\end{IEEEproof}
From Lemma \ref{Lem:diag_Q_inv} and \eqref{Eqn:C_Q}, we have $C_{\rm MIMO} (\bm{Q}_{\rm diag}) \geq C_{\rm MIMO} (\bm{Q})$, where ${\rm tr}\{\bm{Q}\} = {\rm tr}\{\bm{Q}_{\rm diag}\}$ always holds. Note that the entries in $\bm{Q}_{\rm diag}$ are the same as those in $\bm{\Sigma}^2_{P}$. Thus, choosing $\bm{U}_P = \bm{V}_H$ with $\bm{V}_P$ arbitrary is optimal to maximize the constrained capacity of the system in \eqref{Eqn:GPA}.

\subsection{Proof of Lemma \ref{Lem:diag_Q}}\label{App:diag_Q}
Let $g(v, \bm{Q}) = \mr{tr}\big\{\big(v^{-1}\bm{I}+\sigma^{-2}{\bm{\Sigma}}_{H}\bm{Q}{\bm{\Sigma}}_{H}^{\rm H}\big)^{-1}\big\}$. It is equivalent to prove that $g(v, \bm{Q}_{\rm diag}) \leq g(v, \bm{Q})$. First, we show the convexity of $g(v, \bm{Q})$ w.r.t $\bm{Q}$, which is omitted in \cite{EST-EQ}.
\begin{lemma}[convexity]\label{Pro:conv_f}
    For $v > 0$ and Hermitian positive semidefinite matirx $\bm{Q}$,
    \begin{align}
        g(v, \bm{Q}) = \mr{tr}\big\{(v^{-1}\bm{I}+\sigma^{-2}{\bm{\Sigma}}_{H}\bm{Q}{\bm{\Sigma}}_{H}^{\rm H})^{-1}\big\}
    \end{align}
    is convex w.r.t $\bm{Q}$.
\end{lemma}
\begin{IEEEproof}
    Let $\bm{B} = v^{-1}\bm{I}+\sigma^{-2}{\bm{\Sigma}}_{H}\bm{Q}{\bm{\Sigma}}_{H}^{\rm H}$, which preserves convexity since it is an affine function of $\bm{Q}$. Thus, we only need to show that ${\rm tr}\{\bm{B}^{-1}\}$ is convex w.r.t Hermitian positive definite $\bm{B}$. It is known that $x^{-1}$ is a operator convex function w.r.t $x \in (0, \infty)$ \cite{simon2019loewner}, i.e., for any Hermitian positive definite $\bm{B}_1$ and $\bm{B}_2$, and $\alpha \in [0, 1]$,
    \begin{align}
        [\alpha \bm{B}_1 + (1-\alpha \bm{B}_2)]^{-1} \preceq \alpha \bm{B}_1^{-1} + (1-\alpha) \bm{B}_2^{-1}.
    \end{align}
    Subsequently, we have
    \begin{align}
        {\rm tr}\{[\alpha \bm{B}_1 + (1-\alpha \bm{B}_2)]^{-1}\} \leq {\rm tr}\{\alpha \bm{B}_1^{-1} + (1-\alpha) \bm{B}_2^{-1}\},
    \end{align}
    which implies that ${\rm tr}\{\bm{B}^{-1}\}$ is convex w.r.t Hermitian positive definite $\bm{B}$. Hence, we completed the proof.
\end{IEEEproof}

Next, the proof is the same as that in \cite{EST-EQ}. Let $\bm{J}_1, \cdots\!, \bm{J}_T$ denotes all the possible $M \times M$ diagonal matrices with diagonal entries of $1$ or $-1$, where $T = 2^M$. For $1 \leq i \leq T$, it can be verified that 
\begin{align}
    g(v, \bm{Q}) = g(v, \bm{J}_i\bm{Q}\bm{J}_i).
\end{align}
Moreover, from \cite{wang2009worst}, we have
\begin{align}
    \bm{Q}_{\rm diag} = \frac{1}{T} \textstyle\sum\limits_{i=1}^T \bm{J}_i\bm{Q}\bm{J}_i.
\end{align}
Therefore,
\begin{align}
    g(v, \bm{Q}) &= \frac{1}{T} \textstyle\sum\limits_{i=1}^T g(v, \bm{J}_i\bm{Q}\bm{J}_i) \\
    &\mathop \geq \limits^{(a)} g\Big(v, \frac{1}{T} \textstyle\sum\limits_{i=1}^T\bm{J}_i\bm{Q}\bm{J}_i\Big) \\
    &= g(v, \bm{Q}_{\rm diag}),
\end{align}
where (a) holds due to Lemma \ref{Pro:conv_f} and Jensen’s inequality. Hence, we have completed the proof.

\bibliographystyle{IEEEtran}
\bibliography{reference}

\end{document}